\documentclass[useAMS,usenatbib,usegraphicx]{mn2e}
\usepackage{graphicx}

\usepackage[breaklinks,colorlinks,citecolor=black,linkcolor=black,urlcolor=black]{hyperref}

\newcommand*{\mysub}[2]{\ensuremath{#1_{\mathrm{#2}}}}
\newcommand*{\unit}[1]{\ensuremath{\mathrm{\, #1}}}
\newcommand*{\phmin}{\hspace{1.9ex}} 

\newcommand*{\Msun}{\ensuremath{\, M_{\odot}}}
\newcommand*{\keV}{\unit{keV}}
\newcommand*{\cm}{\unit{cm}}
\newcommand*{\kpc}{\unit{kpc}}
\newcommand*{\second}{\unit{s}}

\newcommand*{\Mgas}{\mysub{M}{gas}}
\newcommand*{\fgas}{\mysub{f}{gas}}
\newcommand*{\Lce}{\mysub{L}{ce}}
\newcommand*{\Lcore}{\mysub{L}{core}}
\newcommand*{\Yx}{\mysub{Y}{X}}

\newcommand*{\LCDM}{\ensuremath{\Lambda}CDM}
\newcommand*{\Omegam}{\mysub{\Omega}{m}}
\newcommand*{\Omegal}{\mysub{\Omega}{\Lambda}}
\newcommand*{\rhocr}{\mysub{\rho}{cr}}

\newcommand*{\E}[1]{\ensuremath{\times 10^{#1}}}
\newcommand*{\ltsim}{\ {\raise-.75ex\hbox{$\buildrel<\over\sim$}}\ }
\newcommand*{\gtsim}{\ {\raise-.75ex\hbox{$\buildrel>\over\sim$}}\ }
\newcommand*{\proptosim}{\ {\raise-.75ex\hbox{$\buildrel\propto\over\sim$}}\ }

\newcommand*{\Chandra}{{\it Chandra}}
\newcommand*{\Planck}{{\it Planck}}

\defcitealias{Mantz1502.06020}{I}
\newcommand*{\morphpaper}{\citetalias{Mantz1502.06020}}
\defcitealias{Mantz1402.6212}{II}
\newcommand*{\cosmopaper}{\citetalias{Mantz1402.6212}}
\defcitealias{Applegate1509.02162}{IV}
\newcommand*{\calpaper}{\citetalias{Applegate1509.02162}}
\defcitealias{Navarro9611107}{NFW}
\newcommand*{\NFW}{\citetalias{Navarro9611107}}

\begin{document}

\title[Relaxed Clusters: Profiles and Scaling Relations]{Cosmology and Astrophysics from Relaxed Galaxy Clusters III: Thermodynamic Profiles and Scaling Relations}

\author[A. B. Mantz et al.]{A. B. Mantz,$^{1,2,3,4}$\thanks{E-mail: \href{mailto:amantz@slac.stanford.edu}{\tt amantz@slac.stanford.edu}} {} 
  S. W. Allen,$^{3,4,5}$
  R. G. Morris,$^{3,5}$
  R. W. Schmidt$^6$\\
  $^1$Department of Astronomy and Astrophysics, University of Chicago, 5640 South Ellis Avenue, Chicago, IL 60637, USA\\
  $^2$Kavli Institute for Cosmological Physics, University of Chicago, 5640 South Ellis Avenue, Chicago, IL 60637, USA\\
  $^3$Kavli Institute for Particle Astrophysics and Cosmology, Stanford University, 452 Lomita Mall, Stanford, CA 94305, USA\\
  $^4$Department of Physics, Stanford University, 382 Via Pueblo Mall, Stanford, CA 94305, USA\\
  $^5$SLAC National Accelerator Laboratory, 2575 Sand Hill Road, Menlo Park, CA 94025, USA\\
  $^6$Astronomisches Rechen-Institut, Zentrum f\"ur Astronomie der Universit\"at Heidelberg, M\"onchhofstrasse 12-14, D-69120 Heidelberg, Germany
}
\date{Accepted 2015 December 07}

\pagerange{\pageref{firstpage}--\pageref{lastpage}} \pubyear{2016}
\maketitle
\label{firstpage}

\begin{abstract}
  This is the third in a series of papers studying the astrophysics and cosmology of massive, dynamically relaxed galaxy clusters. Our sample comprises 40 clusters identified as being dynamically relaxed and hot (i.e., massive) in Papers~\morphpaper{} and \cosmopaper{} of this series. Here we consider the thermodynamics of the intracluster medium, in particular the profiles of density, temperature and related quantities, as well as integrated measurements of gas mass, average temperature, total luminosity and center-excluded luminosity. We fit power-law scaling relations of each of these quantities as a function of redshift and cluster mass, which can be measured precisely and with minimal bias for these relaxed clusters using hydrostatic arguments. For the thermodynamic profiles, we jointly model the density and temperature and their intrinsic scatter as a function of radius, thus also capturing the behavior of the gas pressure and entropy. For the integrated quantities, we also jointly fit a multidimensional intrinsic covariance. Our results reinforce the view that simple hydrodynamical models provide a good description of relaxed clusters outside their centers, but that additional heating and cooling processes are important in the inner regions (radii $r \ltsim 0.5\,r_{2500} \approx 0.15\,r_{500}$). The thermodynamic profiles remain regular, with small intrinsic scatter, down to the smallest radii where deprojection is straightforward ($\sim20$\,kpc); within this radius, even the most relaxed systems show clear departures from spherical symmetry. Our results suggest that heating and cooling are continuously regulated in a tight feedback loop, allowing the cluster atmosphere to remain stratified on these scales.
\end{abstract}


\section{Introduction}
\label{sec:intro}

Highly dynamically relaxed galaxy clusters represent a minority of the cluster population, but provide a key laboratory for learning about the physics of the intracluster medium (ICM) and its interaction with cluster galaxies and active galactic nuclei (AGN). Uniquely in these morphologically regular systems, the three dimensional, radial profiles of the thermodynamic properties of the ICM can be reconstructed with minimal uncertainties from projection effects. In addition, X-ray data can provide precise constraints on the mass profiles of these clusters under the assumption of hydrostatic equilibrium. The latter feature comes at the expense of introducing some systematic uncertainty due to non-thermal support, but both simulations and direct calibration using weak gravitational lensing data show that the overall bias in \Chandra{} X-ray mass estimates for relaxed clusters is small ($\ltsim 10$ per cent; \citealt{Nagai0609247, Applegate1509.02162}).

Constraints on mass profiles, in comparison with estimates of the mass within a single characteristic radius, enable a wider range of investigations. In general, they provide a way to identify dynamically comparable radii across clusters, yielding the most meaningful comparison of interesting features. More specifically, they allow the scaling relations of thermodynamic quantities to be investigated as a function of radius. Violations of the self-similar scaling predicted from purely gravitational spherical collapse shed light on the astrophysics of the ICM, in particular the heating mechanism that prevents classical cooling flows from forming in relaxed clusters. While numerous studies have investigated the scaling relations of global cluster properties (for a review, see \citealt{Giodini1305.3286}), there has not yet been a complete study of the scaling of cluster thermodynamic profiles, including the radial structure of their intrinsic scatter (though see \citealt{Sayers1211.1632}).

Other papers in this series have focused on identifying a sample of massive, relaxed clusters \citep{Mantz1502.06020}, exploiting measurements of their gas mass fractions at intermediate radii to constrain cosmology \citep{Mantz1402.6212}, and measuring the average bias of X-ray hydrostatic masses for these clusters (\citealt{Applegate1509.02162}; hereafter, respectively, Papers~\morphpaper{}, \cosmopaper{} and \calpaper{}). Here we assume a concordance \LCDM{} cosmology (with parameters $h=0.7$, $\Omegam=0.3$ and $\Omegal=0.7$) and present the thermodynamic profiles of our relaxed cluster sample, their scaling properties, and some related astrophysical results. We briefly review aspects of our sample selection and spectral analysis methods in Section~\ref{sec:data}. Section~\ref{sec:methods} discusses the additional statistical methodology used in this work, in particular for performing multivariate regression with intrinsic scatter and/or arbitrary measurement error distributions. In Section~\ref{sec:profiles}, we present the thermodynamic profiles of the cluster sample and discuss some of the immediate consequences of these measurements that are available without further model fitting. Section~\ref{sec:scaling} presents scaling relations of the profiles, including the radial structure of their intrinsic scatter, as well as ``traditional'' scaling relations of globally measured quantities. We summarize these results in Section~\ref{sec:conclusions}.

In this paper, we follow the convention of defining characteristic cluster masses and radii in terms of the critical density of the Universe at the cluster's redshift,
\begin{equation} \label{eq:MDelta}
  M_\Delta = \frac{4}{3} \pi \Delta \rhocr(z) r_\Delta^3,
\end{equation}
where $\Delta$ is often referred to as the ``overdensity''. With this definition, the self-similar predictions for the scaling of the thermodynamic ICM properties can be written \citep{Kaiser1986MNRAS.222..323K}
\begin{eqnarray}
  \label{eq:selfsim}
  n(r_\Delta) &\propto& E(z)^2, \\
  kT(r_\Delta) &\propto& \left[\Delta^{1/2} E(z) M_\Delta\right]^{2/3}, \nonumber
\end{eqnarray}
where $n$ and $kT$ are the number density and temperature of the ICM. Here $E(z) = H(z)/H_0$ is the Hubble parameter $\left(\propto \sqrt{\rhocr(z)}\right)$, normalized to its value at $z=0$ for convenience. Note that, using $\Delta$ as a surrogate for cluster radius, the shape of the self-similar density profile is completely specified by the expression above, whereas the shape of the temperature profile is determined by the shape of the underlying mass profile. Expressions for the self-similar scaling of pressure, $P=nkT$, and pseudo-entropy, $K=kT/n^{2/3}$, follow directly from those above, as do the predicted scalings of quantities integrated within a given overdensity (for which we drop the explicit dependence on $\Delta$):
\begin{eqnarray}
  \label{eq:selfsimbulk}
  \Mgas &\propto& M_\Delta, \\
  kT &\propto& \left[E(z) M_\Delta\right]^{2/3}, \nonumber\\
  L &\propto& E(z)^{2+2\alpha/3} M_\Delta^{1+2\alpha/3}. \nonumber
\end{eqnarray}
Here and throughout this work, we define $L$ to be the intrinsic rest-frame \emph{soft-band} X-ray luminosity of the ICM (specifically, energies of 0.1--2.4\,keV). For the $kT > 5$\,keV clusters in our data set, the soft-band luminosity within a fixed radius has a relatively mild dependence on temperature, $L\propto (kT)^\alpha$ with $\alpha\approx -0.13$. This is in contrast to the bolometric luminosity ($\mysub{\alpha}{bol}\approx 1/2$), for which the corresponding self-similar scaling is $L_\mathrm{bol} \propto E(z)^{7/3} M_\Delta^{4/3}$.\footnote{Note that the $2/3$ terms in the exponents of the $L$--$M_\Delta$ relation are identified with the slope of the $kT$--$M_\Delta$ relation, making these two scaling relations less than independent where departures from self-similarity are concerned (see discussion by \citealt{Maughan1212.0858}).}

Where we employ log-normal probability distributions in this work, these are defined as Gaussian in the \emph{natural} log of the argument, such that their standard deviations can be approximately thought of as a fractional scatter.

\section{Data Set and Spectral Analysis}
\label{sec:data}
\label{sec:spectral}

This work employs the same sample of 40 massive, dynamically relaxed clusters used to constrain cosmological models in Paper~\cosmopaper{}. The identification of this sample is described in Papers~\morphpaper{} and \cosmopaper{}. Briefly, we searched the \Chandra{} archive for morphologically regular clusters, as defined by a suite of image statistics designed specifically for the task. These measurements probe the sharpness of the cluster surface brightness peak, the alignment of a series of standard isophotes with one another, and the symmetry of those isophotes about a globally defined cluster center. In addition, a minimum temperature of 5\,keV (excluding the core) is required.\footnote{This temperature threshold is motivated by the expectation from simulations that the bias and scatter introduced by the assumption of hydrostatic equilibrium should increase for low-mass clusters and groups, even after selecting the most relaxed systems \citep{Nagai0609247}. Simulations also predict a mass dependence of the gas mass fraction for less massive clusters, which would complicate the cosmological analysis of Paper~\cosmopaper{}.} The cluster sample is summarized in Table~\ref{tab:sample}.

\begin{table*}
  \centering
  \caption[]{
    Names, redshifts and masses of the relaxed clusters in our sample (see Papers~\morphpaper{} and \cosmopaper{}). Masses are from this work, using an updated \Chandra{} calibration compared with Paper~\cosmopaper{}. Also shown are the clean exposure time from this re-processing, as well as the specific \Chandra{} observations (OBSIDs) used.
  }
  \label{tab:sample}
  \begin{tabular}{lcr@{ $\pm$ }lr@{ $\pm$ }lrc}
    \hline
    Cluster & $z$ & \multicolumn{2}{c}{$r_{2500}$} &\multicolumn{2}{c}{$M_{2500}$} & exp. & OBSIDs \\
    & & \multicolumn{2}{c}{ (kpc)} & \multicolumn{2}{c}{ ($10^{14}\Msun$)} & (ks) & \\
    \hline
    Abell~2029 & 0.078 & 648 & 4 & 4.17 & 0.08 & 118.9 & 891,4977,6101,10434,10435,10436,10437 \\
    Abell~478 & 0.088 & 620 & 8 & 3.69 & 0.15 & 129.4 & 1669,6102,6928,6929,7217,7218,7222,7231,7232,7233,7234,7235 \\
    PKS~0745$-$191 & 0.103 & 671 & 7 & 4.74 & 0.14 & 148.8 & 2427,6103,7694,12881 \\
    RX~J1524.2$-$3154 & 0.103 & 483 & 9 & 1.77 & 0.10 & 40.9 & 9401 \\
    Abell~2204 & 0.152 & 689 & 13 & 5.41 & 0.30 & 90.1 & 499,6104,7940 \\
    RX~J0439.0+0520 & 0.208 & 496 & 15 & 2.14 & 0.19 & 35.5 & 527,9369,9761 \\
    Zwicky~2701 & 0.214 & 472 & 7 & 1.85 & 0.08 & 111.3 & 3195,7706,12903 \\
    RX~J1504.1$-$0248 & 0.215 & 699 & 15 & 6.02 & 0.38 & 39.9 & 4935,5793 \\
    Zwicky~2089 & 0.235 & 452 & 12 & 1.66 & 0.13 & 47.0 & 7897,10463 \\
    RX~J2129.6+0005 & 0.235 & 558 & 13 & 3.13 & 0.22 & 36.7 & 552,9370 \\
    RX~J1459.4$-$1811 & 0.236 & 554 & 17 & 3.07 & 0.29 & 39.6 & 9428 \\
    Abell~1835 & 0.252 & 657 & 7 & 5.21 & 0.16 & 183.6 & 496,6880,6881,7370 \\
    Abell~3444 & 0.253 & 549 & 11 & 3.04 & 0.19 & 35.7 & 9400 \\
    MS~2137.3$-$2353 & 0.313 & 476 & 10 & 2.11 & 0.13 & 50.9 & 928,5250 \\
    MACS~J0242.5$-$2132 & 0.314 & 532 & 31 & 2.96 & 0.52 & 7.7 & 3266 \\
    MACS~J1427.6$-$2521 & 0.318 & 449 & 16 & 1.79 & 0.19 & 41.3 & 3279,9373 \\
    MACS~J2229.7$-$2755 & 0.324 & 471 & 15 & 2.08 & 0.20 & 25.8 & 3286,9374 \\
    MACS~J0947.2+7623 & 0.345 & 608 & 15 & 4.58 & 0.34 & 49.0 & 2202,7902 \\
    MACS~J1931.8$-$2634 & 0.352 & 573 & 12 & 3.86 & 0.24 & 104.0 & 3282,9382 \\
    MACS~J1115.8+0129 & 0.355 & 561 & 14 & 3.63 & 0.27 & 44.3 & 3275,9375 \\
    MACS~J1532.8+3021 & 0.363 & 554 & 11 & 3.53 & 0.20 & 102.4 & 1649,1665,14009 \\
    MACS~J0150.3$-$1005 & 0.363 & 432 & 16 & 1.67 & 0.18 & 26.1 & 11711 \\
    MACS~J0011.7$-$1523 & 0.378 & 526 & 16 & 3.07 & 0.29 & 49.2 & 3261,6105 \\
    MACS~J1720.2+3536 & 0.391 & 529 & 20 & 3.18 & 0.35 & 51.7 & 3280,6107,7718 \\
    MACS~J0429.6$-$0253 & 0.399 & 537 & 36 & 3.35 & 0.67 & 19.3 & 3271 \\
    MACS~J0159.8$-$0849 & 0.404 & 617 & 18 & 5.12 & 0.45 & 62.6 & 3265,6106,9376 \\
    MACS~J2046.0$-$3430 & 0.423 & 427 & 16 & 1.73 & 0.20 & 42.8 & 5816,9377 \\
    IRAS~09104+4109 & 0.442 & 518 & 19 & 3.17 & 0.35 & 69.0 & 10445 \\
    MACS~J1359.1$-$1929 & 0.447 & 463 & 27 & 2.27 & 0.40 & 54.2 & 5811,9378 \\
    RX~J1347.5$-$1145 & 0.451 & 809 & 28 & 12.19 & 1.28 & 206.5 & 506,507,3592,13516,13999,14407 \\
    3C~295 & 0.460 & 452 & 22 & 2.15 & 0.31 & 90.9 & 578,2254 \\
    MACS~J1621.3+3810 & 0.461 & 501 & 13 & 2.92 & 0.23 & 129.3 & 3254,6109,6172,7720,9379,10785 \\
    MACS~J1427.2+4407 & 0.487 & 477 & 19 & 2.60 & 0.32 & 50.8 & 6112,9380,9808,11694 \\
    MACS~J1423.8+2404 & 0.539 & 482 & 11 & 2.86 & 0.19 & 122.7 & 1657,4195 \\
    SPT-CL~J2331$-$5051 & 0.576 & 425 & 22 & 2.05 & 0.32 & 31.8 & 9333,11738 \\
    SPT-CL~J2344$-$4242 & 0.596 & 568 & 29 & 5.01 & 0.78 & 10.7 & 13401 \\
    SPT-CL~J0000$-$5748 & 0.702 & 420 & 34 & 2.29 & 0.56 & 28.4 & 9335 \\
    SPT-CL~J2043$-$5035 & 0.723 & 385 & 14 & 1.81 & 0.19 & 73.3 & 13478 \\
    CL~J1415+3612 & 1.028 & 316 & 11 & 1.44 & 0.15 & 349.1 & 4163,12255,12256,13118,13119 \\
    3C~186 & 1.063 & 330 & 13 & 1.71 & 0.21 & 213.8 & 3098,9407,9408,9774,9775 \\
 \hline
  \end{tabular}
\end{table*}

Our procedure for reducing and cleaning of the \Chandra{} data is described in detail in Paper~\cosmopaper{}. The only difference is that in this work we employ a more recent version of the \Chandra{} analysis software and calibration files (specifically {\sc ciao}\footnote{\url{http://cxc.harvard.edu/ciao/}
} 4.6.1 and {\sc caldb}\footnote{\url{http://cxc.harvard.edu/caldb/}} 4.6.2). This update, from {\sc caldb} 4.4.10 to 4.6.2, spans a relatively important change to the model of the contaminant affecting the \Chandra{}-ACIS detectors that, at some level, affects the analysis of observations as far back as 2004.\footnote{\url{http://cxc.cfa.harvard.edu/caldb/downloads/Release_notes/CALDB_v4.5.9.html}} The update is, however, minor during the period when most of our data were observed. Comparing results from both versions of the analysis, we find a small ($5\pm5$ per cent) reduction in temperatures on average, and negligible change in density measurements.\footnote{Note that the cosmological results presented in Paper~\cosmopaper{} are insensitive to an overall bias in temperature, since in that work the overall normalization of cluster mass profiles is calibrated using weak gravitational lensing data (see also Paper~\calpaper{}). Subsequent CALDB updates as of this writing (through 4.6.10a) should have no effect on our analysis, given that the data employed here were all observed prior to 2013 (see Paper~\morphpaper).}

Paper~\cosmopaper{} also describes in detail our procedure for analysing X-ray spectra for the clusters in our sample. We perform deprojections of the intracluster medium under two sets of assumptions, respectively using the {\sc nfwmass} and {\sc projct} models in {\sc xspec}.\footnote{\url{http://heasarc.gsfc.nasa.gov/docs/xanadu/xspec/}} In the first analysis, we assume that the cluster mass (dark plus baryonic) is described by a \citet[][hereafter NFW]{Navarro9611107} profile. The ICM is modeled as a series of concentric, isothermal shells, and is assumed to be in hydrostatic equilibrium. This analysis provides simultaneous constraints on the mass, temperature and gas density profiles of a given cluster, but has the disadvantage of over-constraining the temperature profile. That is, the shape and especially the precision of the resulting temperature profile is largely driven by the shape of the surface brightness profile, which has much higher signal-to-noise than the temperature information available intrinsically in the spectra. Consequently, we use these results only for determining cluster mass profiles, as well as the covariance between mass and temperature measurements (see below).

For the thermodynamic measurements which are the focus of this work, we instead use the ``non-parametric'' analysis described in Paper~\cosmopaper{}. This analysis also assumes spherical symmetry, but does not require hydrostatic equilibrium. The results therefore contain no information about the cluster mass, but by the same token they are independent of assumptions regarding the mass profiles, and hence do not over-constrain the temperature profiles. Note that there is negligible difference between the gas density profiles recovered from the two fitting methods compared to the uncertainties.

The annuli in which we extract spectra to analyse are chosen to provide good signal-to-noise for measurements of the gas density. To constrain temperatures, a more demanding task, it is necessary to assume isothermality across groups of a few adjacent spherical shells. The effective resolution of our temperature profiles is thus lower than that of our density profiles from this analysis.\footnote{We also leave metallicities free, although these must be binned even more coarsely than temperatures.} Figure~\ref{fig:allprofiles} in the appendix shows the temperature and density profiles measured for each of our clusters, from both of the analyses described above. As already mentioned, the temperature profiles can be more finely binned when hydrostatic equilibrium and a form for the mass profile are assumed. Note that the good agreement between temperature profiles from the two analyses verifies that the \NFW{} profile-hydrostatic equilibrium model is a good description of the data, within measurement uncertainties.

Our primary results are based on the profiles shown in Figure~\ref{fig:allprofiles}. However, we also consider integrated (i.e. not radially resolved) thermodynamic measurements in Section~\ref{sec:bulk}. For these results, we use the same apparatus as the non-parametric model described above, but with some specializations. Specifically, this model has only two free temperatures, corresponding to the cluster volumes at radii $<0.15\,r_{500}$ and $>0.15\,r_{500}$, with $r_{500}$ estimated from the \NFW{} fit. The temperature in the outer radial bin is essentially identical to what we would obtain from a typical analysis of the spectrum in projection, but this approach allows the measurement covariance between this ``center-excised'' temperature, the projected luminosity, and the spherically integrated gas mass to be fully captured.

In Paper~\cosmopaper{}, we identified for each cluster a central region that was excluded from the \NFW{} model fits. By default, this was a circle of radius 50\,kpc, although in some cases the excluded region is larger in order to encompass visible structure in the ICM.\footnote{ Point sources are always masked, regardless of their position within the cluster.} The motivation for excluding these data from the fit is to avoid biases due to local violations of hydrostatic equilibrium, gas inhomogeneities, asphericity, and any offset between the densest gas and the center used for deprojection (which was chosen based on the large-scale emission). Equilibrium is not a concern for our non-parametric fits, but the other issues above can potentially still bias the results of a spherical deprojection at small radii. In particular, we generically expect non-zero ellipticity or an offset of the deprojection center from the brightest pixel to result in an underestimate of the gas densities and/or overestimate of the temperatures at the smallest radii. In the following sections, we will present results from non-parametric fits which extend all the way into the cluster centers alongside (more robust) results where the central regions are discarded. Fits to the thermodynamic profiles will always use center-excluded data.

In Paper~\cosmopaper{}, we calibrated our X-ray mass measurements to a standard provided by the Weighing the Giants weak lensing analysis \citep{von-der-Linden1208.0597, Kelly1208.0602, Applegate1208.0605}. As detailed further in Paper~\calpaper{}, this correction factor is consistent with unity ($0.96 \pm 0.08$), and is consistent with being constant as a function of mass. For simplicity, we therefore do not apply any lensing-based correction to the X-ray \NFW{} masses in this work.

\section{Additional Methodology}
\label{sec:methods}

At various points in the following sections, we fit model profiles and trends with mass and/or redshift. This section summarizes the approaches we employ in these analyses.

In Section~\ref{sec:bulk}, we constrain the multivariate scaling relations of X-ray luminosity, temperature and gas mass as a function of total mass and redshift for our sample. For this analysis, we approximate the measurement uncertainties on all quantities as log-normal, and the intrinsic scatter as a multi-dimensional log-normal distribution, so that the power law model for the scaling relations becomes a linear model in the transformed (logarithmic) variables. Covariances among the measured quantities are accounted for (see below), as well as covariances in the intrinsic scatter. Our fitting method is inspired by the work of \citet{Kelly0705.2774}, who proposed an efficient Gibbs sampling algorithm for fitting a similar model with only one response variable. An associated paper \citep{Mantz1509.00908} describes our implementation of a multivariate version of the \citet{Kelly0705.2774} algorithm, which is publically available as a package for {\sc r}\footnote{ \url{http://www.r-project.org}} called {\sc lrgs}.\footnote{ \url{https://github.com/abmantz/lrgs}} Note that generic Markov Chain Monte Carlo algorithms have significant difficulty navigating the complex parameter space of the elements of a covariance matrix, at least in our experience, making the Gibbs sampling approach of {\sc lrgs} potentially useful for multivariate regression in general.

The thermodynamic results that we present in this paper stem from the non-parametric spectral analysis described in the previous section. The uncertainties from that analysis directly reflect the statistical power of the data. However, in order to constrain scaling relations with mass for the entire sample of clusters, most of which do not have weak lensing data, we must use mass measurements from the hydrostatic \NFW{} analysis. This introduces a dilemma without a perfect solution: we would like to use masses from the \NFW{} analysis but, to avoid spuriously tight constraints, temperatures from the non-parametric analysis. At the same time, it is crucial to account for the covariance between mass and temperature introduced by the hydrostatic assumption, i.e.\ the fact that statistical fluctuations in temperatures also affect the measured masses. We deal with this issue for each cluster as follows. First, we identify posterior samples of total mass from the \NFW{} analysis, sorted into increasing order, with samples of the other X-ray observables from the non-parametric analysis sorted into increasing-temperature order. At this point, the mass--temperature correlation coefficient from the list of samples is unity. We then randomly permute elements of the mass list until the mass--temperature covariance in this hybrid list of samples matches the mass--temperature covariance internal to the \NFW{} fit (as measured directly from the \NFW{} posterior samples). The result of this procedure is a set of samples with marginal temperature variance matching that of the non-parametric fit, marginal mass variance matching the \NFW{} fit, and with an orientation of the mass--temperature error ellipse that also matches the \NFW{} fit. The internal covariance of gas mass, temperature and luminosity is unaffected by this process. Note that we implicitly assume negligible mass--$\Mgas$ and mass--luminosity correlations at fixed radius; certainly we do not expect strong correlations to be introduced by the hydrostatic assumption apart from that between mass and temperature (given the use of a soft-band luminosity; see Section~\ref{sec:intro}).

In Section~\ref{sec:profscale}, we simultaneously fit a model for the mean scaled cluster density and temperature profiles, the intrinsic scatter between these profiles and internal to each as a function of radius, and the overall scaling of each profile with mass and redshift. The model is similar to the ``Gaussian process'' pressure profile model used by \citet{Sayers1211.1632}, consisting of a value for the mean profile at a set of scaled radii, the intrinsic covariance among profiles at those radii, and overall scalings with mass/redshift.\footnote{ Pedantically speaking, neither work employs a Gaussian \emph{process}, since the models are parametrized at a finite number of points, between which we interpolate. Nevertheless, we perpetuate this incorrect terminology in the hope that future work will move beyond these relatively limited, discrete models.} We expand on the \citet{Sayers1211.1632} approach by marginalizing over uncertainty in the mass profile of each cluster. We use the same approach as above to account for the mass--temperature covariance, but in this case the mass affects not just the scaling of the density and temperature, but also the scaling of the ordinate that the mean profile is a function of, i.e.\ $\Delta$ or $r/r_{2500}$. As above, the intrinsic and measurement covariances are modelled as log-normal (now a function of radius). In practice, we interleave Metropolis steps for the global scaling parameters (the exponents in the scaling relations) with {\sc lrgs} Gibbs samples of the mean profiles and intrinsic covariance.

\section{Thermodynamic Profiles}
\label{sec:profiles}

\begin{figure*}
  \centering
  \includegraphics[scale=0.85]{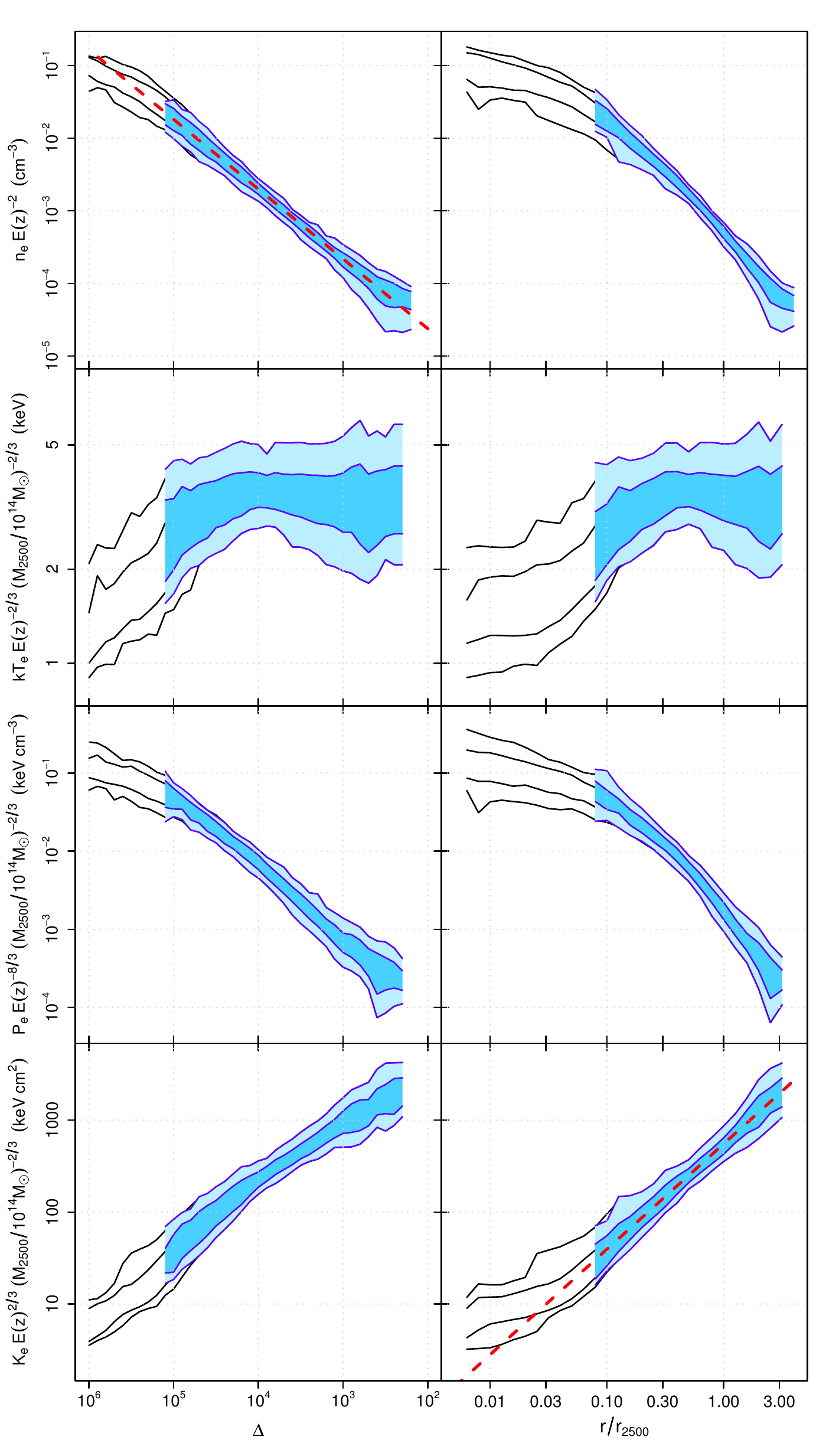}
  \caption[]{
    Profiles of electron number density, temperature, pressure and entropy, scaled according to the \citet{Kaiser1986MNRAS.222..323K} model, as a function of overdensity and scaled radius. The 68.3 and 95.4 per cent confidence regions at each radius are shown, where these probabilities encompass both measurement uncertainties and intrinsic scatter among clusters in our sample. Blue lines and shading show the results when excluding the central region of each cluster (as defined in Section~\ref{sec:spectral}), while the results shown with black lines include the centers (where we expect larger systematic uncertainties due asphericity). The two sets of results do not match precisely at all radii where they overlap because the central exclusion radius corresponds to different values of $\Delta$ or $r/r_{2500}$ from cluster to cluster; hence, the two profiles differ at radii where only a subset of clusters contributes to the center-excluded results. In the upper-left and lower-right panels, the dashed lines are power-laws fit to the center-excised data, $n\propto\Delta^{0.96}$ and $K\propto(r/r_{2500})^{1.15}$.
  }
  \label{fig:Qprofiles}
\end{figure*}

To summarize the results of our non-parametric analysis for the entire cluster sample, Figure~\ref{fig:Qprofiles} shows ``ensemble'' profiles of electron density, temperature, pressure ($nkT$) and entropy ($kTn^{-2/3}$), after applying scalings from the \citet{Kaiser1986MNRAS.222..323K} model. Dark and light shaded regions show, at each radius, the 68.3 and 95.4 per cent limits of the distribution of results for all clusters that provide data at that radius, reflecting both measurement uncertainty and intrinsic scatter. These figures thus represent the results in a minimally processed form, whereas estimates of the intrinsic, cluster-to-cluster scatter of the profiles require us to adopt and fit a model (see Section~\ref{sec:profscale}). The shaded part of each profile shows the results when the central region of each cluster is excluded, while unshaded lines show the continuation of the profiles into the cluster centers (see Section~\ref{sec:data}). Note that the entire cluster sample contributes to the ensemble at overdensities $\Delta > 10^3$; however, the number of clusters where we can confidently measure temperatures drops quickly at larger radii, and results for $\Delta \ltsim 500$ should be particularly treated with caution.

In Figure~\ref{fig:Qprofiles}, the thermodynamic profiles are plotted against two ordinates: the overdensity (Equation~\ref{eq:MDelta}), and the physical radius in units of $r_{2500}$, both determined from the best fitting \NFW{} profile for each cluster. The former is a natural choice for examining the scaling relations given in Equation~\ref{eq:selfsim}, which in the self-similar model would hold at all $\Delta$, and their intrinsic scatter.

The following subsections describe some of the qualitative features of these thermodynamic profiles, and compare them to results in the literature.

\subsection{Density and Surface Brightness}
\label{sec:density}

The self-similar assumption of a constant gas-mass fraction with radius leads to a simple, mass-independent scaling for gas density at a given overdensity, $n \propto E(z)^2 \Delta$. Figure~\ref{fig:Qprofiles} shows that this relation approximately holds over a wide range in $\Delta$; the dashed line in the top-left panel corresponds to a power law $n E^{-2} \propto \Delta^{0.96}$, fit to the median center-excised profile. Qualitatively, we see that the self-similar expectation is approximately met over most of the volume probed, in particular in the center-excised region. The departure from a precise $\Delta^{1.0}$ scaling can be seen more clearly in the $\fgas = \Mgas/M$ profiles presented in Paper~\cosmopaper{} (Figures~2 and 3), specifically as a shallow increase of \fgas{} with radius. 

\begin{figure*}
 \centering
 \includegraphics[scale=0.9]{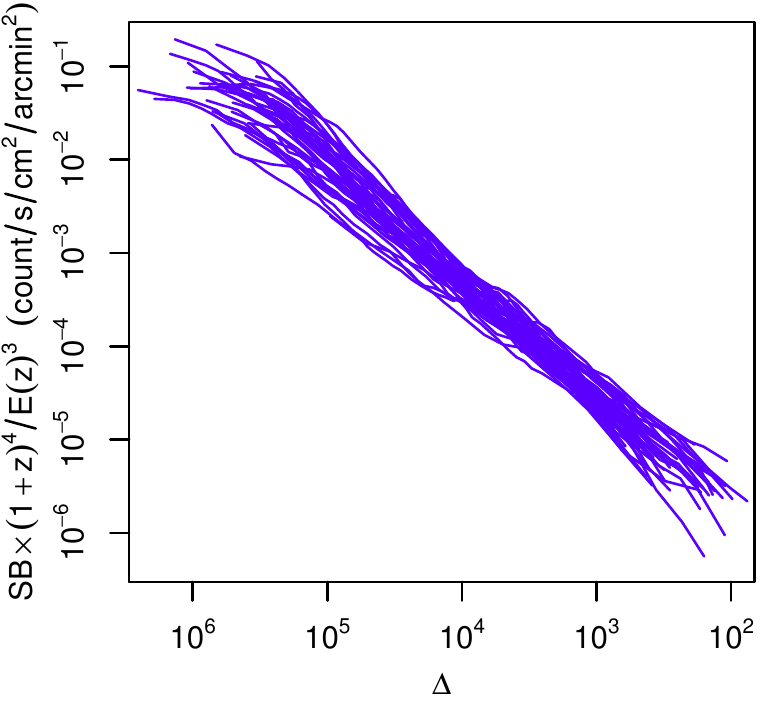}
 \hspace{1cm}
 \includegraphics[scale=0.9]{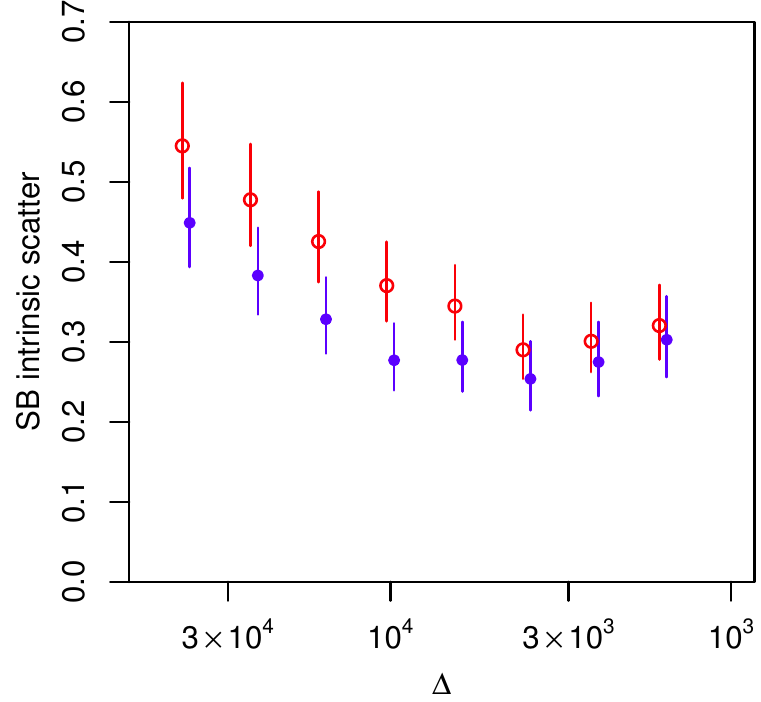}
 \caption[]{
   Left: surface brightness in the 0.6--2.0\,keV band for each of our clusters, scaled with redshift as indicated. For clarity, measurement uncertainties are not shown. Right: intrinsic scatter in surface brightness at fixed overdensity. Open circles show the scatter when the profiles are scaled only by the redshift-dependent factor $E(z)^3/(1+z)^4$, as in the left panel, while filled circles show the slightly smaller scatter achievable by including the temperature-dependent scaling terms.
 }
 \label{fig:sbscat}
\end{figure*}

The intrinsic scatter in scaled density profiles has immediate consequences for mass proxies based on the soft X-ray luminosity, the most straightforward quantity to estimate from shallow data. To demonstrate, we extract (observer frame) 0.6--2.0\,keV band surface brightness profiles for each cluster. The expected self-similar scaling for surface brightness at fixed $\Delta$ is $S \propto kT\,K(z,T)E(z)^3/(1+z)^4$ (e.g., Paper~\morphpaper{}), where $K$ is the redshift- and temperature-dependent K-correction. Adopting this scaling, we fit a mean profile and log-normal intrinsic scatter (including covariance terms) as a function of $\Delta$. The results are shown in Figure~\ref{fig:sbscat}. We find a log-normal intrinsic scatter of $\sim 0.25$--0.3 at intermediate radii, $10^4 \gtsim \Delta \gtsim 10^3$, with increasing scatter at smaller radii. Despite the relatively large scatter, the simplicity of the mean model, which is well approximated by a power law, is appealing; in particular, these results imply that constraints on the shape of a cluster's mass profile can in principle be obtained from surface brightness data only, by statistically associating isophotes with overdensities (as Figure~\ref{fig:sbscat}  shows, neglecting the temperature-dependent components of the scaling increases the intrinsic scatter only marginally).\footnote{ Strictly speaking, this is true only of clusters that are morphologically equivalent to those in our sample, although we expect the surface brightness template for relaxed clusters to roughly hold for the general population at intermediate-to-large radii, where earlier studies have shown approximate self-similarity of the density profiles (e.g.\ \citealt{Vikhlinin0507092, Croston0801.3430}; Paper~\morphpaper{}).} In practice, the relatively large scatter at fixed $\Delta$ makes this approach less effective than traditional X-ray proxies when one only wants to estimate a total mass, which is usually the case.

\newpage

It has long been known that luminosities measured excluding cluster centers (by some reasonable definition) have a smaller intrinsic scatter with mass than the total luminosity (e.g., \citealt{Fabian1994MNRAS.267..779F, Markevitch9802059, Maughan0703504, Mantz0909.3099}). For the cluster population in general this is intuitive, since the presence or absence of a cool cores can mean as much as factor of two difference in total luminosity. The increase in surface brightness scatter at small radii, in part reflecting the evolution in the central brightness of the most relaxed clusters noted in Paper~\morphpaper{} (see also \citealt{Santos1008.0754, McDonald1305.2915}), shows that a center-excised luminosity should display reduced scatter with mass even within our morphologically similar cluster sample. As it happens, for realistic \NFW{} concentration parameters, the inner boundary of the lowest-scatter range noted above, $\Delta \sim 10^4$, roughly corresponds to the inner radius adopted by \citet{Maughan0703504} and \citet{Mantz0909.3099} when defining center-excised luminosity, $0.15\,r_{500}$. We return to the subject of scaling relations using integrated X-ray observables in Section~\ref{sec:bulk}.

\subsection{Pressure}
\label{sec:pressure}

\begin{figure}
 \centering
 \includegraphics[scale=0.9]{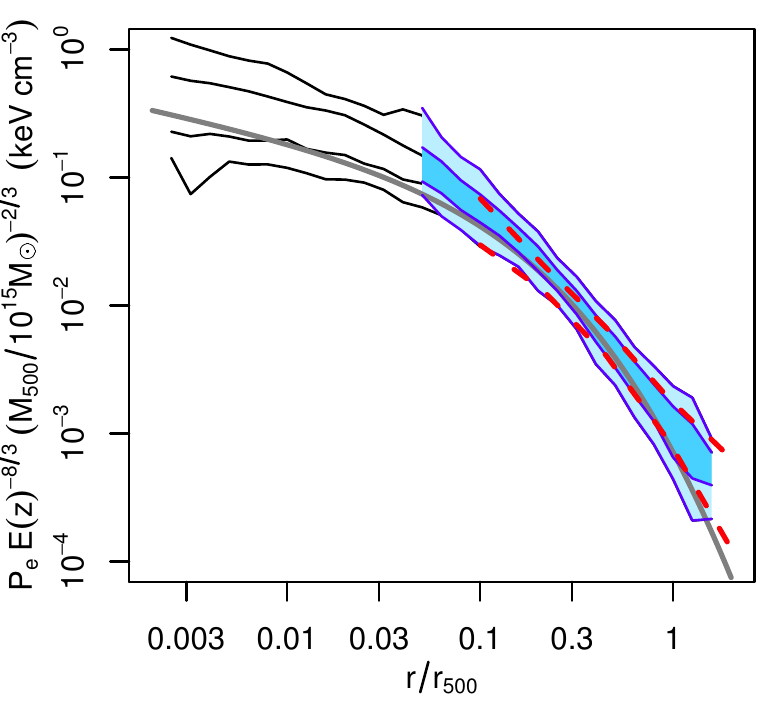}
 \caption[]{
   Scaled ensemble pressure profile as a function of $r/r_{500}$. The thick, gray line shows a generalized NFW fit to XMM and \Planck{} data \citep{Planck1207.4061}, while the dashed, red lines show the $1\sigma$ intrinsic scatter about the mean profile from a Gaussian process fit to Bolocam data \citep{Sayers1211.1632}.
 }
 \label{fig:pressure}
\end{figure}

The shape of the ensemble pressure profile is relevant for reconstructing the integrated Compton-$Y$ signal from Sunyaev-Zel'dovich (SZ) measurements with limited resolution \citep{Planck1303.5089}, field of view \citep{Czakon1406.2800}, or sensitivity to sufficiently large scales \citep{Mantz1401.2087}. To facilitate comparison with results in the literature, Figure~\ref{fig:pressure} shows our ensemble pressure profile as a function of $r/r_{500}$. The dashed red lines in the figure show the $1\sigma$ intrinsic scatter region from a Gaussian process fit to Bolocam data (\citealt{Sayers1211.1632}; similar to the fit we perform in Section~\ref{sec:profscale}). The slope of the Bolocam Gaussian process profile is in excellent agreement with ours at large radii ($\gtsim 0.5\,r_{500}$), but appears to flatten earlier at small radii; this difference can plausibly be explained by that fact that the Bolocam fit was not limited to relaxed/cool-core clusters. The gray, solid line in the figure is the generalized NFW pressure profile for cool-core clusters from the \citet{Planck1207.4061}, which is fit to a combination of XMM-Newton X-ray data at small radii and \Planck{} SZ data at large radii. There is an overall offset of order 10 per cent between our median pressure profile and the XMM/\Planck{} profile, although they agree within our statistical+systematic uncertainties. The agreement in shape between our profile and the XMM/\Planck{} results is qualitatively good over the entire radial range.

For reference, we note that our median profile is well described by 
\begin{equation}
  P(r) = 0.68\keV\cm^{-3}  E(z)^{8/3}\left(\frac{M_{500}}{10^{15}\Msun}\right)^{2/3}  G\left(\frac{r}{r_{500}}\right),
\end{equation}

\newpage\noindent
where $G$ is a generalized NFW function (see \citealt{Nagai0703661}) with parameters $(P_0, c_{500}, \gamma, \alpha, \beta) = (1, 1.19, -0.01, 0.51, 4.37)$. We stress that this parametrized fit should be used only as a simple description of our results, and only over the radial range shown in the figure. In particular, the scaled profile and its intrinsic scatter from cluster to cluster are modeled in greater depth in Section~\ref{sec:profscale}.

\subsection{Entropy}
\label{sec:entropy}

The ensemble entropy profile in Figure~\ref{fig:Qprofiles} displays features in common with numerous other studies of clusters and elliptical galaxies, namely a power law with radius (dashed line in the figure), flattening to a ``floor'' at small radii \citep{Lloyd-Davies0002082, Ponman0304048, Piffaretti0412233, Donahue0511401, Morandi0706.2971, Cavagnolo0902.1802, Pratt0909.3776,  Werner1205.1563, Werner1310.5450, Panagoulia1312.0798, McDonald1404.6250}. We note, however, that the departure from power-law behavior is limited to radii which are excluded from our main results due to the considerations discussed in Section~\ref{sec:data}, specifically the possibility that the choice of center, ellipticity, or presence of local density fluctuations (sloshing, cavities, etc.) may bias the density profiles inferred from a spherical deprojection. The ensemble profiles in Figure~\ref{fig:Qprofiles} are relatively conservative in this regard, excluding at least the central 50\,kpc, even when no disturbances in the ICM are visible.

\begin{figure}
 \centering
 \includegraphics[scale=0.9]{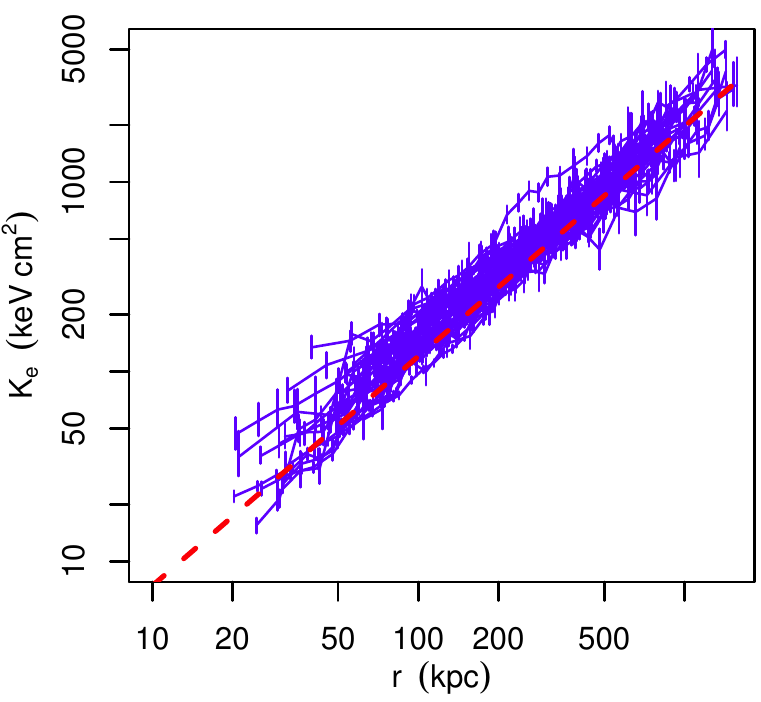}
 \caption[]{Profiles of entropy as a function of metric radius in our cluster sample. In this figure, the profile for each cluster is extended to the smallest radius where the ICM is undisturbed and where a spherical deprojection appears valid. With this restriction, each of the profiles shown appears consistent with a power law. The dashed line corresponds to the predicted behavior from simple hydrodynamical simulations (without cooling or feedback) scaled according to the median mass, redshift and gas mass fraction of the observed sample.}
 \label{fig:entropy}
\end{figure}

To better probe the central regions of the clusters, we do away with this default exclusion in this section, extending the entropy profiles inwards until the radii where the specific features mentioned above appear to become significant. The resulting individual profiles (blue lines with error bars) are shown as a function of metric radius in Figure~\ref{fig:entropy}. Although the ``undisturbed'' region of several of our cluster profiles reaches as low as $\sim 20$\,kpc ($\sim$0.01--0.02\,$r_{200}$), less than half the radius of the previous default exclusion, there are no strong indications of departures from a power law. Our results do not exclude the possibility of an entropy floor of radius $\sim 30$\,kpc in cool-core clusters, as generally agreed upon in the literature. However, they do call into question the validity of this concordance, given that systematics that may affect the determination of deprojected density are present at these radii in even the most relaxed and massive clusters. We note that both \citet{Panagoulia1312.0798} and \citet{Werner1205.1563, Werner1310.5450} present measured entropy profiles that display power-law behavior down to radii of $\ltsim 1$\,kpc ($\ltsim0.01\,r_{200}$), respectively for samples of groups and highly relaxed elliptical galaxies.

The dashed, red line in Figure~\ref{fig:entropy} shows the ``baseline'' entropy power law due to purely gravitational structure formation found from the simulations of \citet{Voit0511252}, which is in good agreement with more recent simulations using several different codes (e.g.\ \citealt{Sembolini1503.06065}). Here we have scaled the baseline profile according to the median mass and redshift of our cluster sample (see \citealt{Voit0511252}), and to account for the difference between the cluster gas-mass fraction in the simulations and that measured from our data at $r_{2500}$. There is good agreement between the theoretical curve and the data at radii $\gtsim 500$\, kpc, although the power-law slope of the observed profiles is somewhat shallower than the model ($\sim1.15$ compared with 1.21), suggesting that we may be seeing some excess entropy near cluster centers.

We emphasize that a power-law entropy profile extending down to relatively small radii in this sample of dynamically relaxed clusters is compatible with the notion that cooling and feedback processes significantly affect the gas at small radii in groups and clusters. Indeed, our results in Sections~\ref{sec:bulk} and \ref{sec:fgas2500} indicate an important role for cooling and feedback within $\sim0.15\,r_{500}$ for the massive clusters studied here. The measured entropy profiles suggest that heating and cooling in the central regions of these systems proceeds via a continuous feedback cycle that allows the central entropy profiles to remain stratified, preserving the similarity and relatively low intrinsic scatter evident in Figure~\ref{fig:entropy}, rather than violent episodic outbursts. Within the central 20\,kpc, clear departures from spherical symmetry prevent robust thermodynamic profiles from being extracted from the data.

\section{Scaling Relations}
\label{sec:scaling}

We now turn to the scaling of the measured thermodynamic quantities with mass and redshift. First, in Section~\ref{sec:profscale}, we consider the scaling of the thermodynamic profiles, including their intrinsic covariance as a function of radius. Section~\ref{sec:bulk} then considers the scaling relations of traditional integrated or global cluster measurements.

In previous work, we have stressed that scaling relations of the cluster population as a whole can only be reliably constrained when the sample selection and the underlying cosmological mass function are properly incorporated into the fitted model (\citealt{Mantz0909.3098, Allen1103.4829}; see also \citealt{Evrard1403.1456, Czakon1406.2800}). This is impossible in the present case, since the sample selection is complex (Paper~\morphpaper{}) and depends on features which are not yet accurately reproduced in simulations (e.g., cool cores). Among the selection criteria, the $kT>5$\,keV cut is clearly important in the context of scaling relations. Most of our clusters are also initially selected from X-ray flux limited surveys, and must satisfy a redshift-dependent cut on central surface brightness, so we might expect selection effects to influence the density or luminosity scaling relations at some level. It is less clear that the selection should affect gas masses or center-excised luminosity measurements, which are dominated by the fainter parts of a cluster. While it has been argued that marginalizing over a flexible prior on the underlying distribution of masses and redshifts, as we do, may partially mitigate these issues \citep{Sereno1502.05413}, our results in this section should be interpreted as strictly empirical, and certainly only apply to relaxed clusters.

\subsection{Scaling of Profiles}
\label{sec:profscale}

Using the method described in Section~\ref{sec:methods}, we fit a joint model for the mean density and temperature profiles, their intrinsic covariance, and their scaling with mass and $E(z)$. In detail, this means that at a fixed set of $N_\Delta$ ordinates, $\{\Delta_i\}$, we constrain the scaled density and temperature, $n_i [E(z)/E(0.35)]^{-\beta_{nz}} (M_{2500}/3\E{14}\Msun)^{-\beta_{nm}}$ and $kT_i [E(z)/E(0.35)]^{-\beta_{tz}} (M_{2500}/3\E{14}\Msun)^{-\beta_{tm}}$.\footnote{ Note that there is no particular reason to expect departures from self-similar evolution to manifest themselves as a power-law in $E(z)$, as opposed to some other function of redshift. However, this parametrization does provide a simple and convenient null test of whether such departures exist.} The slopes, $\beta$, are also free parameters of the fit. The model also includes a $2N_\Delta \times 2N_\Delta$ log-normal intrinsic covariance matrix, encoding the marginal cluster-to-cluster scatter in density and temperature at each $\Delta$, the covariance of each of these quantities at each pair of overdensities (e.g., the covariance of departures from the mean density at small and large radii), and the density--temperature scatter covariance as a function of overdensity. Constraining this model requires the full set of 40 clusters to provide data at each overdensity. To ensure that this is the case, even though the mass profile of each cluster (effectively, $\Delta$ as a function of angular radius) is being marginalized over, we model only the overdensity range $10^{4.6} \geq \Delta \geq 10^{3.4}$. Table~\ref{tab:profscale} summarizes the results. Note that it is straightforward to convert this model for the joint density--temperature scaling and scatter to a model for pressure, entropy, or some other product of powers of $n$ and $kT$.

\begin{table*}
  \centering
  \caption[]{
    Constraints on the mean scaled electron density and temperature profiles of our sample, the power-law scaling indices of each profile with $M_{2500}$ and $E(z)$, and the intrinsic covariance about the mean profiles. The model is given in terms of a discrete set of overdensities (as a proxy for radius), and the scaling is relative to $z=0.35$ and $M_{2500}=3\E{14}\Msun$. The covariance is broken down into the density--density, temperature--temperature and density--temperature blocks, and expressed as marginal standard deviations along the diagonal and correlation coefficients off of the diagonal.
  }
  \label{tab:profscale}
  \begin{minipage}[b]{0.63\textwidth}
    \centerline{Mean}
    \centering
    \begin{tabular}{cccccc}
      \hline
      $\Delta$ & $10^{4.6}$ & $10^{4.3}$ & $10^{4.0}$ & $10^{3.7}$ & $10^{3.4}$ \\
      $\ln (n/\mathrm{cm}^{-3})$ & $-4.51 \pm 0.05$ & $-5.18 \pm 0.04$ & $-5.83 \pm 0.03$ & $-6.51 \pm 0.03$ & $-7.18 \pm 0.03$ \\
      $\ln (kT/\mathrm{keV})$ & $2.007 \pm 0.030$ & $2.077 \pm 0.023$ & $2.107 \pm 0.014$ & $2.097 \pm 0.014$ & $2.073 \pm 0.017$ \\
      \hline
    \end{tabular}
  \end{minipage}
  \hspace{0.09\textwidth}
  \begin{minipage}[b]{0.27\textwidth}
    \centerline{Slopes}
    \centering
    \begin{tabular}{ccc}
      \hline
      & $E(z)$ & $M_{2500}$ \\
      $n$ & $2.0 \pm 0.2$ & $0.03 \pm 0.06$ \\
      $kT$ & $0.64 \pm 0.08$ & $0.67 \pm 0.02$ \\
      \hline
    \end{tabular}
  \end{minipage}
  \\~\\
  \centerline{Covariance}
  \begin{tabular}{ccccccccc}
    \hline
    & & & $kT(\Delta_1)$ & $kT(\Delta_2)$ & $kT(\Delta_3)$ & $kT(\Delta_4)$ & $kT(\Delta_5)$ & \\
     $n(\Delta_1)$ & $0.32^{+0.05}_{-0.04}$  &  ~                       &  $0.15^{+0.03}_{-0.02}$  &  $0.78^{+0.09}_{-0.13}$  &  $0.46^{+0.22}_{-0.30}$  &  $0.33^{+0.25}_{-0.30}$  &  $0.13^{+0.27}_{-0.29}$ & $kT(\Delta_1)$  \\
    $n(\Delta_2)$ & $0.93^{+0.03}_{-0.04}$  &  $0.25^{+0.04}_{-0.03}$  &  ~                       &  $0.11^{+0.02}_{-0.02}$  &  $0.58^{+0.19}_{-0.28}$  &  $0.45^{+0.22}_{-0.28}$  &  $0.23^{+0.26}_{-0.29}$ & $kT(\Delta_2)$  \\
    $n(\Delta_3)$ & $0.87^{+0.05}_{-0.07}$  &  $0.92^{+0.03}_{-0.05}$  &  $0.18^{+0.03}_{-0.03}$  &  ~                       &  $0.05^{+0.02}_{-0.01}$  &  $0.44^{+0.21}_{-0.28}$  &  $0.29^{+0.25}_{-0.29}$ & $kT(\Delta_3)$  \\
    $n(\Delta_4)$ & $0.70^{+0.10}_{-0.14}$  &  $0.79^{+0.07}_{-0.10}$  &  $0.84^{+0.06}_{-0.09}$  &  $0.17^{+0.03}_{-0.02}$  &  ~                       &  $0.06^{+0.01}_{-0.01}$  &  $0.39^{+0.21}_{-0.26}$ & $kT(\Delta_4)$  \\
    $n(\Delta_5)$ & $0.37^{+0.19}_{-0.22}$  &  $0.47^{+0.17}_{-0.20}$  &  $0.55^{+0.15}_{-0.19}$  &  $0.70^{+0.11}_{-0.14}$  &  $0.16^{+0.03}_{-0.03}$  &  ~                       &  $0.07^{+0.02}_{-0.02}$ & $kT(\Delta_5)$  \\
    &$n(\Delta_1)$ & $n(\Delta_2)$ & $n(\Delta_3)$ & $n(\Delta_4)$ & $n(\Delta_5)$ \\
    \\
    &$n(\Delta_1)$ & $n(\Delta_2)$ & $n(\Delta_3)$ & $n(\Delta_4)$ & $n(\Delta_5)$ \\
    $kT(\Delta_1)$ & $-0.82^{+0.11}_{-0.07}$  &  $-0.77^{+0.14}_{-0.09}$  &  $-0.68^{+0.17}_{-0.12}$  &  $-0.46^{+0.22}_{-0.18}$  &  $-0.11^{+0.25}_{-0.24}$  \\
    $kT(\Delta_2)$ & $-0.88^{+0.09}_{-0.05}$  &  $-0.86^{+0.10}_{-0.06}$  &  $-0.81^{+0.13}_{-0.08}$  &  $-0.63^{+0.19}_{-0.14}$  &  $-0.29^{+0.25}_{-0.22}$  \\
    $kT(\Delta_3)$ & $-0.59^{+0.29}_{-0.20}$  &  $-0.60^{+0.30}_{-0.20}$  &  $-0.57^{+0.31}_{-0.21}$  &  $-0.50^{+0.33}_{-0.22}$  &  $-0.30^{+0.33}_{-0.26}$  \\
    $kT(\Delta_4)$ & $-0.44^{+0.30}_{-0.24}$  &  $-0.46^{+0.30}_{-0.24}$  &  $-0.47^{+0.30}_{-0.24}$  &  $-0.42^{+0.32}_{-0.25}$  &  $-0.23^{+0.33}_{-0.28}$  \\
    $kT(\Delta_5)$ & $-0.21^{+0.29}_{-0.27}$  &  $-0.21^{+0.30}_{-0.28}$  &  $-0.23^{+0.30}_{-0.28}$  &  $-0.20^{+0.32}_{-0.30}$  &  $-0.09^{+0.31}_{-0.31}$  \\
    \hline
  \end{tabular}
\end{table*}

\begin{figure*}
 \centering
 \includegraphics[scale=0.7]{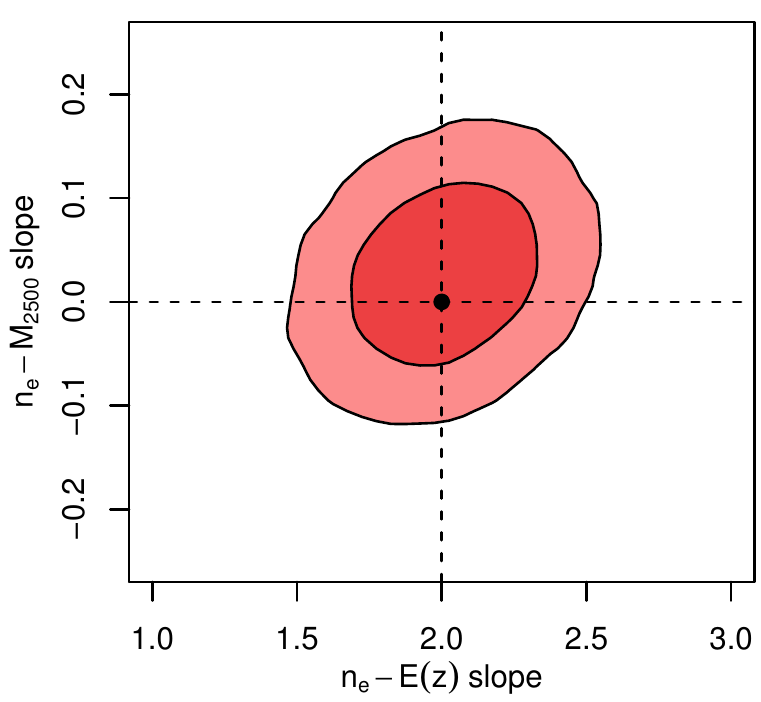}
 \hspace{2mm}
 \includegraphics[scale=0.7]{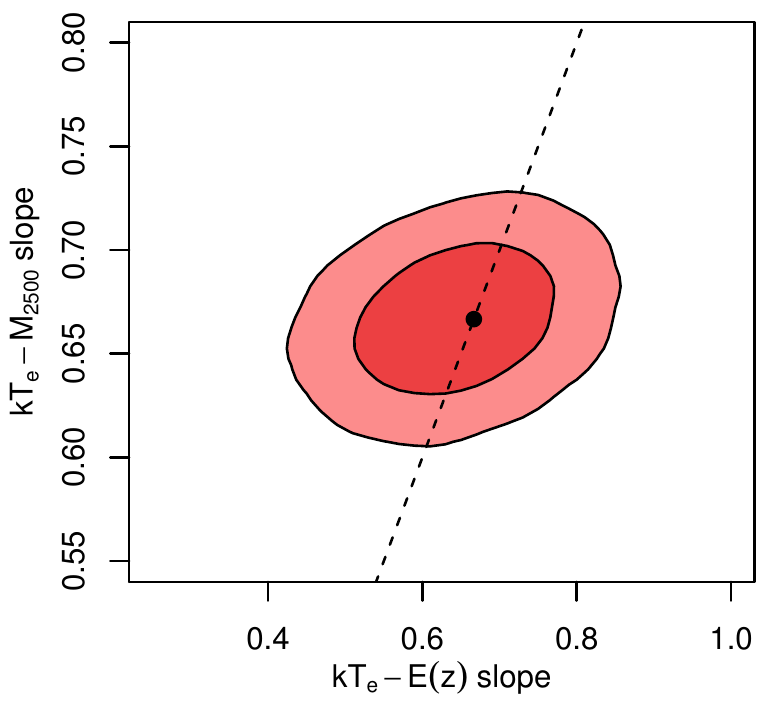}
 \hspace{2mm}
 \includegraphics[scale=0.7]{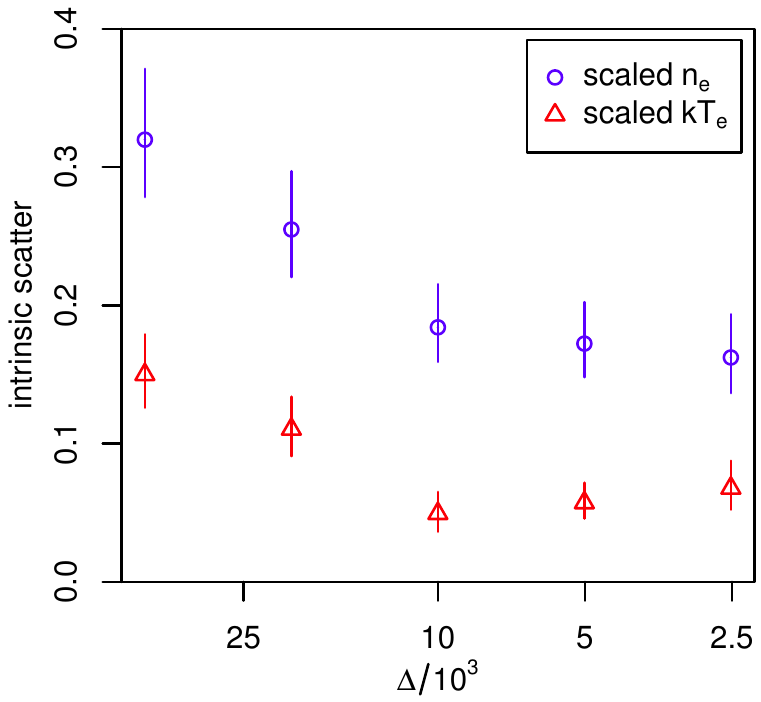}
 \caption[]{
   Left and center: constraints (68.3 and 95.4 per cent confidence) on power-law slopes of the overall scaling of gas density and temperature profiles with mass and $E(z)$. The self-similar expectation is indicated by filled black circles, and dashed lines show the effect of changing a single exponent at a time in Equation~\ref{eq:selfsim}.
   Right: marginal intrinsic scatters of density and temperature at the modelled set of overdensities (see Table~\ref{tab:profscale}).
 }
 \label{fig:profslopes}\label{fig:profscat}
\end{figure*}

Figure~\ref{fig:profslopes} shows the 68.3 and 95.4 per cent confidence constraints on the power-law slopes of density and temperature with mass and $E(z)$. Black circles show the predictions of the self-similar model, and dashed lines correspond to varying a single exponent at a time in Equation~\ref{eq:selfsim} (including the implicit $M^0$ in the expression for $n$). All of the measured slopes are consistent with the self-similar model.

The right panel of Figure~\ref{fig:profscat} shows the marginal intrinsic scatter in density and temperature at each overdensity. Consistent with our analysis of the surface brightness in Section~\ref{sec:density}, the scatter in density is largest at small radii, and levels off at $\Delta \ltsim 10^4$. The temperature scatter shows a similar trend, and is smaller than both the scatter in the density profile and the scatter in average temperature determined in Section~\ref{sec:bulk}. The latter might be expected, given that in this section we are taking advantage of the radially resolved, 3-dimensional temperature profiles. However, we caution that the individual temperature measurements are still effectively averaged over a relatively large volume (i.e.\ the resolution in radius is poor) compared to the density; this might result in intrinsic scatter on smaller scales, which can be probed in density, being suppressed in the temperature profiles. A comparison to realistic hydrodynamical simulations and/or very deep, high-resolution observations could shed light on this question.

\begin{figure*}
 \centering
 \includegraphics[scale=0.75]{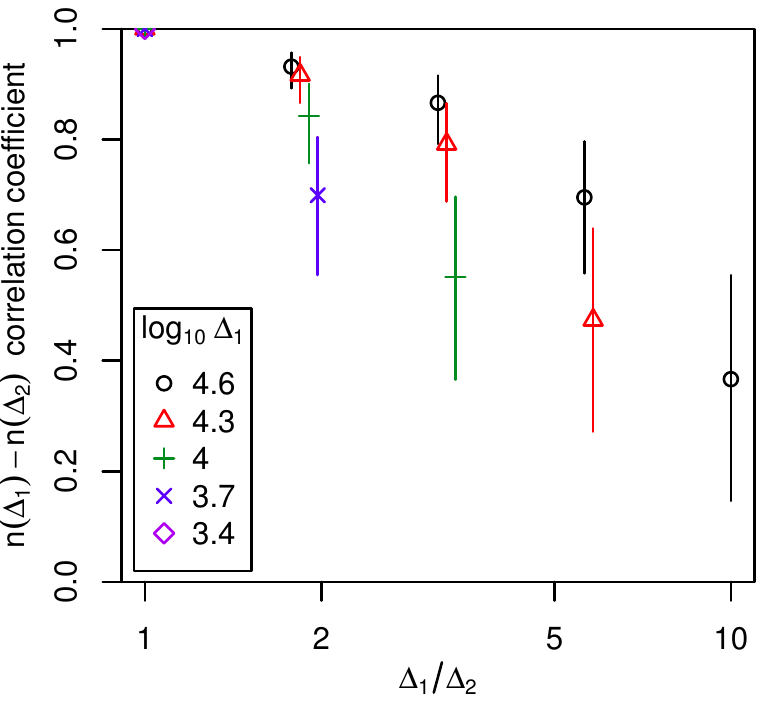}
 \includegraphics[scale=0.75]{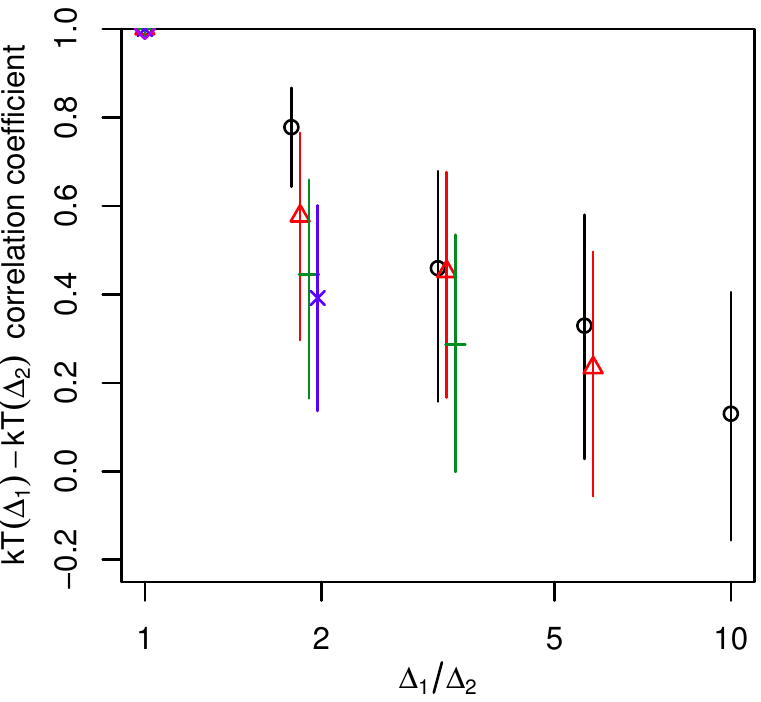}
 \includegraphics[scale=0.75]{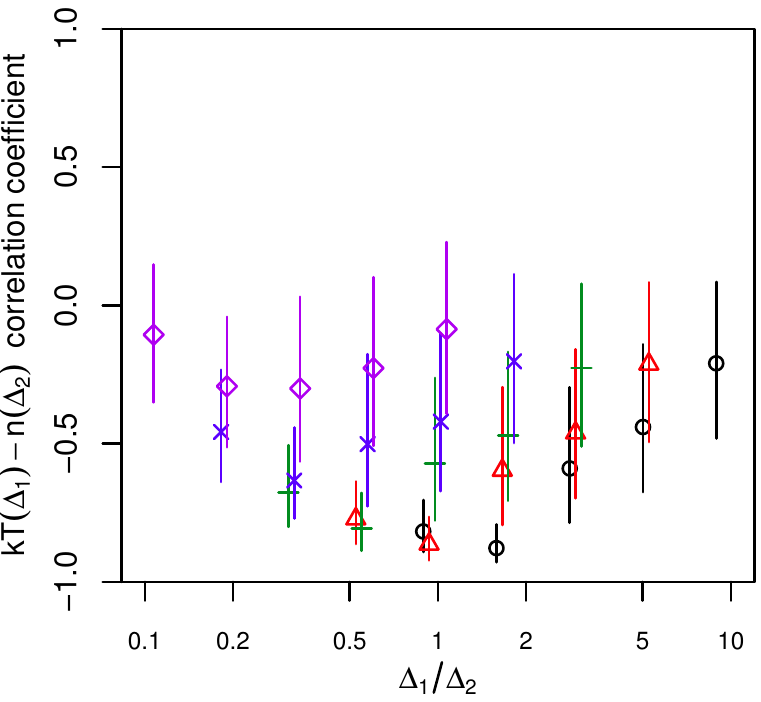}
 \caption[]{
   Correlation coefficients of the joint density--temperature intrinsic covariance, plotted as a function of separation in overdensity units. $n$--$n$ and $kT$--$kT$ correlations appear to drop with separation in a roughly translation-invariant way. The $kT$--$n$ correlations are all approximately consistent with zero except at small radii (large $\Delta_1 \approx \Delta_2$). Small horizontal offsets are introduced to make the plots clearer.
 }
 \label{fig:profcov}
\end{figure*}

Constraints on the off-diagonal covariances of the model, in the form of correlation coefficients, are shown in Figure~\ref{fig:profcov}. The correlation of density at different radii appears to drop monotonically with their separation, approximately independent of their absolute position within the cluster (at least, within the modeled radial range). The same is true of temperature, albeit with larger uncertainties. This indicates that the overall scalings with mass and $E(z)$ are broadly successful in standardizing the density and temperature profiles, whereas strong correlations or anticorrelations at large separations would suggest that perhaps the scalings should be radius-dependent. The density--temperature correlations are consistent with zero with the exception of anticorrelations at small separation and small radius; this presumably reflects local pressure equilibrium even in the presence of variations in the spatial extent and magnitude of cool cores.

Overall, these results are consistent with what we would expect for the most relaxed clusters in the Universe, i.e.\ simple mean scalings with mass and redshift and similar profile shapes (outside the core). Expanding this kind of analysis to a more dynamically diverse cluster selection, most likely in simulations given the difficulty in measuring precise mass profiles for unrelaxed clusters, could prove interesting in the future.

\subsection{Scaling of Integrated Measurements}
\label{sec:bulk}

\begin{figure*}
 \centering
 \includegraphics[scale=0.7]{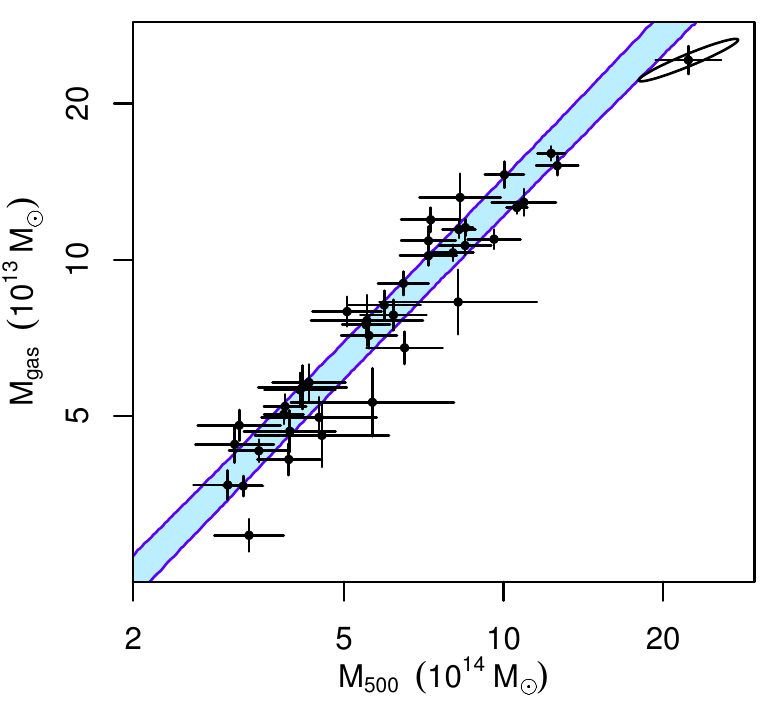}
 \hspace{0.5cm}
 \includegraphics[scale=0.7]{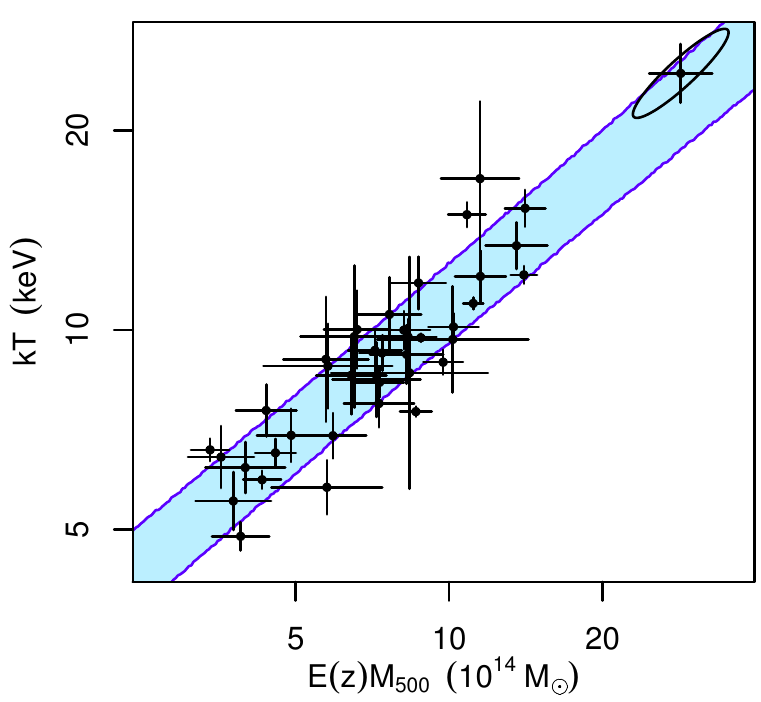}\medskip\\
 \includegraphics[scale=0.7]{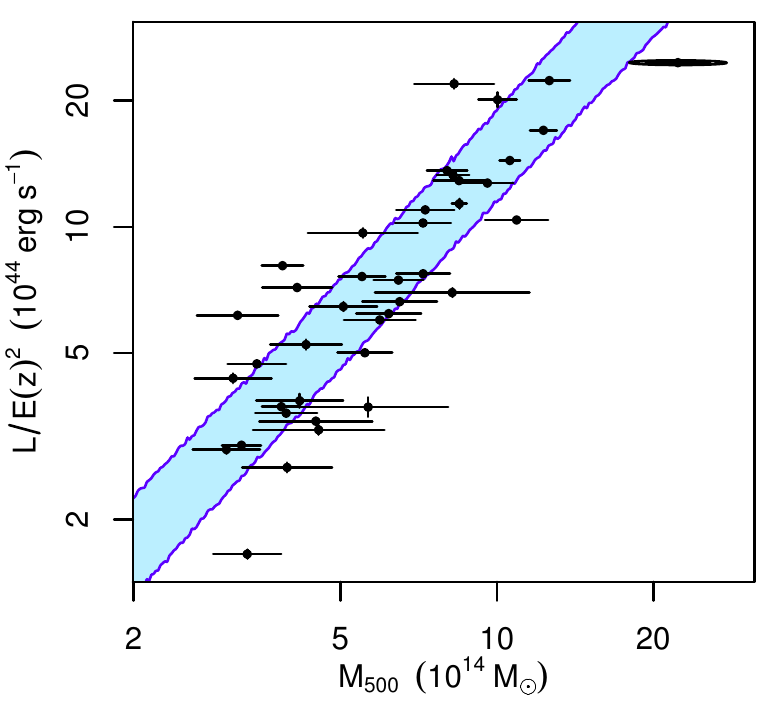}
 \hspace{0.5cm}
 \includegraphics[scale=0.7]{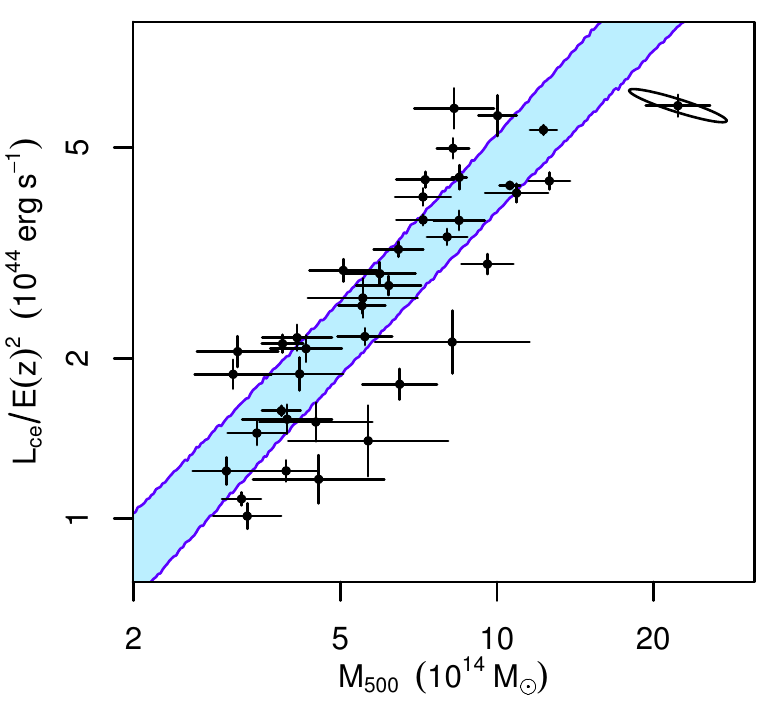}
 \caption[]{
   Scatter plots summarizing the integrated thermodynamic quantities for which we fit scaling relations with $M_{500}$ and $E(z)$. In each panel, the measurement covariance ellipse is shown for the most massive cluster in the sample. Shaded regions show the $1\sigma$ predictions for a subset of the model space we explore, specifically with the power of $E(z)$ fixed to 0.0 (for $\Mgas$) or 2.0 (for $L$ and $\Lce$), or required to be equal to the power of $M_{500}$ (for $kT$).
 }
 \label{fig:bulkdat}
\end{figure*}

We next turn to the more traditional scaling relations of global or integrated quantities. For this analysis, we adopt $r_{500}$, the most common choice in the literature, as the characteristic radius within which to make these measurements. Here the values of $r_{500}$ are determined by our \NFW{} analysis of each cluster, and we account for the correlation of mass and temperature at fixed radius imparted by the hydrostatic assumption as described in Section~\ref{sec:data}, as well as the straightforward correlation in measurement uncertainty between mass and other quantities measured within a mass-dependent aperture. The thermodynamic quantities we consider are the spherically integrated gas mass within $r_{500}$, the average temperature measured in an annulus spanning $0.15 <r/r_{500} < 1$ (as determined from a 2-temperature-bin non-parametric deprojection; see Section~\ref{sec:data}), the soft-band (0.1--2.4\,keV) intrinsic luminosity projected within $r_{500}$, and the ``center-excised'' luminosity projected onto an annulus spanning 0.15--1\,$r_{500}$, denoted $\Lce$. These measurements are plotted in Figure~\ref{fig:bulkdat}, along with a representation of their covariance with mass. Using the regression technique described in Section~\ref{sec:methods} and by \citet{Mantz1509.00908}, we simultaneously fit for the scaling of these 4 variables as a function of $M_{500}$ and $E(z)$, including their $4 \times 4$ intrinsic covariance matrix. For completeness, we also include below some results for the luminosity projected within a radius of $0.15\,r_{500}$ ($\Lcore = L - \Lce$), although the computationally singular correlations among the three luminosity measurements prevent us from fitting all 5 scaling relations simultaneously.

\begin{figure*}
 \centering
 \includegraphics[scale=0.7]{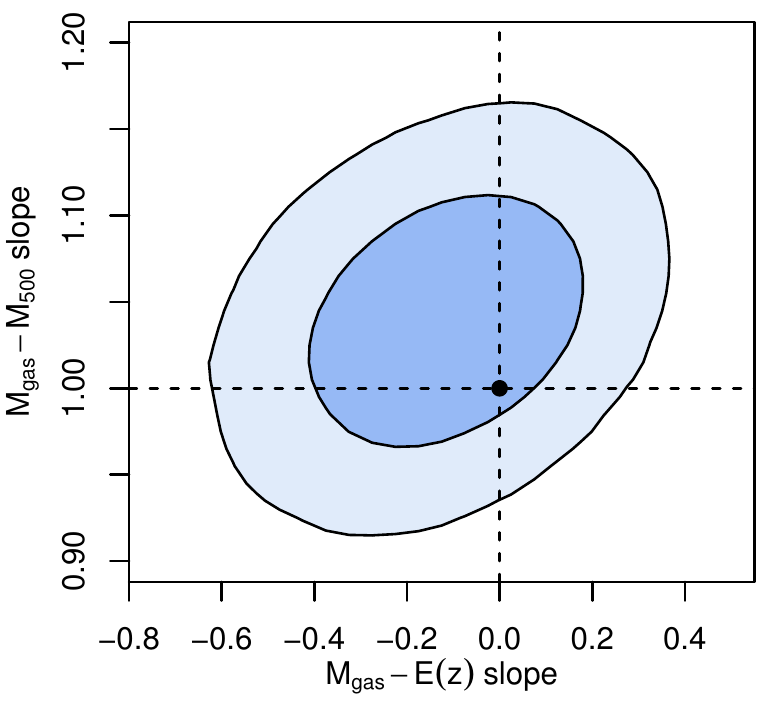}
 \hspace{2mm}
 \includegraphics[scale=0.7]{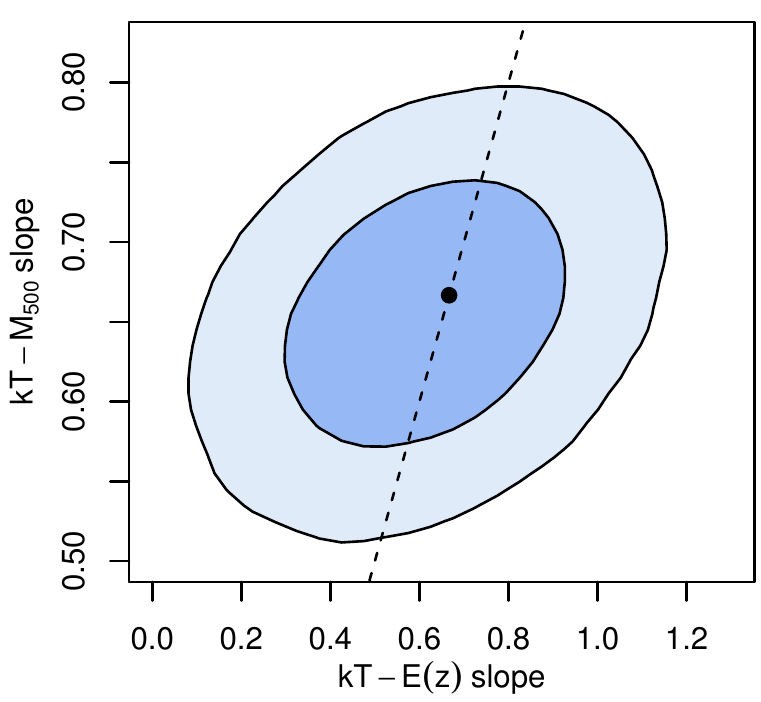}
 \hspace{2mm}
 \includegraphics[scale=0.7]{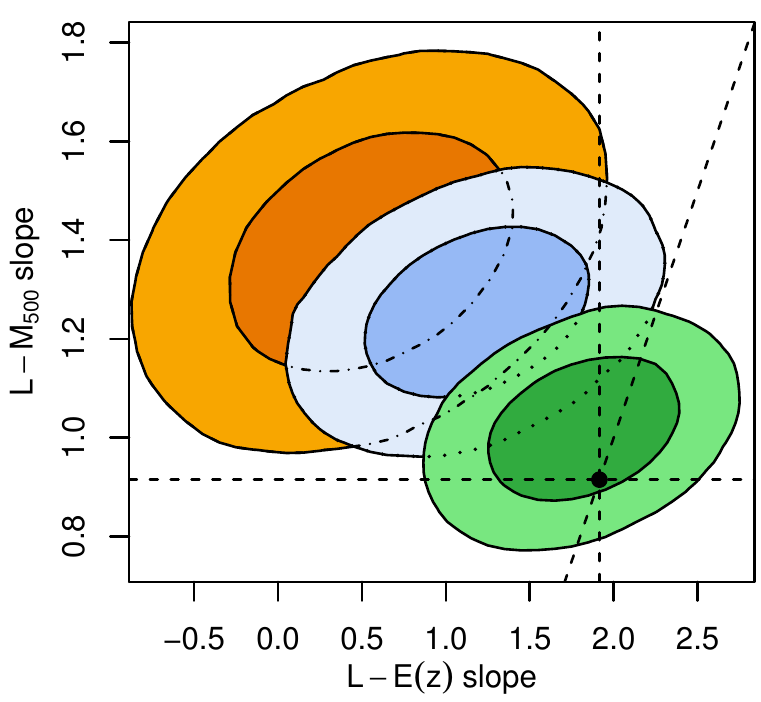}
 \caption[]{
   Constraints on the power-law slopes of the scaling relations of integrated gas mass, temperature and luminosity as a function of mass and $E(z)$. The self-similar expectation is indicated by filled black circles, and dashed lines show the effect of changing a single exponent at a time in Equation~\ref{eq:selfsimbulk}. In the right panel, blue shading corresponds to the slopes for total luminosity ($L$), green shading corresponds to those for $\Lce$, and orange shading corresponds to $\Lcore$.
 }
 \label{fig:bulkslopes}
\end{figure*}

Figure~\ref{fig:bulkslopes} and Table~\ref{tab:bulkslopes} show constraints on each of the scaling relation slopes. The gas mass and temperature slopes are all consistent with self-similarity, as expected from our analysis of the density and temperature profile scalings in Section~\ref{sec:profscale}.\footnote{ We note in particular that the steep $\Mgas$--$M$ slopes inferred from some analyses of dynamically heterogeneous systems, extending down to the group scale, e.g., \citealt{Pratt0809.3784, Sun0805.2320}) are not consistent with the results reported here for the most massive, dynamically relaxed systems.} However, the scaling of total luminosity is not consistent with self-similarity, preferring either a weaker dependence on redshift, a stronger dependence on mass, or both, consistent with a number of previous results (e.g., \citealt{Reiprich0111285, Zhang0702739, Zhang0802.0770, Mantz0709.4294, Mantz0909.3099, Rykoff0802.1069, Pratt0809.3784, Vikhlinin0805.2207, Leauthaud0910.5219, Reichert1109.3708, Sereno1502.05413}; see also the review of \citealt{Giodini1305.3286}). This is a clearly radius-dependent phenomenon, with $\Lce$ being perfectly consistent with self-similarity, $\Lcore$ preferring stronger departures from self-similarity, and the total luminosity occupying a middle ground. While the uncertainty on the evolution of $\Lcore$ is substantial, milder than self-similar evolution is qualitatively consistent with the decreasing central surface brightness of cool-core clusters with redshift seen by \citet{Santos0802.1445, Santos1008.0754} and \citet{McDonald1305.2915}, and in Paper~\morphpaper{}.

\begin{table}
  \centering
  \caption[]{
    Constraints on the power-law slopes of the scaling relations of gas mass, temperature and luminosity (integrated within $r_{500}$) as a function of mass and $E(z)$. Normalizations are given as the natural logarithm of the listed quantities at $z=0.35$ and $M_{500}=6\E{14}\Msun$.
  }
  \label{tab:bulkslopes}
  \begin{tabular}{cccc}
    \hline
     & normalization & $E(z)$ slope & $M_{500}$ slope \\
     \hline
     $\Mgas/M_{\odot}$ & \hspace{1.3ex}$31.98 \pm 0.02$ & $-0.11 \pm 0.18$ & $1.04 \pm 0.05$ \\
     $kT/\mathrm{keV}$ & \hspace{1em}$2.18 \pm 0.02$ & \phmin$0.61 \pm 0.20$ & $0.66 \pm 0.05$ \\
     $L/\mathrm{erg}\second^{-1}$ & $103.70 \pm 0.05$ & \phmin$1.20 \pm 0.43$ & $1.26 \pm 0.11$ \\
     $\Lce/\mathrm{erg}\second^{-1}$ & $102.66 \pm 0.04$ & \phmin$1.82 \pm 0.35$ & $1.02 \pm 0.09$ \\
     \hline
  \end{tabular}
\end{table}

\begin{figure*}
 \centering
 \includegraphics[scale=0.9]{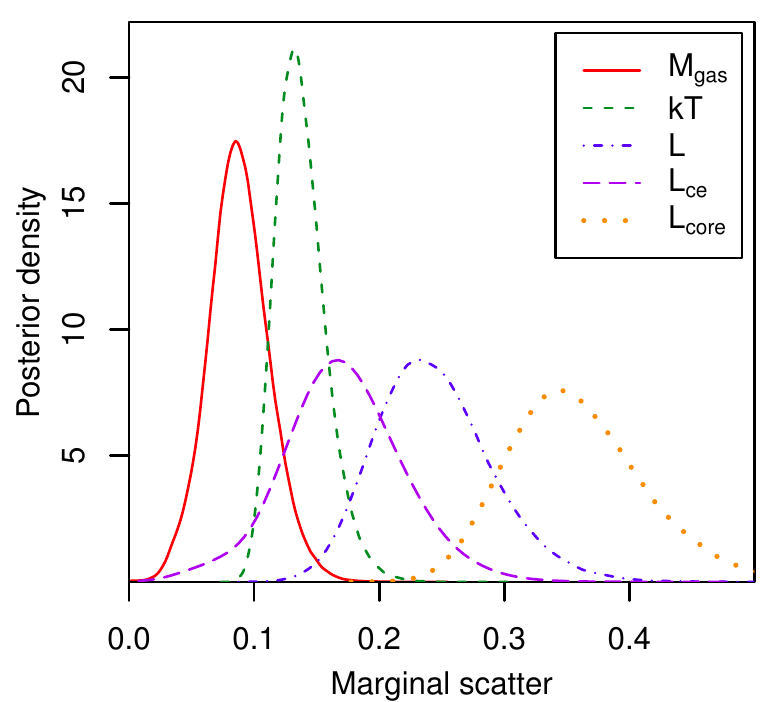}
 \hspace{1cm}
 \includegraphics[scale=0.9]{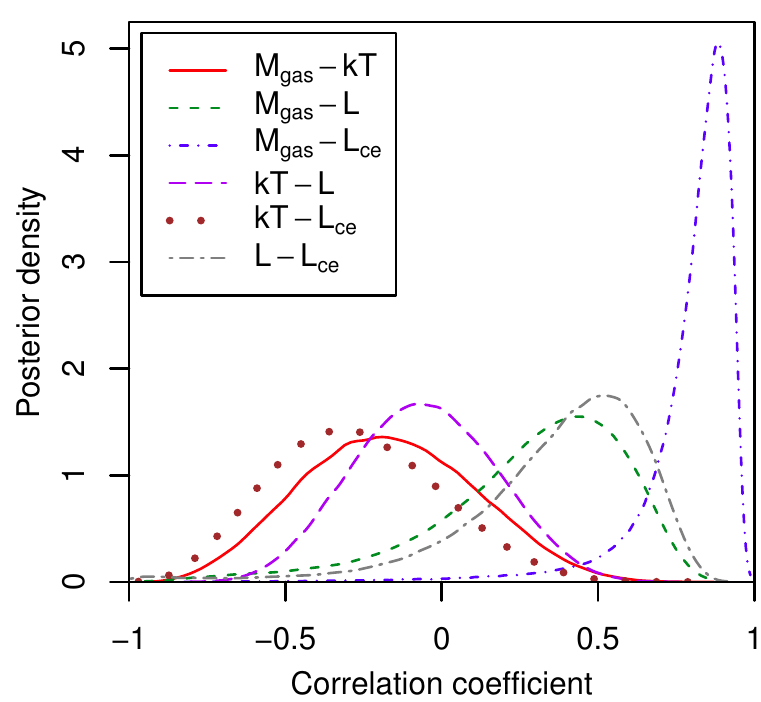}
 \caption[]{
   Marginal scatters (left) and correlation coefficients (right), respectively representing the diagonal and off-diagonal elements of the intrinsic covariance of integrated $\Mgas$, $kT$, $L$ and $\Lce$ at fixed $M_{500}$ and $E(z)$. The marginal scatter of central luminosity is also shown in the left panel.
 }
 \label{fig:bulkscat}
\end{figure*}

Constraints on the intrinsic covariances of the model appear in Figure~\ref{fig:bulkscat} and Table~\ref{tab:bulkscat}. Our constraints on the marginal scatter of $\Mgas$ and $kT$ are largely consistent with earlier work (see \citealt{Allen1103.4829, Giodini1305.3286}, and references therein); in particular the small scatter in $\Mgas$, $0.086 \pm 0.023$ is similar to the scatter found for $\fgas$ in a shell spanning radii 0.8--1.2\,$r_{2500}$ in Paper~\cosmopaper{}, $0.074 \pm 0.023$ (see also Section~\ref{sec:fgas2500}; \citealt{Allen0706.0033}). The $kT$ scatter, $0.134 \pm 0.019$, agrees well with results for the cluster population at large (e.g., \citealt{Arnaud0709.1561, Vikhlinin0805.2207, Mantz0909.3099}). As one would expect, given the intentional morphological similarity of the clusters in our sample, we measure a scatter in total luminosity at fixed mass, $0.24 \pm 0.05$, that is smaller than that of the population at large ($\sim 0.4$; e.g., \citealt{Vikhlinin0805.2207, Mantz0909.3099}). However, even with this selection, the scatter in $\Lcore$ is still significant, $0.34^{+0.06}_{-0.05}$. The scatter in $\Lce$ is smaller than that in the total luminosity, although at low confidence, $0.17 \pm 0.05$.

\begin{table*}
  \centering
  \caption[]{
    Constraints on the marginal scatters (on-diagonal entries) and correlation coefficients (off-diagonal entries), respectively representing the diagonal and off-diagonal elements of the intrinsic covariance of integrated $\Mgas$, $kT$, $L$ and $\Lce$ at fixed $M_{500}$ and $E(z)$.
  }
  \label{tab:bulkscat}
  \begin{tabular}{ccccc}
    \hline
     & \phmin$\Mgas$ & \phmin$kT$ & $L$ & $\Lce$ \\
     $\Mgas$ & \phmin$0.09 \pm 0.02$ \\
     $kT$ & $-0.18 \pm 0.28$ & \phmin$0.13 \pm 0.02$ \\
     $L$ & \phmin$0.43^{+0.22}_{-0.30}$ & $-0.06 \pm 0.24$ & $0.24 \pm 0.05$ \\
     $\Lce$ & \phmin$0.88^{+0.06}_{-0.16}$ & $-0.30 \pm 0.27$ & $0.53^{+0.17}_{-0.29}$ & $0.17 \pm 0.05$ \\
     \hline
  \end{tabular}
\end{table*}

Although the sample employed here is relatively small, we are nevertheless able to place constraints on the off-diagonal terms relating $\Mgas$, $kT$, $L$ and $\Lce$. Our constraint on the $kT$--$L$ correlation coefficient is $-0.06 \pm 0.24$. The $\Mgas$--$kT$ correlation is constrained at a similar level, $-0.18 \pm 0.28$, consistent with zero. In contrast, the $\Mgas$--$L$ correlation prefers positive values, $0.43^{+0.22}_{-0.30}$, reflecting the fact that these quantities are both essentially integrals of the same gas density profile, but with different weighting. The correlation between $\Mgas$ and $\Lce$ is very strong, $0.88^{+0.06}_{-0.16}$; correspondingly, the correlations of $kT$ and $L$ with $\Lce$ are similar to the $\Mgas$--$kT$ and $\Mgas$--$L$ correlations.\footnote{ The strength of the $\Mgas$--$\Lce$ correlation explains why \citet{Mantz0909.3099}, using $\Mgas$  as a proxy for total mass, found a smaller intrinsic scatter in $\Lce$ than we do ($\ltsim 10$ per cent).} There have been few previous measurements of the off-diagonal terms of the cluster scaling relation intrinsic scatter. \citet{Mantz0909.3099} measured the correlation of $L$ and $kT$ at fixed mass to be $0.09 \pm 0.19$ (not restricted to relaxed clusters), consistent with our findings above. \citet{Rozo0902.3702} placed a lower limit on the correlation of $L$ and $M$ at fixed optical richness (as defined for the MaxBCG catalog) of $>0.85$. More recently, \citet{Maughan1212.0858} fit $\Lce$, $kT$ and $\Mgas$ scaling relations, including intrinsic scatter covariances although without modeling measurement covariances, to published data for REXCESS clusters. Those results on the intrinsic correlation coefficients are compatible with, and have similar precision to, our constraints from relaxed clusters.

Overall, the picture that emerges is one in which the gas temperature and density outside of the core, roughly at radii 0.15--1\,$r_{500}$, behaves simply and in accordance with the self-similar model for these massive, relaxed clusters. In particular, the strong correlation in scatter between $\Mgas$ and $\Lce$ requires the gas density profiles to be smooth at these radii (Figure~\ref{fig:Qprofiles}; recall that the absence of substructure on these scales is a requirement of the sample selection in Paper~\morphpaper{}), since scatter in density is reflected disportionately in the luminosity compared to the gas mass. In the centers of relaxed clusters, we see breaking of the self-similar model, specifically in the form of enhanced intrinsic scatter in luminosity, and a preference for scaling relation slopes of core luminosity with mass (redshift) that are steeper (shallower) than the self-similar prediction. These results require the action of a non-self-similar astrophysical process such as AGN feedback (\citealt{Fabian1204.4114, McNamara1204.0006}, and references therein), although in general both heating and cooling processes may contribute to the observed trends.

\begin{figure*}
 \centering
 \includegraphics[scale=0.9]{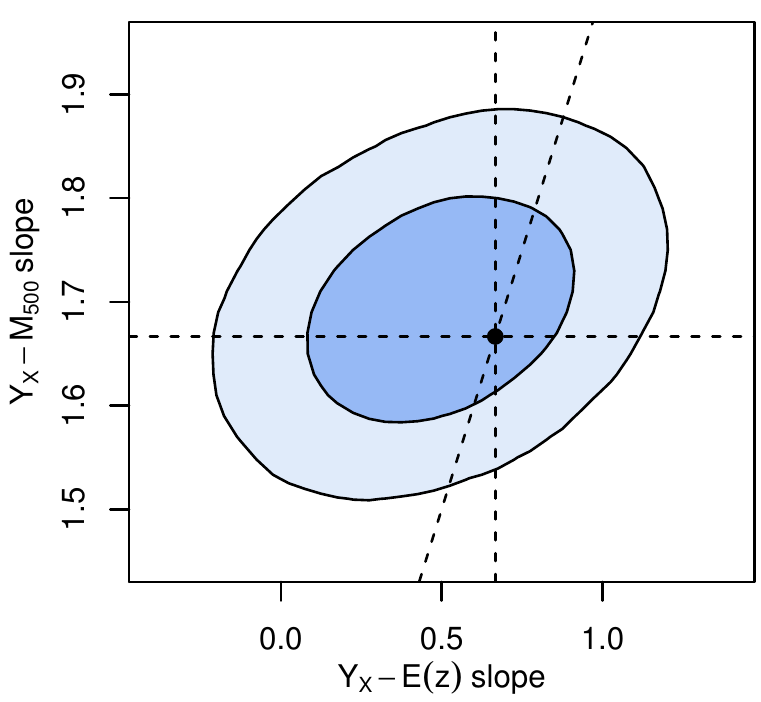}
 \hspace{1cm}
 \includegraphics[scale=0.9]{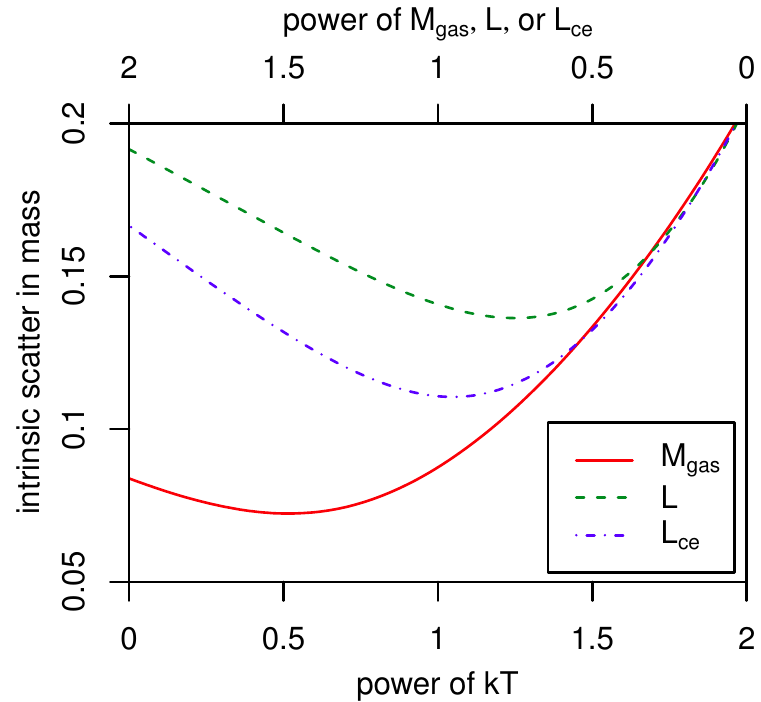}
 \caption[]{
   Left: Constraints on the power-law slopes of $\Yx = \Mgas kT$ as a function of mass and $E(z)$, as in Figure~\ref{fig:bulkslopes}.
   Right: intrinsic scatter in mass from products of observables of the form $kT^\alpha Y^{2-\alpha}$, where $Y$ is one of $\Mgas$, $L$ or $\Lce$.
 }
 \label{fig:Yx}\label{fig:combscatter}
\end{figure*}

The product of gas mass and temperature, $\Yx = \Mgas kT$, has been used extensively as a mass proxy in recent years (e.g., \citealt{Maughan0703504, Vikhlinin0805.2207, Andersson1006.3068, Menanteau1109.0953, Benson1112.5435, Brodwin1504.01397}). Given a power-law plus log-normal scatter model, the $\Yx$ scaling relation and its scatter can be derived directly from our results for the $\Mgas$ and $kT$ scaling relations, and their intrinsic covariance. The constraints on the $\Yx$ slopes appear in the left panel of Figure~\ref{fig:Yx}, and are consistent with self-similarity (since $\Mgas$ and $kT$ are, individually). In principle, masses estimated from such a combination can be no more precise than estimates from the individual measurements, accounting for the full intrinsic covariance matrix.\footnote{ Note, however, that part of the motivation for using $\Yx$ is due to the fact that, in simulations, it appears to be less sensitive to dynamical state and closer to a power-law in mass (extending to poor clusters and groups) than $\Mgas$ or $kT$ individually \citep{Kravtsov0603205}. Our sample, being restricted to massive, relaxed clusters by construction, cannot test these features.} Nevertheless, we show in the right panel of Figure~\ref{fig:combscatter} the intrinsic scatter in mass corresponding to products of observables of the form $kT^\alpha Y^{2-\alpha}$, where $Y$ is one of $\Mgas$, $L$ or $\Lce$, accounting for the intrinsic correlation or anticorrelation of the observables as measured above (e.g., \citealt{Stanek0910.1599}).\footnote{ We have chosen to have the exponents in this combination sum to 2 for convenience, but note that the mass scatter is invariant to this choice, depending only on the relative contributions of $kT$ and $Y$. The minimum of each curve in the figure is the scatter one would obtain by using the individual observables and directly accounting for their intrinsic covariance.} We see that choosing an optimal combination of observables, or simply using their intrinsic covariance as measured above, produces modestly smaller scatter compared to the individual observables involved. Neglecting uncertainties, the intrinsic scatter of mass with the combination $kT^{0.51} \Mgas^{1.49}$ is 0.072, compared with 0.084 with $\Mgas$ alone or 0.21 with $kT$ alone. Combining $kT$ with $\Lce$ yields smaller gains, predictably, with the product $kT^{1.05}\Lce^{0.95}$ yielding a mass scatter of 0.11, and $kT^{1.25} L^{0.75}$ yielding a scatter of 0.14. While it is difficult to improve on the small intrinsic scatter of $\Mgas$ in the regime of our cluster sample, we note that the combination of temperature and luminosity (especially $\Lce$) is competitive. What proxy or proxies are most useful in practice also depends on the size of the statistical and systematic uncertainties of each measurement; being weighted towards smaller radii, $\Lce$ is generally measured with smaller uncertainty than $\Mgas$ at $r_{500}$, which in turn has smaller statistical and systematic uncertainties than the temperature at $r_{500}$ (Figure~\ref{fig:allprofiles}).

\subsection{Scaling of $\fgas$}
\label{sec:fgas2500}

Paper~\cosmopaper{} presented limited results on the scaling of \fgas{}, specifically \fgas{} in a spherical shell spanning 0.8--1.2\,$r_{2500}$, and on the intrinsic scatter of \fgas{} in a few such shells when assuming no scaling with mass. Here we provide a more comprehensive treatment, differing from Paper~\cosmopaper{} in the version of the \Chandra{} calibration used and the consistent use of the same model employed in previous sections, i.e.\ including free power-law slopes with both mass and $E(z)$ as well as intrinsic scatter. These results provide a more complete picture of how the intracluster medium is redistributed, relative to the self-similar model, due to astrophysical processes.

\begin{figure*}
 \centering
 \includegraphics[scale=0.9]{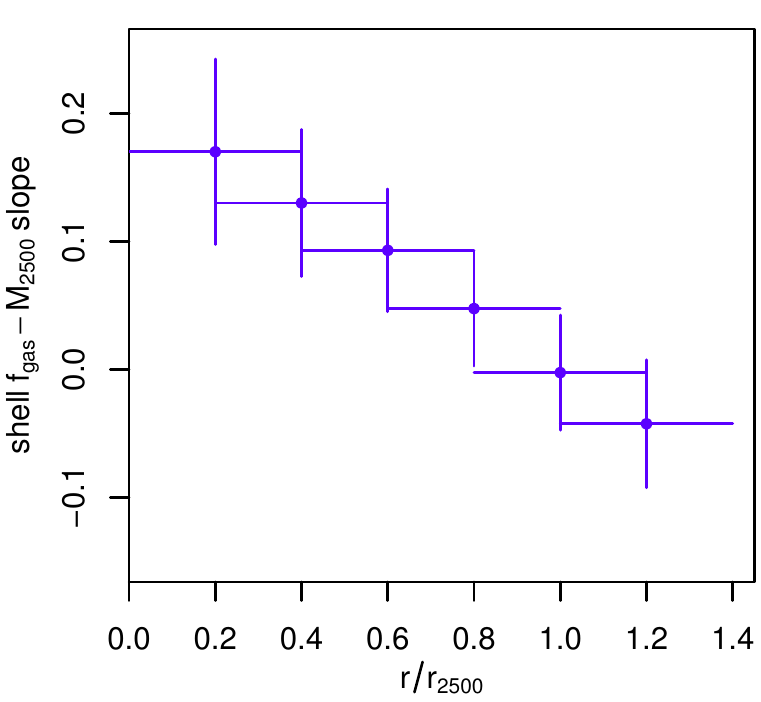}
 \hspace{1cm}
 \includegraphics[scale=0.9]{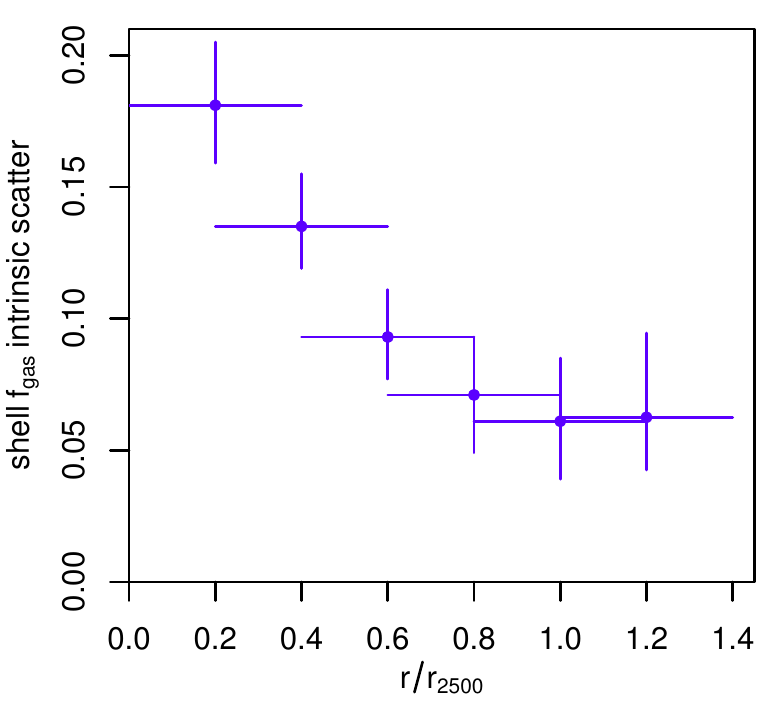}
 \caption[]{
   Constraints on the power-law scaling with mass (left) and intrinsic scatter (right) of \fgas{} measured in spherical shells. Horizontal bars indicate the radial extent of the regions where \fgas{} is measured, while vertical bars show 68.3 per cent confidence intervals. Note that the plotted points are non-trivially correlated; nevertheless, the gross trends are clear. As a function of increasing radius, out to $\sim1.2\,r_{2500}$, the slope of \fgas{} with mass systematically decreases, becoming consistent with zero at $r \gtsim 0.8\, r_{2500}$. The intrinsic scatter decreases to a minimum at $\sim r_{2500}$, and appears to flatten beyond that.
 }
 \label{fig:shellfgas}
\end{figure*}

Figure~\ref{fig:shellfgas} shows constraints on the power-law slope of \fgas{} with $M_{2500}$ and the intrinsic scatter for a series of spherical shells of radial width 0.4\,$r_{2500}$. In interpreting these plots, it must be stressed that the results across shells are non-trivially correlated due to the common $M_{2500}$ values used in the regression, as well as the use of a parametrized mass profile. Nevertheless, it is clear that the gas mass fraction at small cluster radii increases with mass, while at larger radii ($\gtsim 0.8\,r_{2500}$) the slope is consistent with zero. As observed in Paper~\cosmopaper{}, the intrinsic scatter is minimized at radii $\sim r_{2500}$, clearly increasing towards smaller radii. The average \fgas{} in each shell closely traces the differential profiles shown in Paper~\cosmopaper{} (Figure~2), while the power-law slope with $E(z)$ is consistent with zero at the $\sim1\sigma$ level in each shell. Table~\ref{tab:fgas} shows the constraints on the fit parameters for a subset of the shells, as well as for \fgas{} in full spheres bounded by $r_{2500}$ and $r_{500}$ (the latter from Section~\ref{sec:bulk}) fit with the same model.

\begin{table*}
  \centering
  \caption[]{
    Constraints on the gas-mass fraction and its scaling properties in various spherical shells. The first two columns show the inner and outer radii of each shell in units of $r_{2500}$ (however the outer radius for the last line is actually $r_{500}$ rather than a strict multiple of $r_{2500}$). The third column shows the model constraint on \fgas{} in each shell at $z=0.35$ and $M_{2500}=3\E{14}\Msun$, and columns 4--5 show the constraints on power-law indices with mass and $E(z)$. Column 6 shows the intrinsic scatter in $\ln(\fgas)$ at fixed mass and redshift.
  }
  \label{tab:fgas}
  \begin{tabular}{cccccc}
    \hline
    $r_\mathrm{in}$ & \phmin$r_\mathrm{out}$ & \fgas{} & $E(z)$ slope & \phmin$M_{2500}$ slope & scatter \\
    \hline
    0.0 & \phmin0.4 & $0.082 \pm 0.002$ & $-0.38 \pm 0.27$ & \phmin$0.17 \pm 0.07$ & $0.18 \pm 0.02$ \\
    0.4 & \phmin0.8 & $0.105 \pm 0.002$ & $-0.14 \pm 0.17$ & \phmin$0.09 \pm 0.05$& $0.09 \pm 0.02$ \\
    0.8 & \phmin1.2 & $0.129 \pm 0.002$ & $-0.15 \pm 0.17$ & \phmin$0.00 \pm 0.05$ & $0.06 \pm 0.02$ \\
    0.0 & \phmin1.0 & $0.100 \pm 0.002$ & $-0.17 \pm 0.18$ & \phmin$0.10 \pm 0.05$ & $0.11 \pm 0.02$ \\
    0.0 & $\sim2.2$ & $0.129 \pm 0.003$ & $-0.11 \pm 0.18$ & \phmin$0.04 \pm 0.05$ & $0.09 \pm 0.02$ \\
    \hline
  \end{tabular}
\end{table*}

The fact that the gas mass fractions of clusters increase with radius and are well below the cosmic average in their inner regions (e.g.\ \citealt{Allen0205007, Allen0405340, Allen0706.0033, Vikhlinin0507092}; Paper~\cosmopaper{}) provides some of the clearest evidence for feedback (even more so in galaxies; e.g.\ \citealt{McNamara1204.0006}). In contrast, hydrodynamic simulations without cooling or feedback predict a constant gas mass fraction as a function of mass and radius (\citealt{Borgani0906.4370} and references therein). Beyond this simple comparison, the results in Figure~\ref{fig:shellfgas} and Table~\ref{tab:fgas} complement those of Section~\ref{sec:bulk} in demonstrating that the gas distribution at small cluster radii has a larger scatter and greater dependence on mass than the gas properties at larger radii (in this case gas mass rather than luminosity). Interestingly, even though \fgas{} in the shell radially centered at $r_{2500}$ is constant with respect to $M_{2500}$, \fgas{} integrated in a sphere bounded by $r_{2500}$ displays an increasing slope with mass, $0.10 \pm 0.05$, due to the influence of the central regions. When integrating over the larger region bounded by $r_{500}$, this slope is consistent with zero. Note that we do not expect this result to apply to significantly less massive clusters or at the group scale, where the astrophysical processes that break self-similarity in the centers of our clusters may show their influence at larger radii (e.g.\, \citealt{Sun0805.2320}). We also note that both of the spherically integrated measurements retain a larger intrinsic scatter, due to the inclusion of cluster centers, than the asymptotic value of $\sim 0.07$ seen in the shells at $r \gtsim 0.6\,r_{2500}$. As noted in Paper~\cosmopaper{}, the small intrinsic scatter in the 0.8--1.2\,$r_{2500}$ shell, combined with the typically good signal-to-noise at these radii in \Chandra{} data, directly translates into improved cosmological constraints compared with a larger-scatter observable. The independence of mass of the \fgas{} in this shell is appealingly simple, although in principle a mass dependence can easily be incorporated into the full cosmological model (see discussion in Paper~\cosmopaper{}).

\section{Conclusion}
\label{sec:conclusions}

We present profiles of density and temperature (and the derived quantities pressure and entropy) for a sample of 40 massive, dynamically relaxed galaxy clusters, and provide the first constraints on the scaling with mass and redshift of these profiles, including their intrinsic covariance as a function of radius. We also fit scaling relations of traditional, integrated quantities: gas mass, average temperature, total soft-band luminosity, and center-excluded luminosity. This analysis includes the full multivariate intrinsic covariance matrix, and our results represent the first constraints on some of these cross terms.

Whether in terms of profiles or aperture-integrated measurements, our results support a picture in which the ICM in these massive, relaxed clusters follows self-similar scaling laws, with the exception of their innermost centers (radii $\ltsim 0.15\,r_{500}$). The luminosity (i.e., gas density) in these central regions evolves less strongly, and has a stronger dependence on mass, than the self-similar prediction, consistent with the physical effects expected from the development of cool cores and heating by central AGN. Even in this morphologically selected sample of relaxed clusters, cluster centers dominate the intrinsic scatter in X-ray luminosity at fixed mass and redshift. The center-excised luminosity has a smaller scatter, $\sim 15$ per cent, comparable to that of the average temperature. For these clusters, gas mass has the smallest intrinsic scatter, $\sim 8$ per cent, although an optimal combination of gas mass and temperature can also serve well as a total mass proxy. Consistent with the results above and in Paper~\cosmopaper{}, we find that a shell spanning $\sim 0.8$--1.2\,$r_{2500}$ is near optimal for cosmological studies using \fgas{}, due to the small scatter of the gas mass fraction at these radii.

Within the centers of our clusters, the measured trends in luminosity and the gas mass fraction indicate re-distribution of the ICM by non-gravitational processes such as cooling and feedback. At the same time, the measured entropy profiles remain regular, with small scatter, down to the smallest radii where deprojected measurements can be robustly made ($20\,\kpc \sim 0.01\,r_{200}$). The results suggest that the heating and cooling processes at work in the centers of massive clusters are tightly coupled, allowing the gas to remain stratified on these scales.

\section*{Acknowledgements}
We acknowledge support from the U.S.\ Department of Energy under contract number DE-AC02-76SF00515, and from the National Aeronautics and Space Administration (NASA) through \Chandra{} Award Numbers GO8-9118X and TM1-12010X, issued by the \Chandra{} X-ray Observatory Center, which is operated by the Smithsonian Astrophysical Observatory for and on behalf of NASA under contract NAS8-03060. ABM also received support from the National Science Foundation under grant AST-1140019.

\def \aap {A\&A} 
\def \aapr {A\&AR} 
\def \aaps {A\&AS} 
\def \statisci {Statis. Sci.} 
\def \physrep {Phys. Rep.} 
\def \pre {Phys.\ Rev.\ E} 
\def \sjos {Scand. J. Statis.} 
\def \jrssb {J. Roy. Statist. Soc. B} 
\def \pan {Phys. Atom. Nucl.} 
\def \epja {Eur. Phys. J. A} 
\def \epjc {Eur. Phys. J. C} 
\def \jcap {J. Cosmology Astropart. Phys.} 
\def \ijmpd {Int.\ J.\ Mod.\ Phys.\ D} 
\def \nar {New Astron. Rev.} 
\def \araa {ARA\&A}
\def \aj {AJ}
\def \aar {A\&AR}
\def \apj {ApJ}
\def \apjl {ApJL}
\def \apjs {ApJS}
\def \asl {Adv. Sci. Lett.} 
\def \mnras {MNRAS}
\def \nat {Nat}
\def \pasj {PASJ}
\def \pasp {PASP}
\def \science {Sci}
\def \compcom {Comput.\ Phys.\ Commun.}
\def \gca {Geochim.\ Cosmochim.\ Acta}
\def \npa {Nucl.\ Phys.\ A}
\def \plb {Phys.\ Lett.\ B}
\def \prc {Phys.\ Rev.\ C}
\def \prd {Phys.\ Rev.\ D}
\def \prl {Phys.\ Rev.\ Lett.}
\def \ssr {Space Sci.\ Rev.}

\appendix

\section{Individual Cluster Profiles}

Figure~\ref{fig:allprofiles} shows the deprojected temperature and electron density profiles of each individual cluster in our sample, as determined from the non-parametric spectral fits (red/orange crosses in each panel) and the fits that assume an \NFW{} mass profile (blue shaded boxes and lines with errors). Note that the choice of a deprojection center to optimize symmetry of the emission on large scales may result in biases in the density and temperature at very small radii. For this reason, the cluster centers are excluded from the \NFW{} fits, and there are no corresponding solutions for densities and temperatures in those regions (from the \NFW{} model). Confidence intervals at the 68.3 and 95.4 per cent level are shown, with the exception of the density profiles from the \NFW{} model, where only 95.4 per cent intervals are shown. Vertical dashed lines show the best-fitting value of $r_{2500}$ for each cluster. Radii expressed in kpc and densities are dependent on our adopted reference cosmological model. See Section~\ref{sec:spectral} for more details.

\begin{figure*}
 \centering
 \includegraphics[scale=0.75]{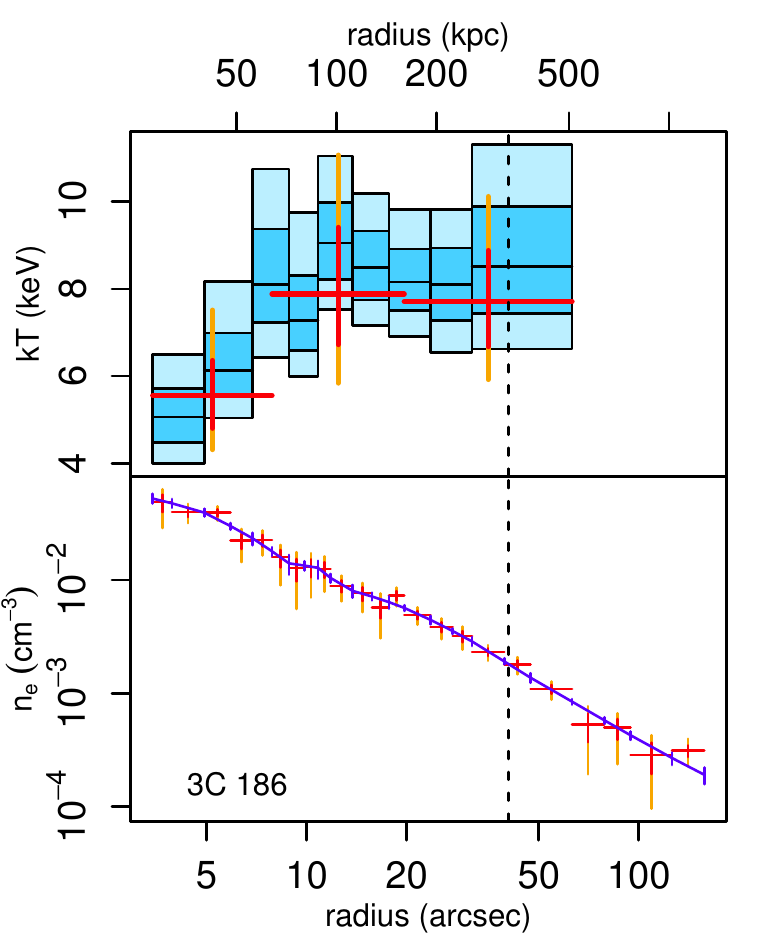}
 \includegraphics[scale=0.75]{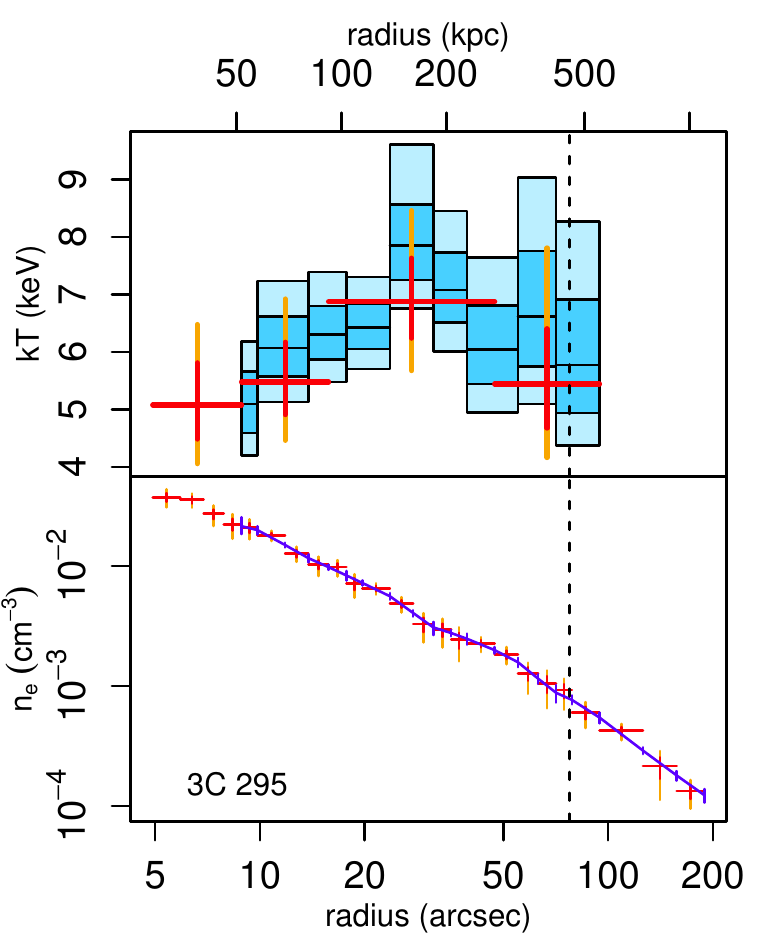}
 \includegraphics[scale=0.75]{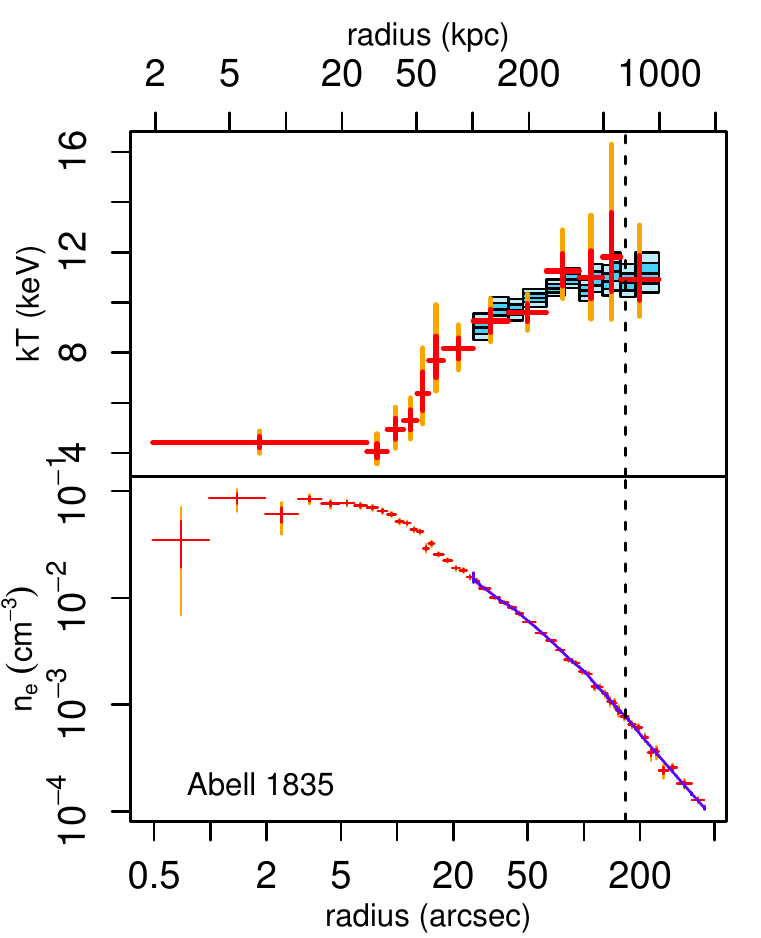}\bigskip\\
 \includegraphics[scale=0.75]{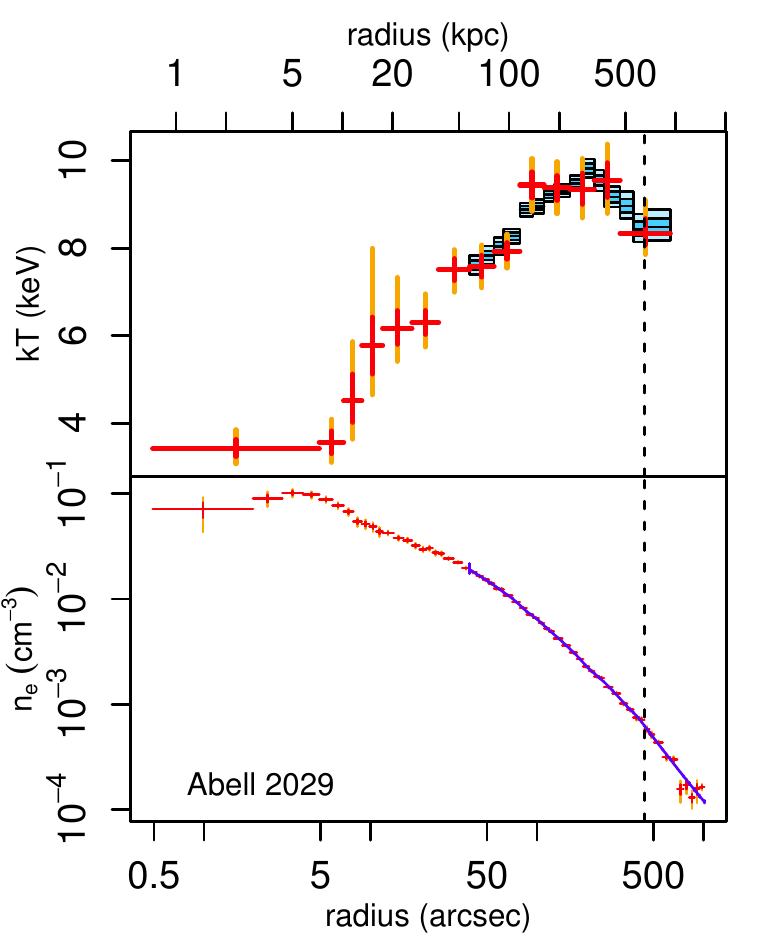}
 \includegraphics[scale=0.75]{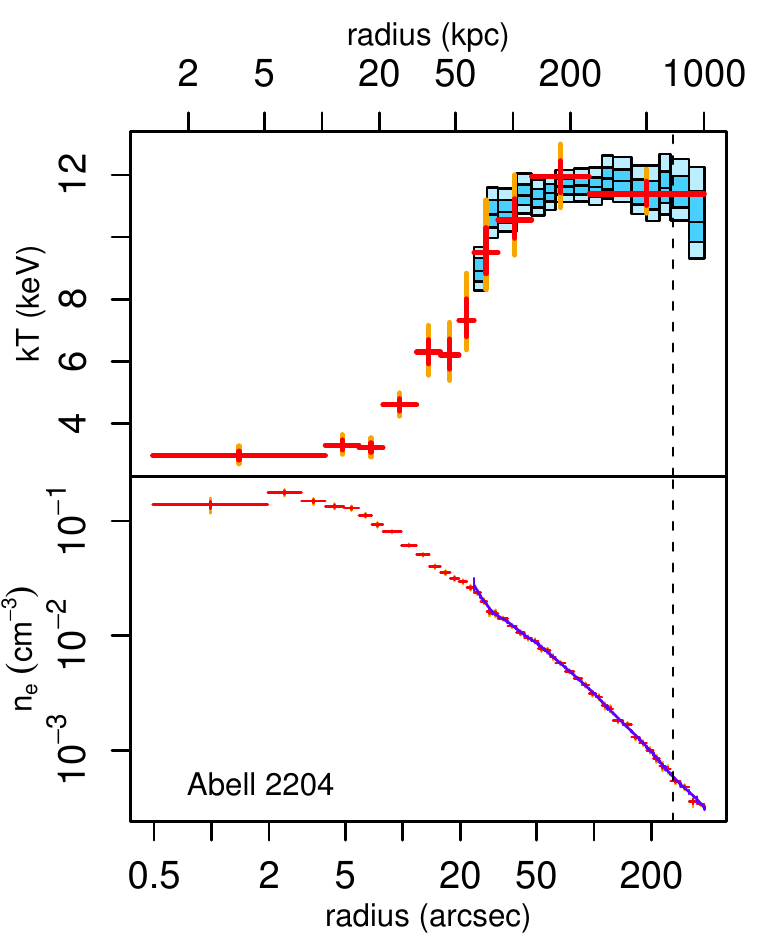}
 \includegraphics[scale=0.75]{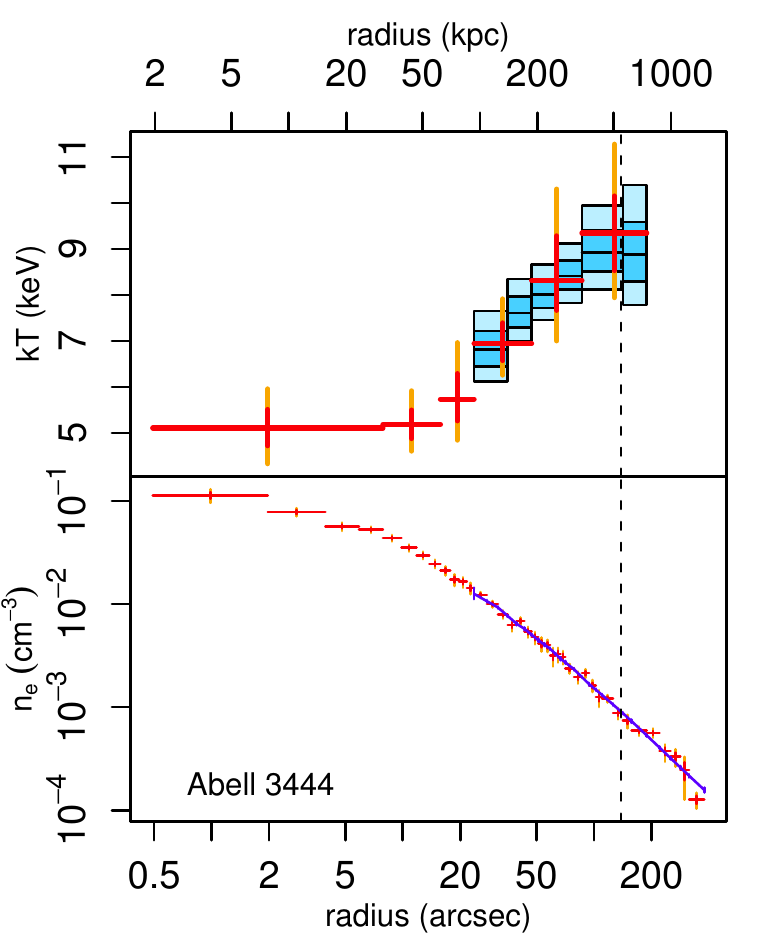}\bigskip\\
 \includegraphics[scale=0.75]{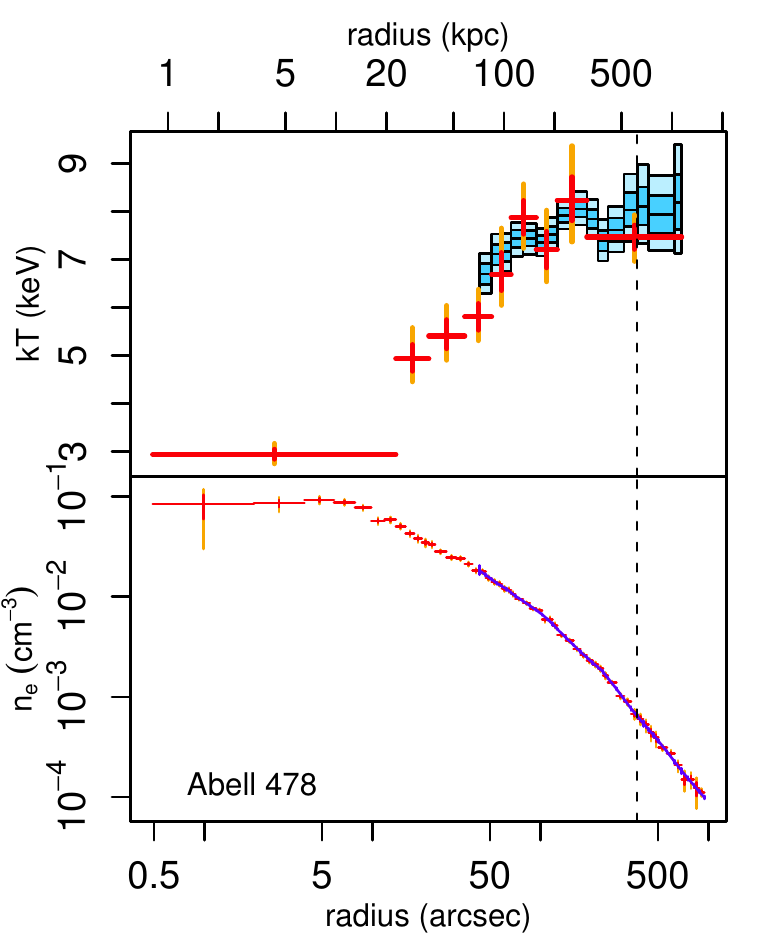}
 \includegraphics[scale=0.75]{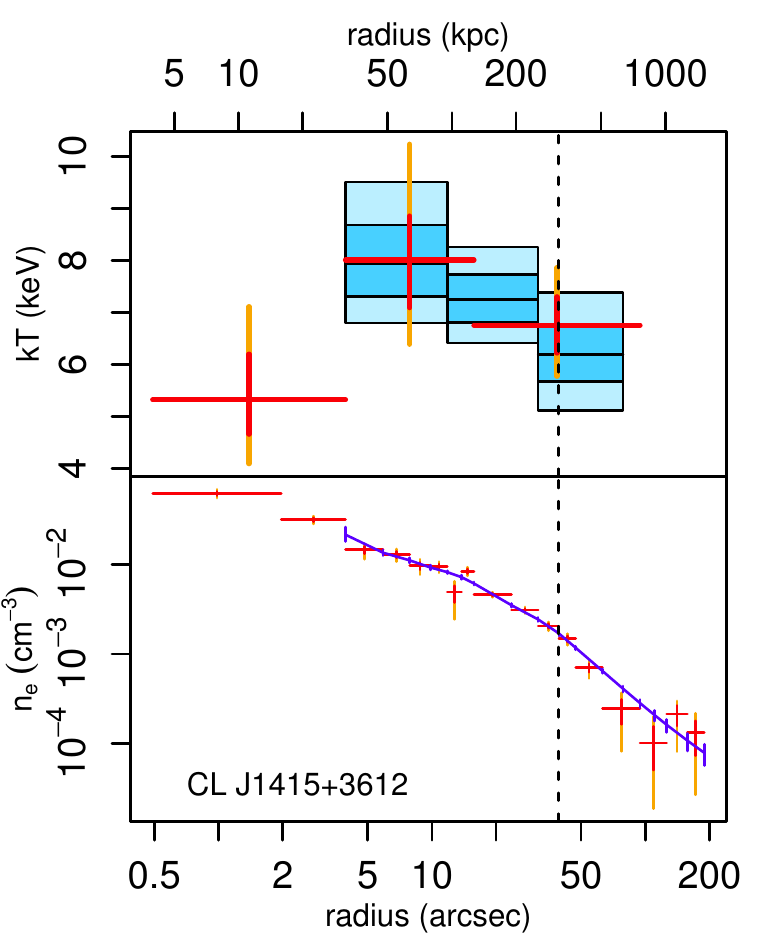}
 \includegraphics[scale=0.75]{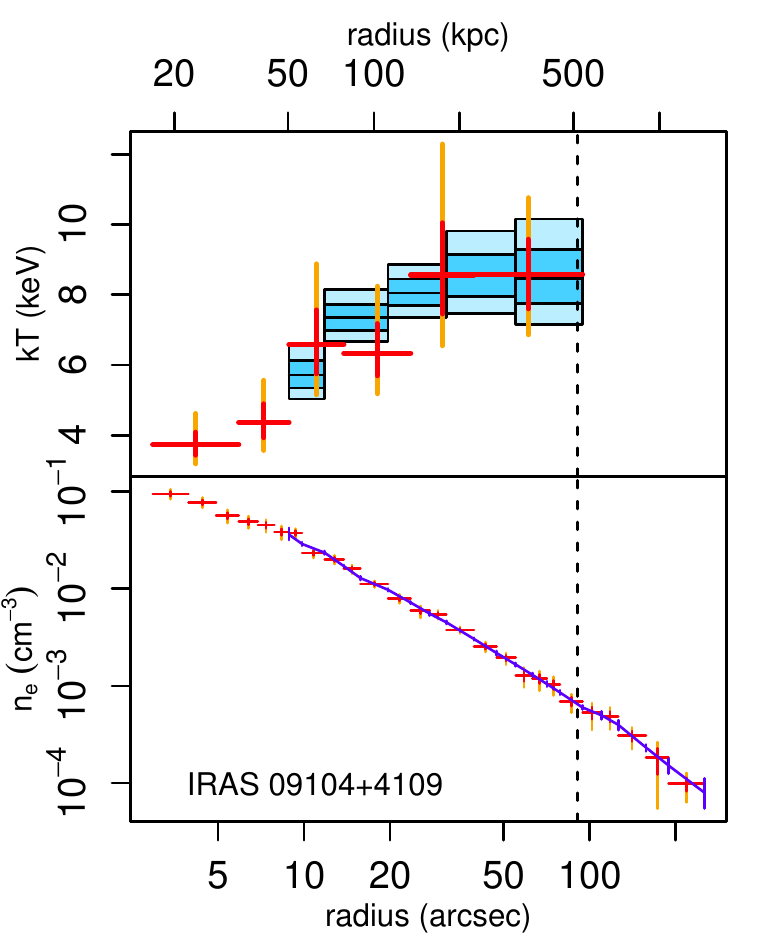}
 \caption{}
 \label{fig:allprofiles}
\end{figure*}

\begin{figure*}
 \centering
 \includegraphics[scale=0.75]{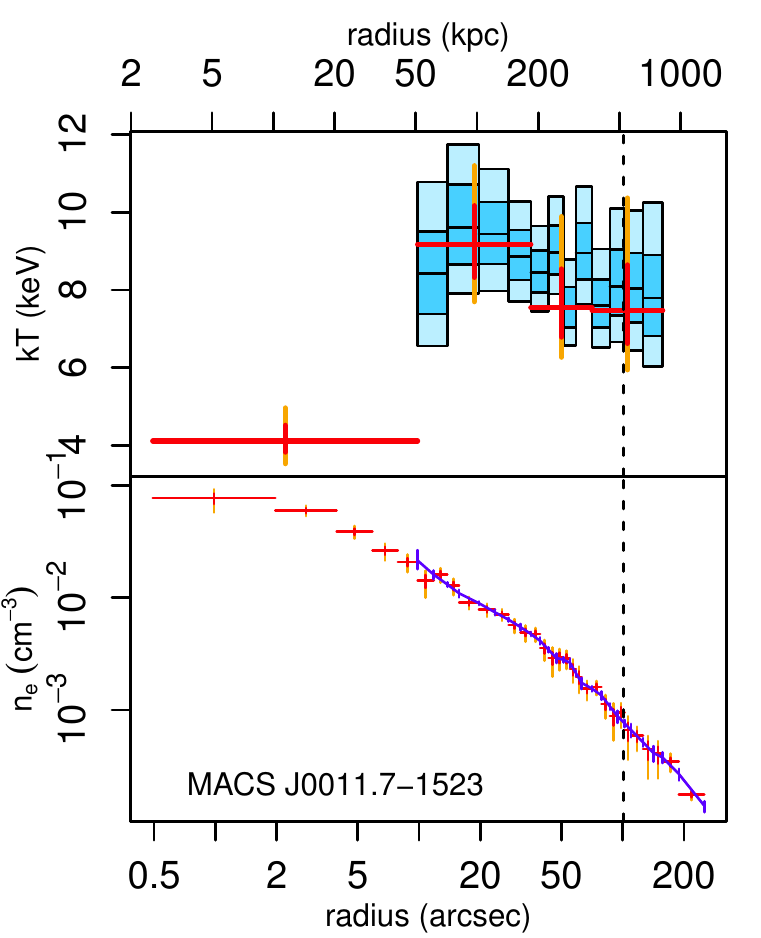}
 \includegraphics[scale=0.75]{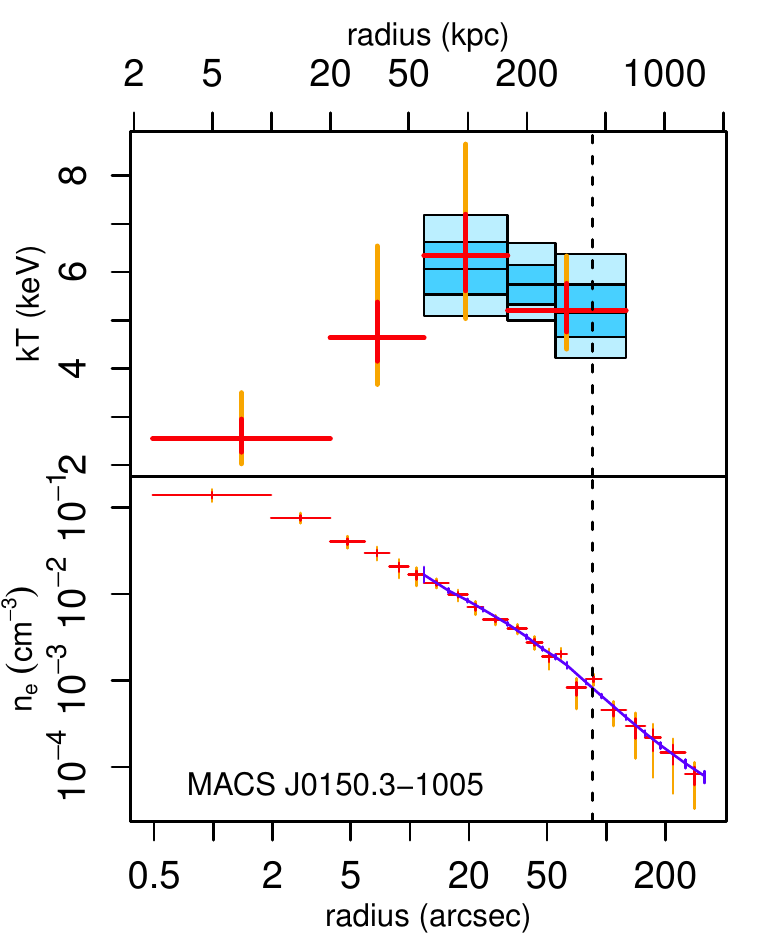}
 \includegraphics[scale=0.75]{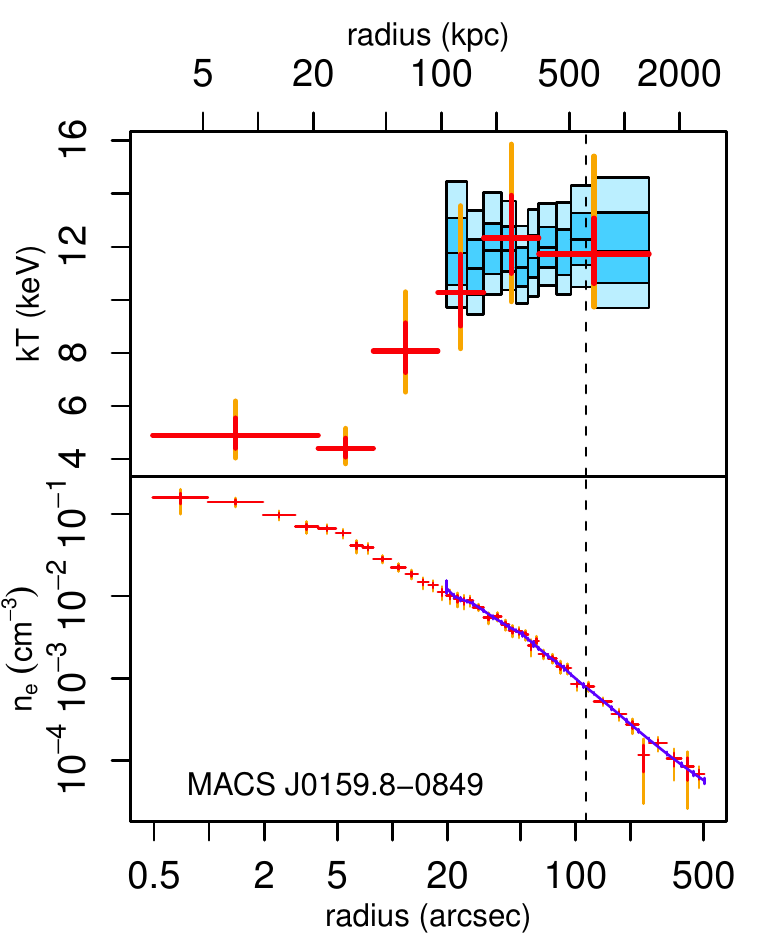}\bigskip\\
 \includegraphics[scale=0.75]{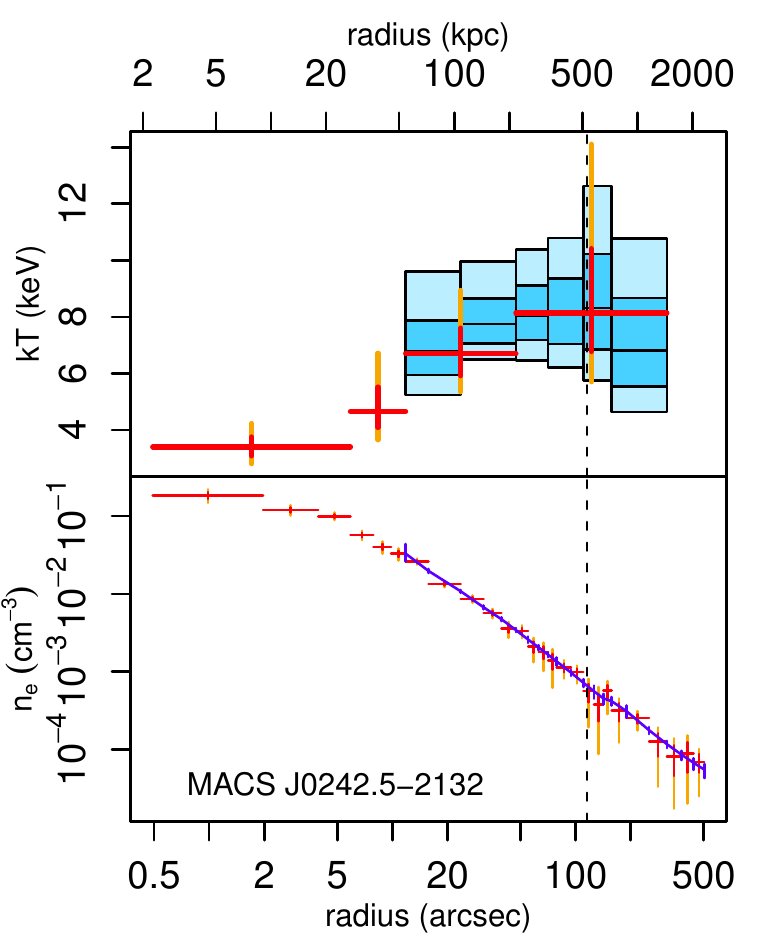}
\includegraphics[scale=0.75]{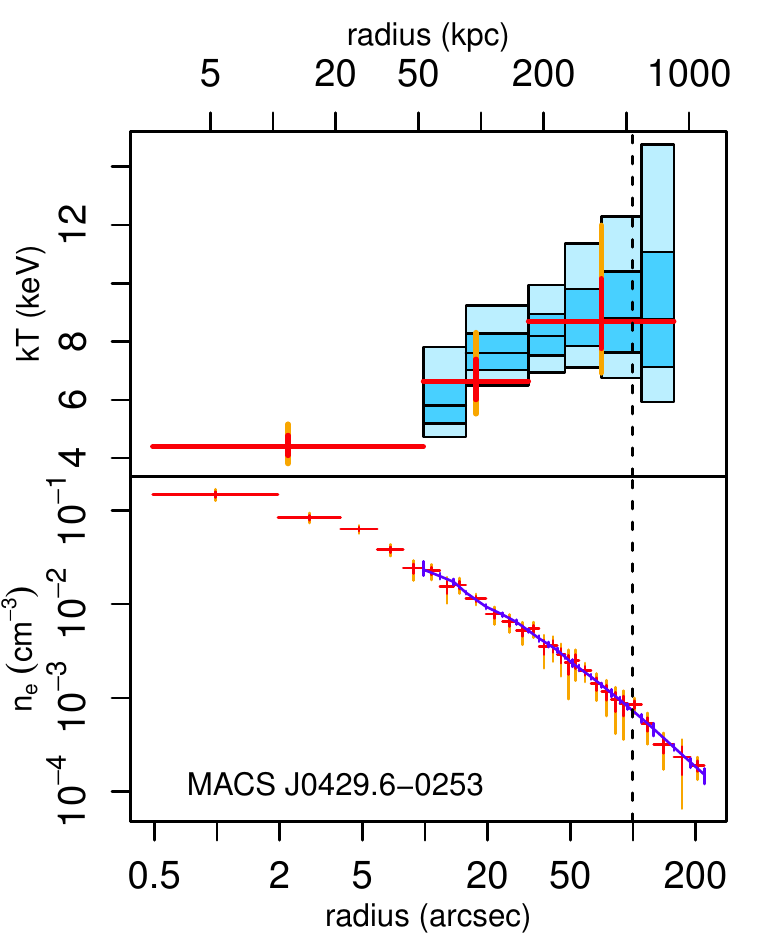}
 \includegraphics[scale=0.75]{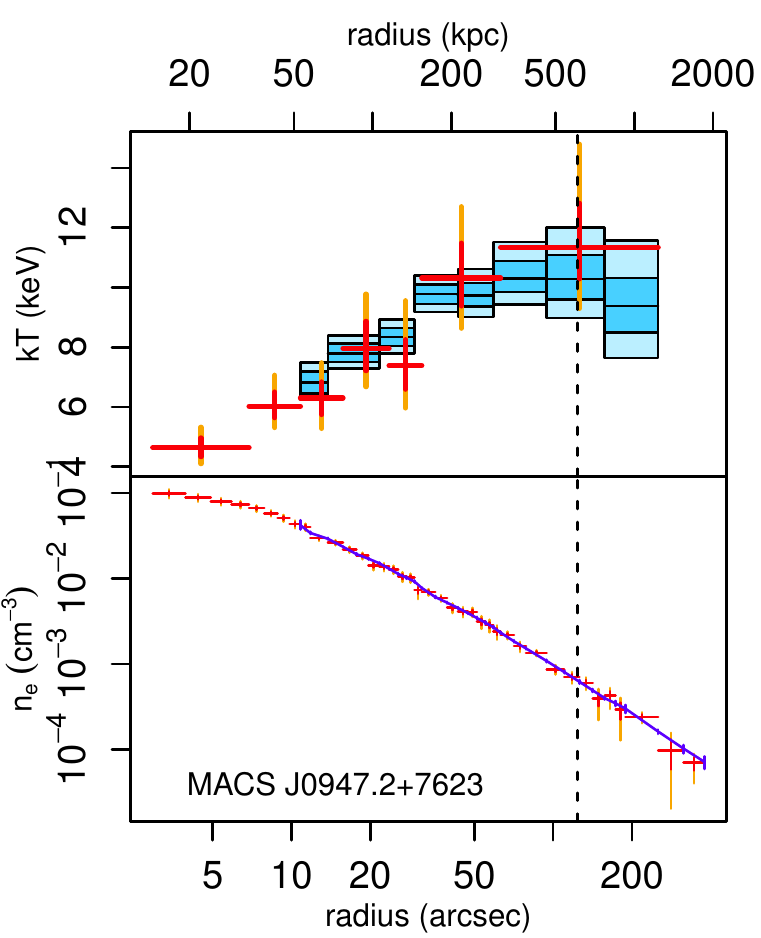}\bigskip\\
 \includegraphics[scale=0.75]{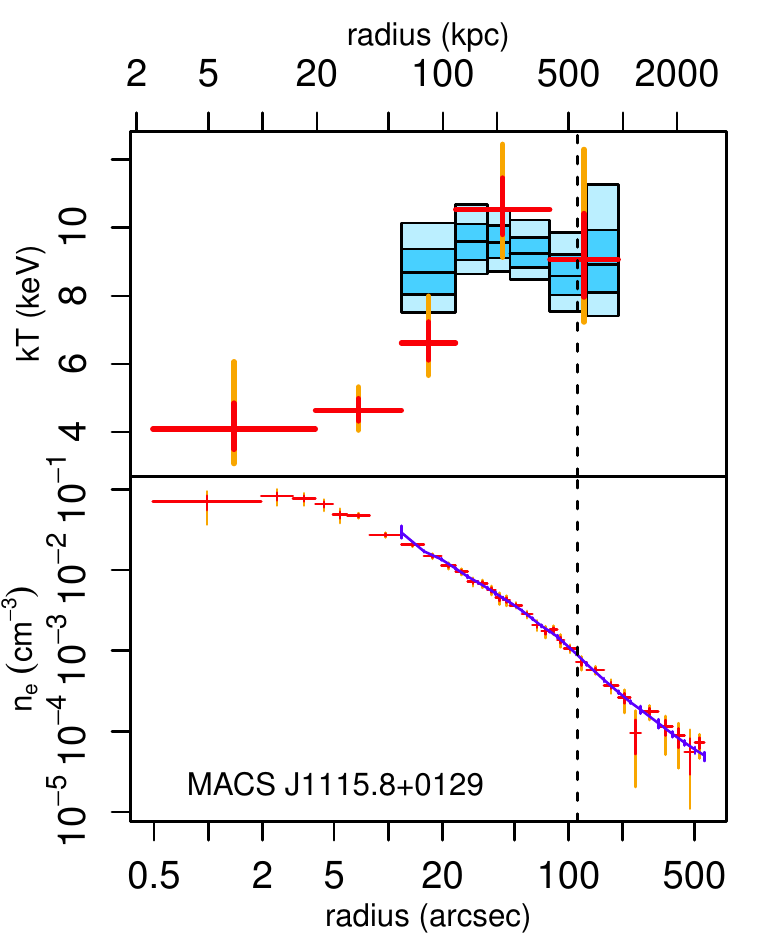}
 \includegraphics[scale=0.75]{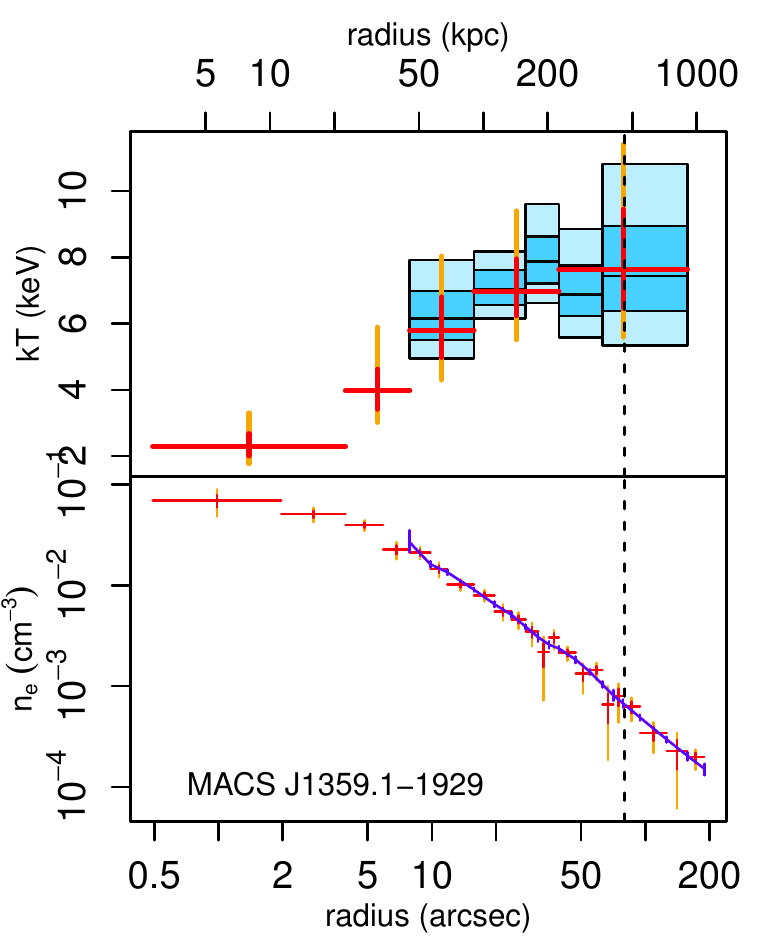}
 \includegraphics[scale=0.75]{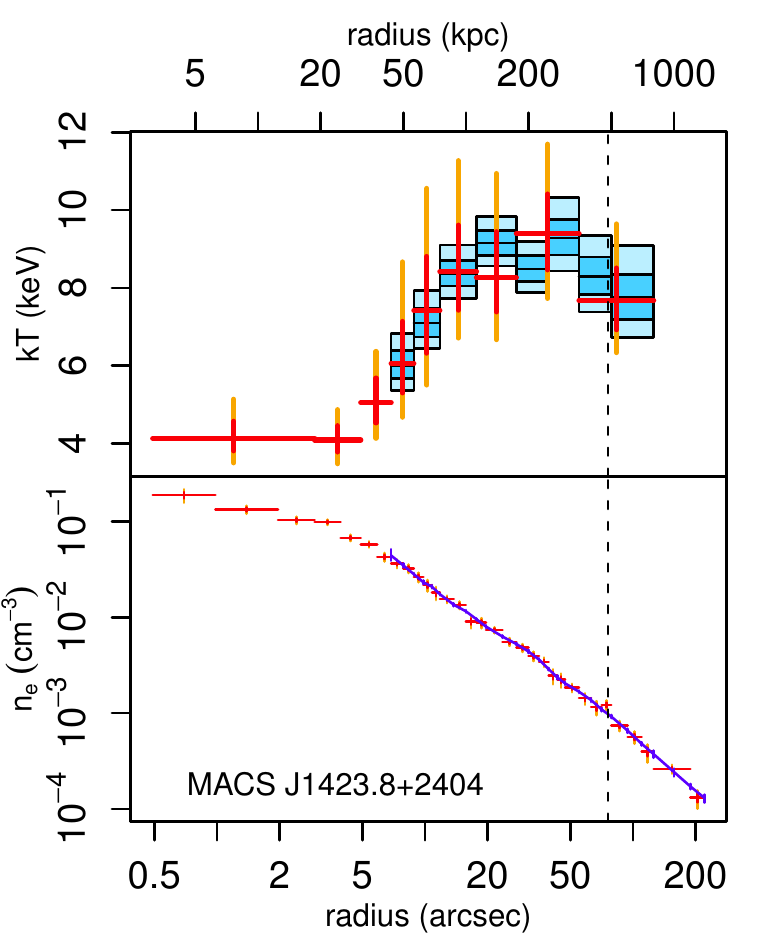}
 \contcaption{}
\end{figure*}

\begin{figure*}
 \centering
 \includegraphics[scale=0.75]{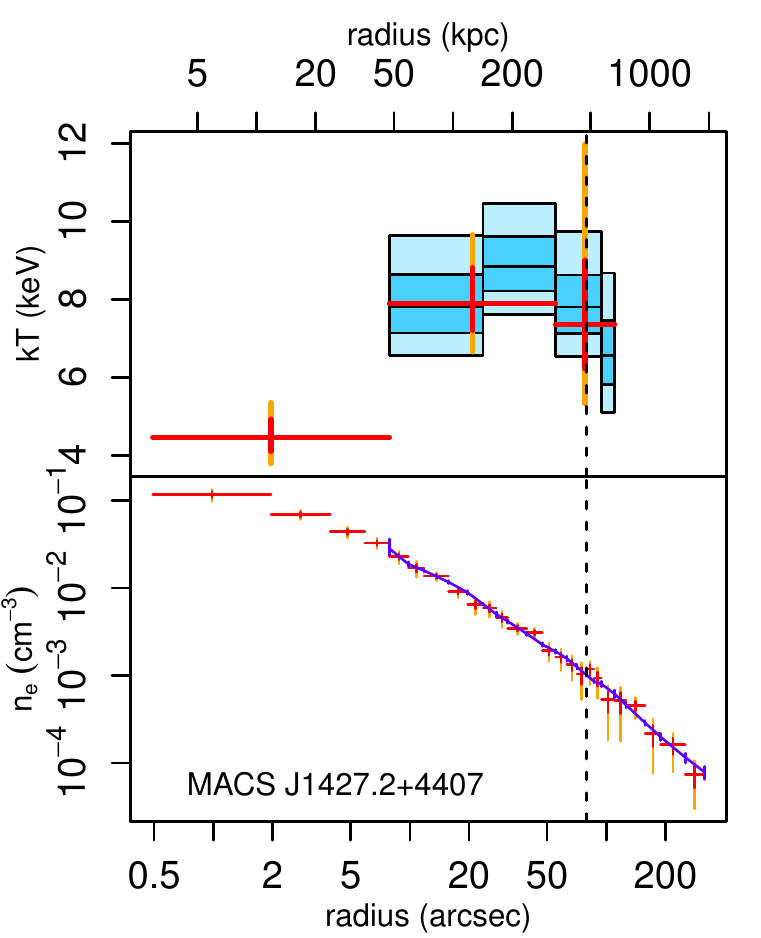}
 \includegraphics[scale=0.75]{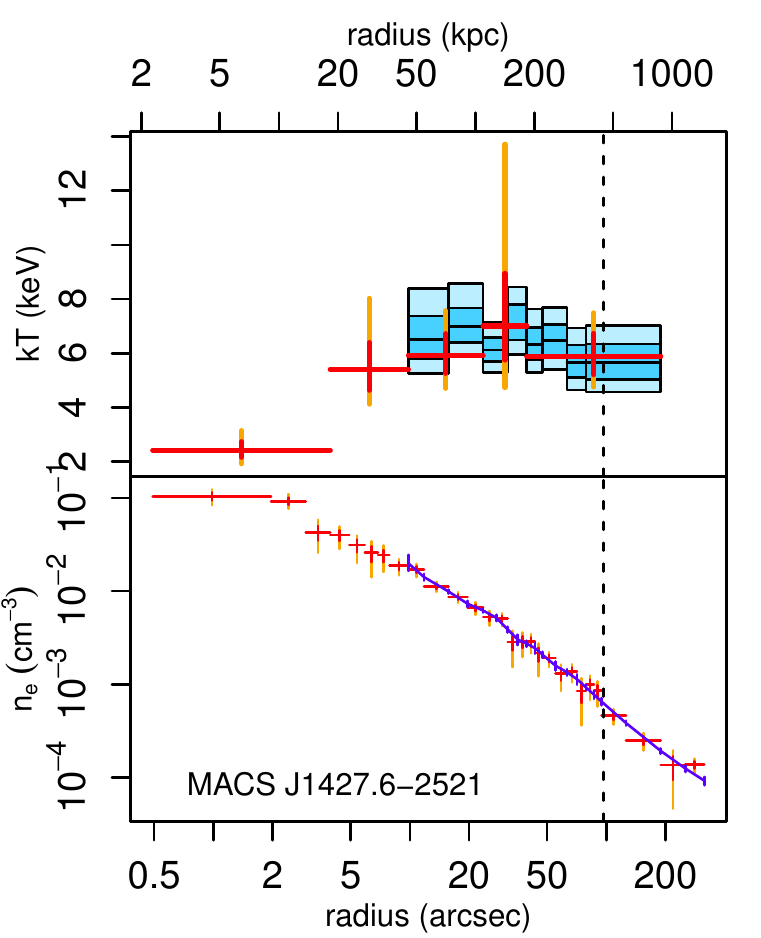}
 \includegraphics[scale=0.75]{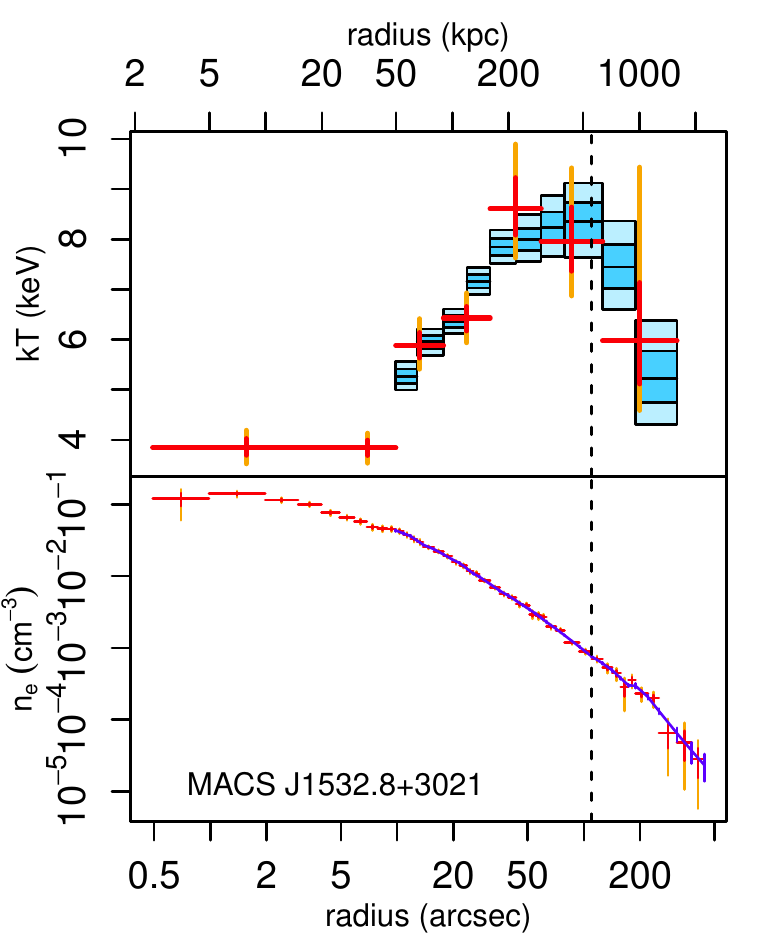}\bigskip\\
 \includegraphics[scale=0.75]{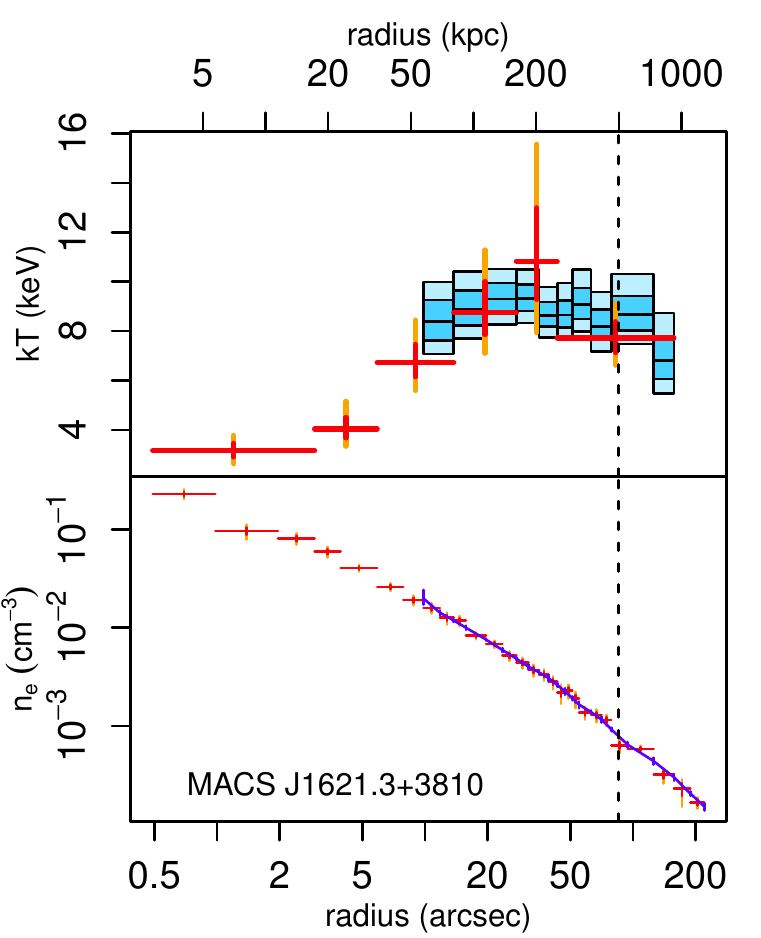}
 \includegraphics[scale=0.75]{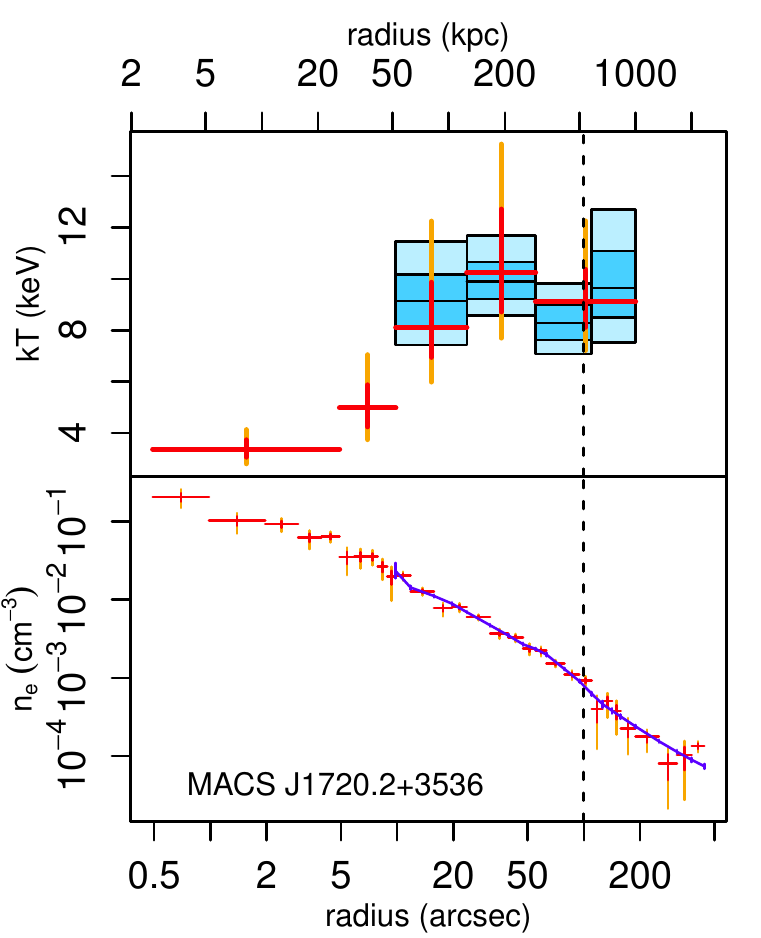}
 \includegraphics[scale=0.75]{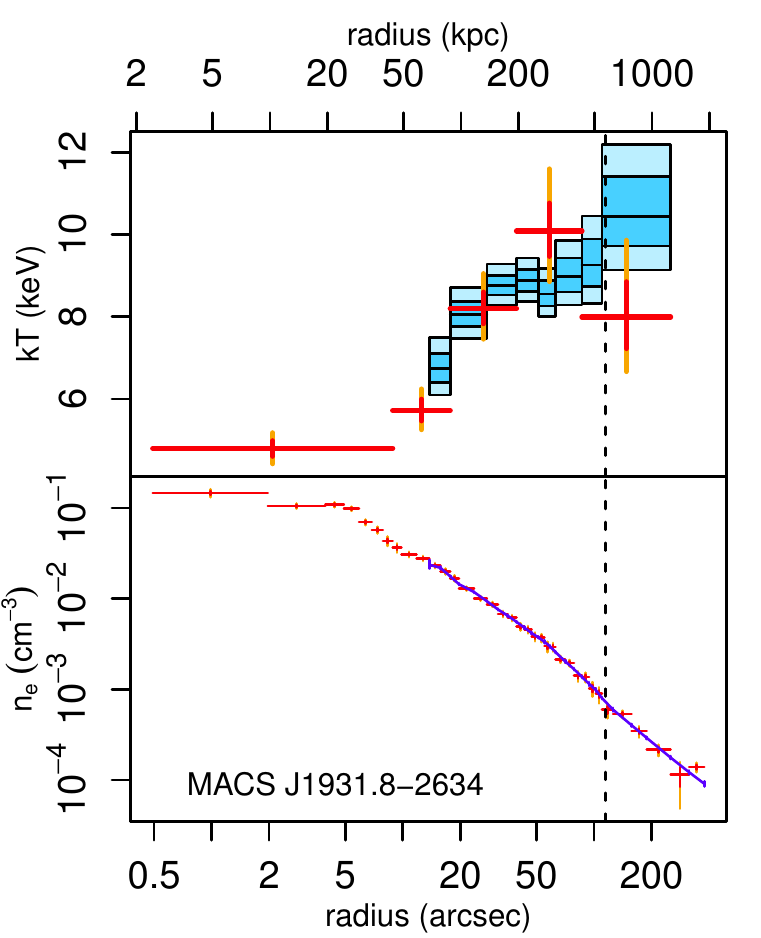}\bigskip\\
 \includegraphics[scale=0.75]{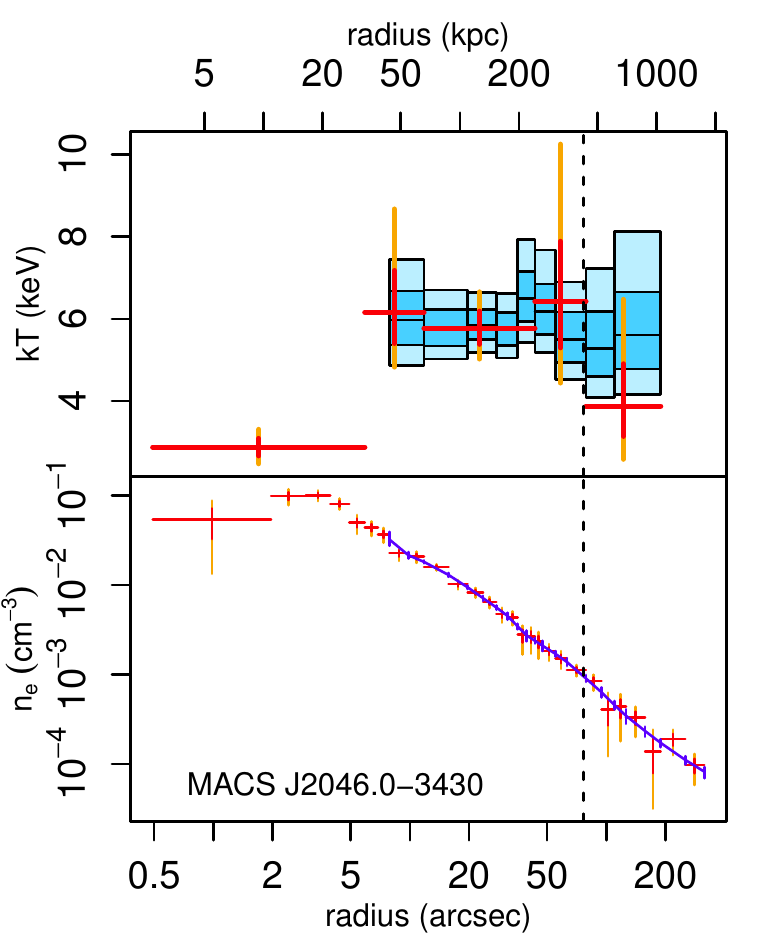}
 \includegraphics[scale=0.75]{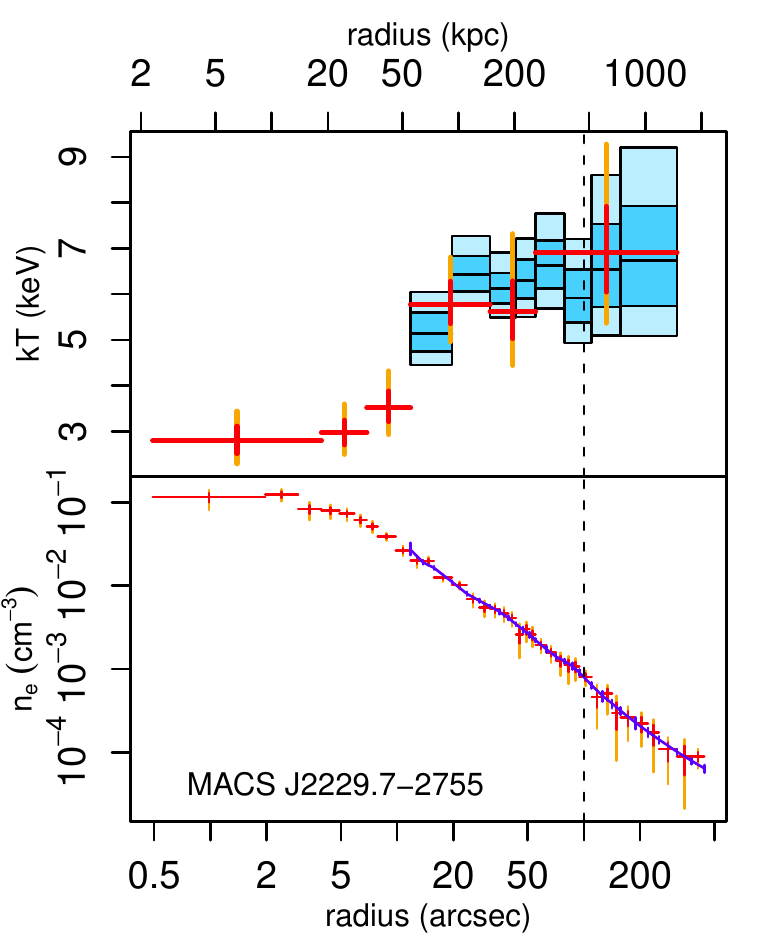}
 \includegraphics[scale=0.75]{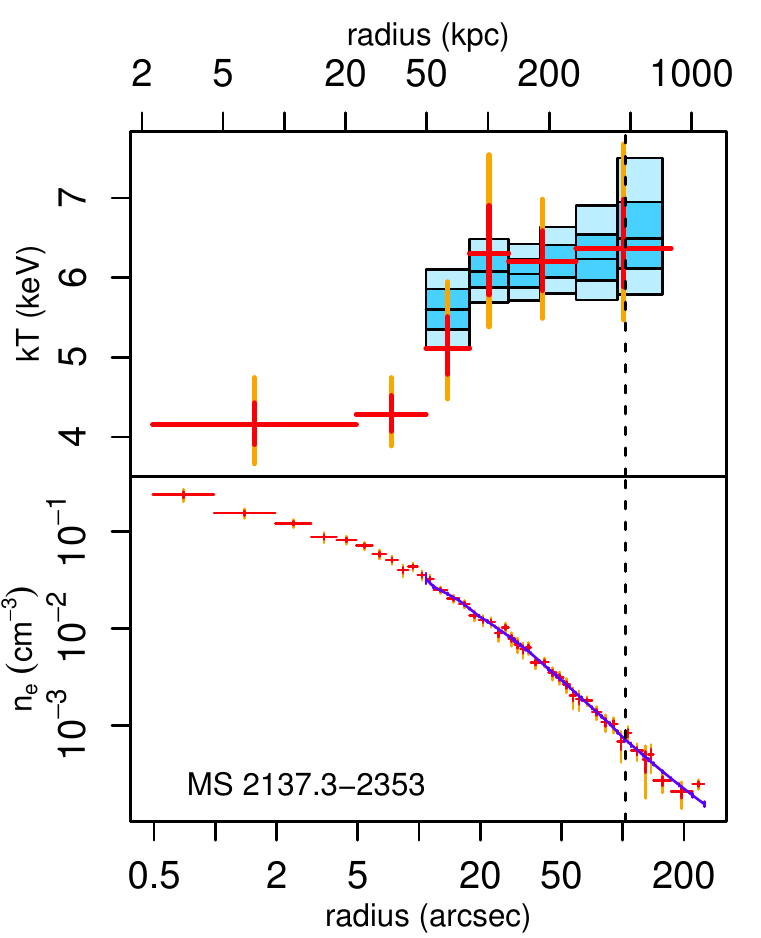}
 \contcaption{}
\end{figure*}

\begin{figure*}
 \centering
 \includegraphics[scale=0.75]{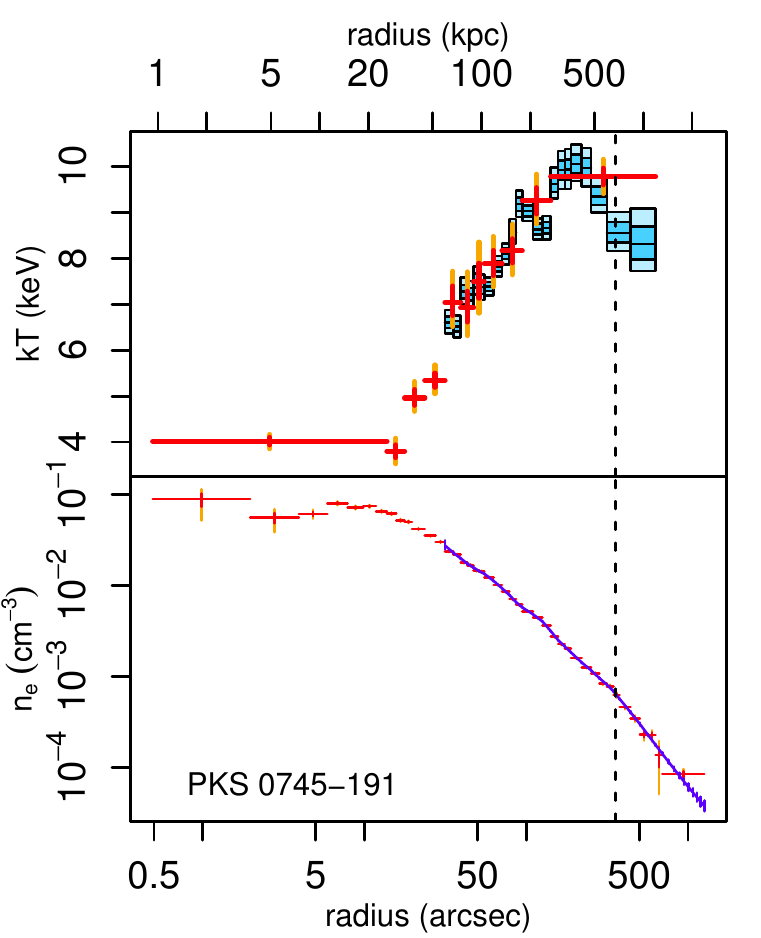}
 \includegraphics[scale=0.75]{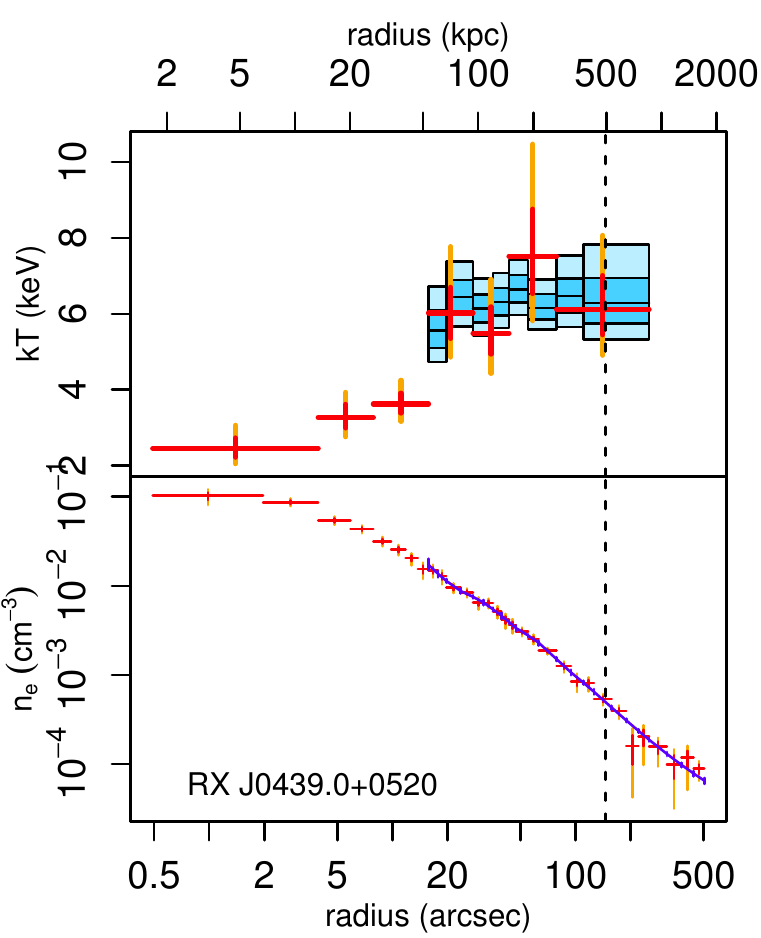}
 \includegraphics[scale=0.75]{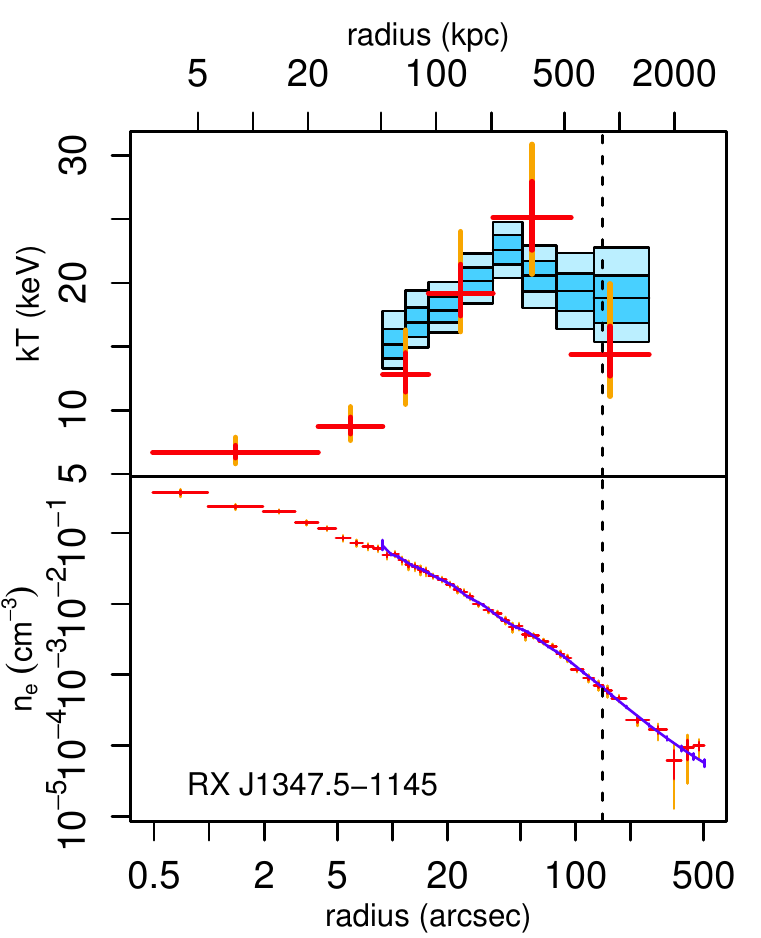}\bigskip\\
 \includegraphics[scale=0.75]{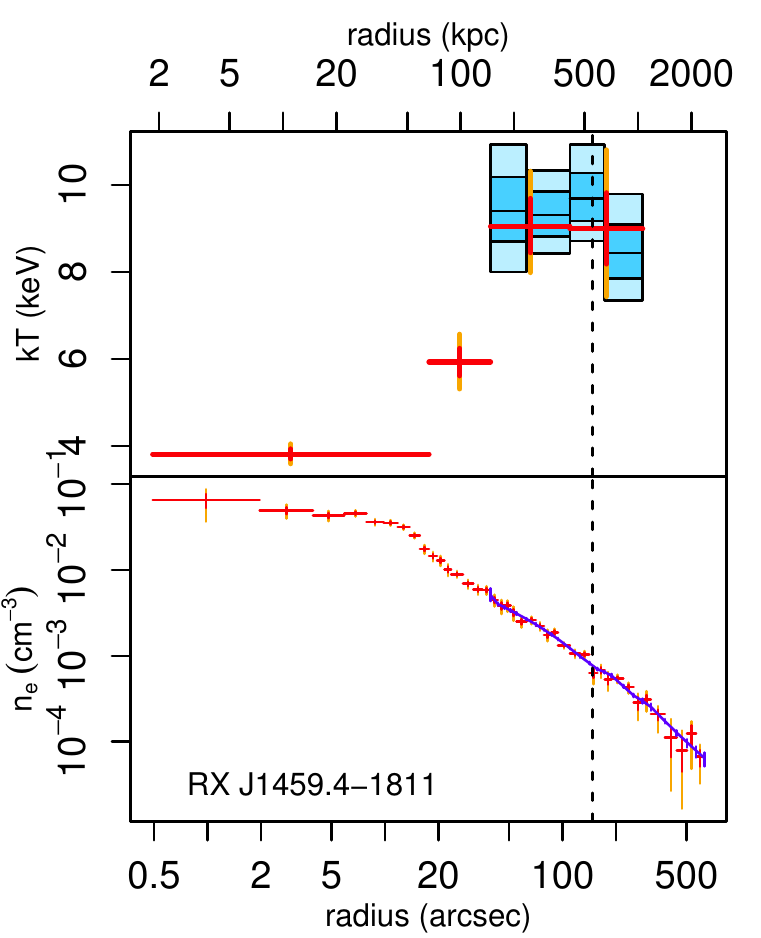}
 \includegraphics[scale=0.75]{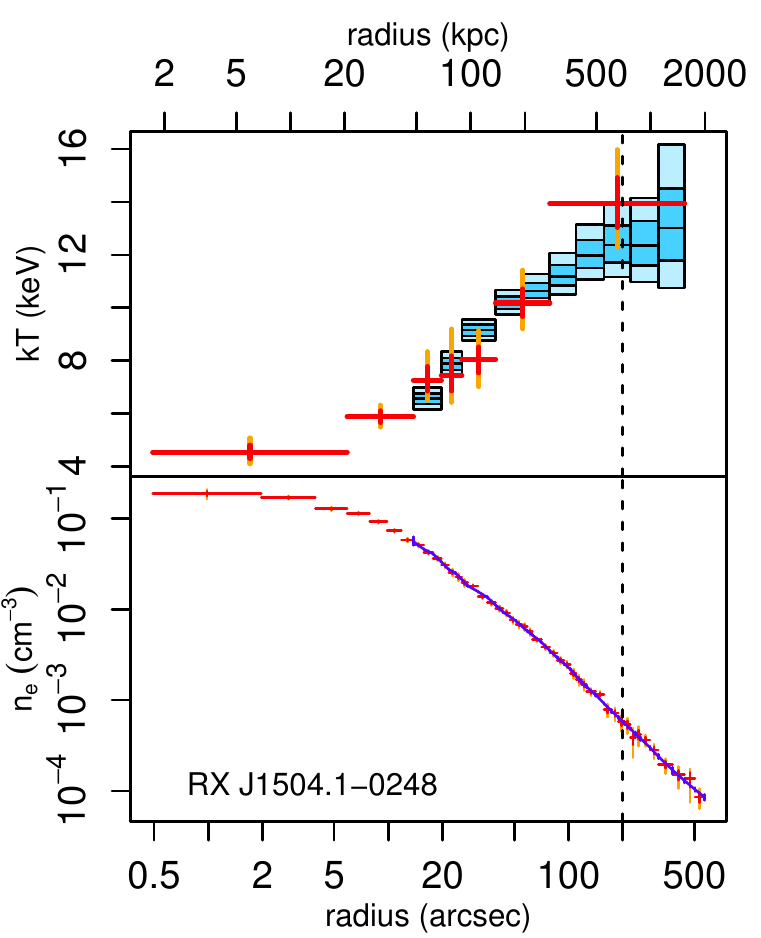}
 \includegraphics[scale=0.75]{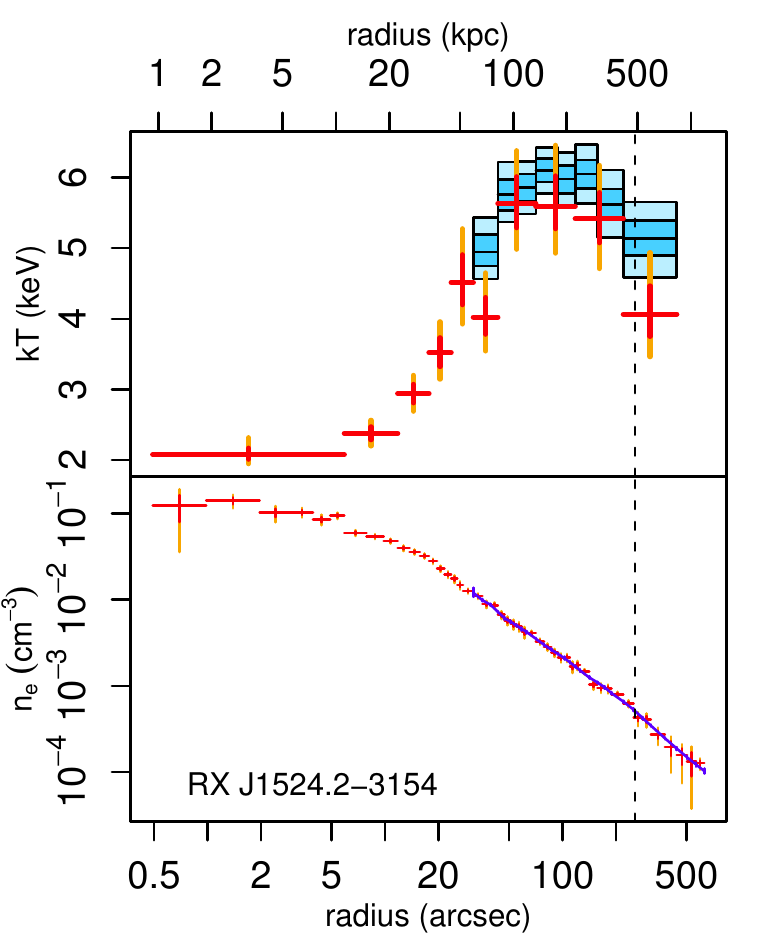}\bigskip\\
 \includegraphics[scale=0.75]{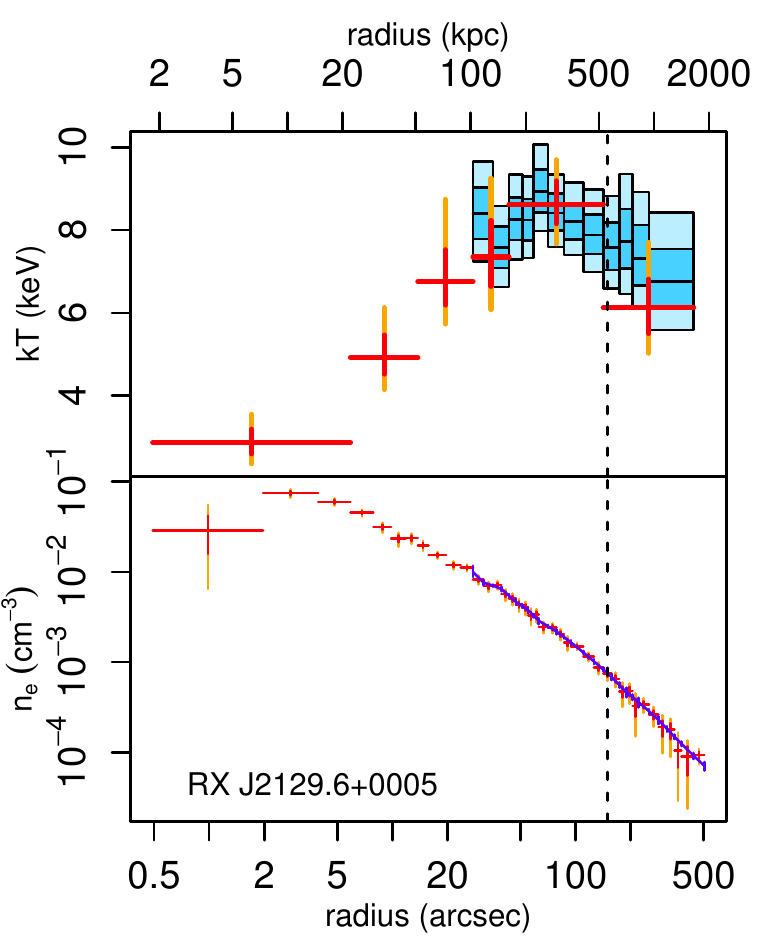}
 \includegraphics[scale=0.75]{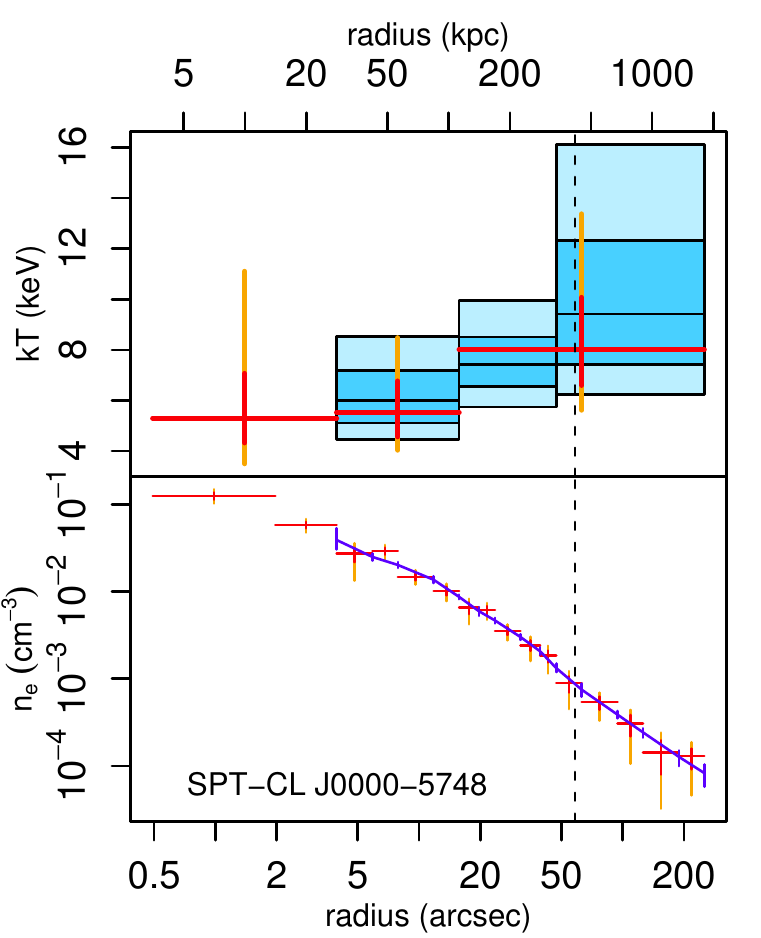}
 \includegraphics[scale=0.75]{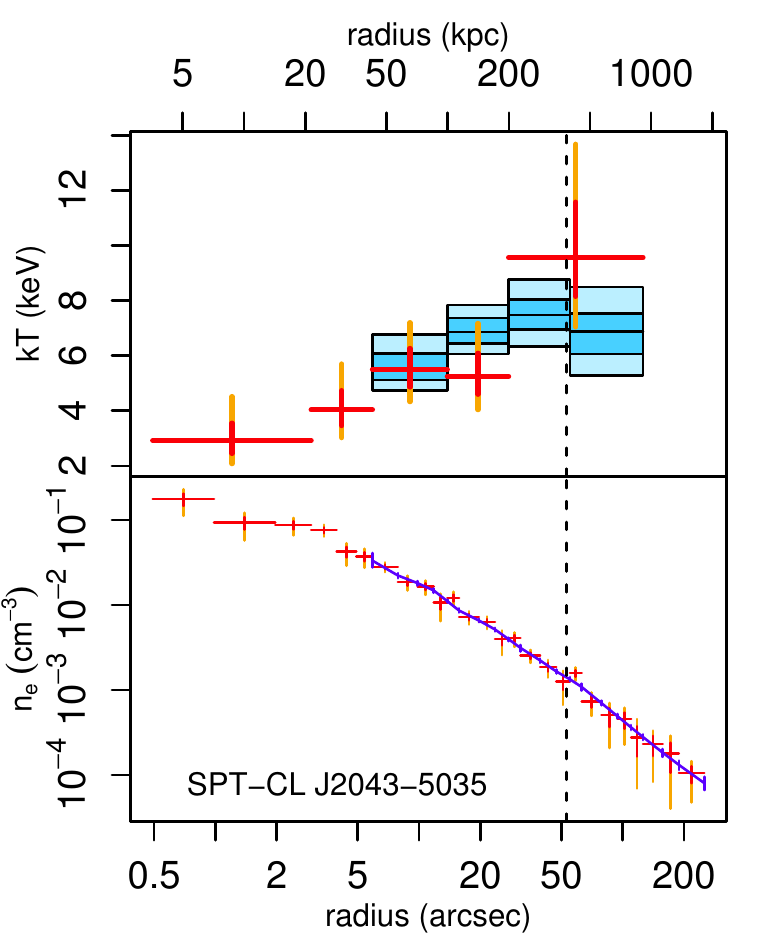}
 \contcaption{}
\end{figure*}

\begin{figure*}
 \centering
 \includegraphics[scale=0.75]{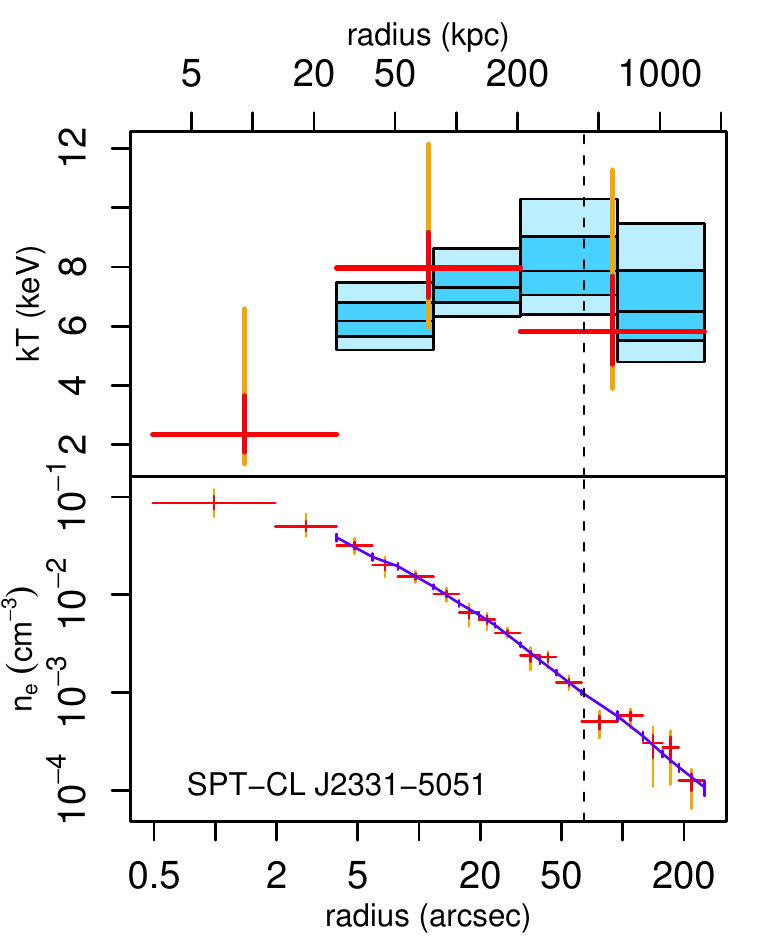}
 \includegraphics[scale=0.75]{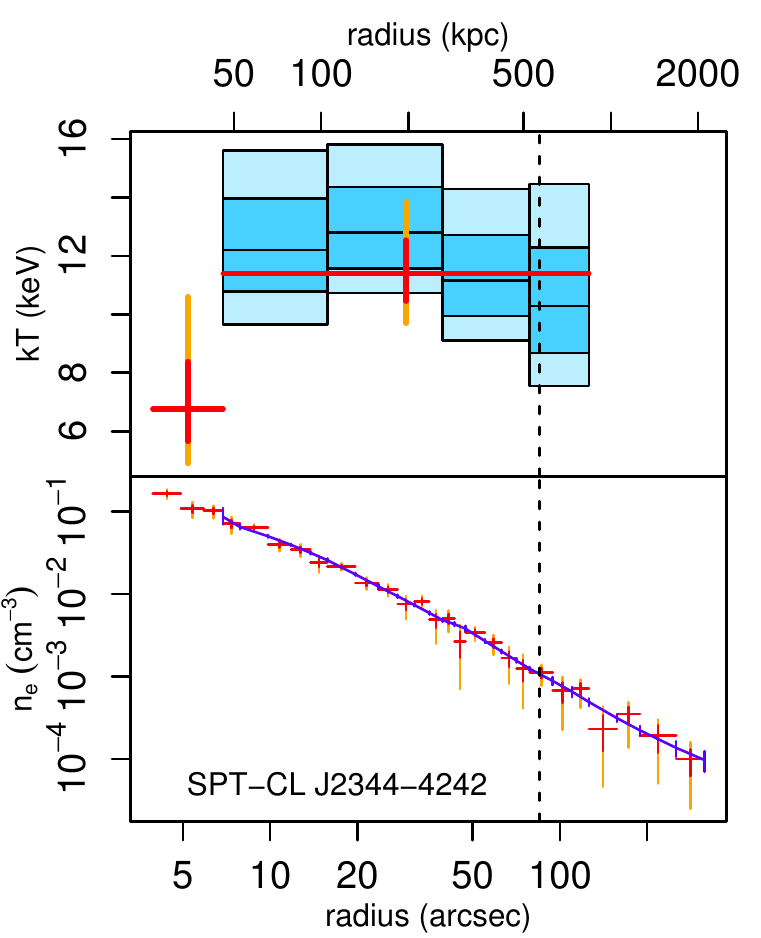}
 \includegraphics[scale=0.75]{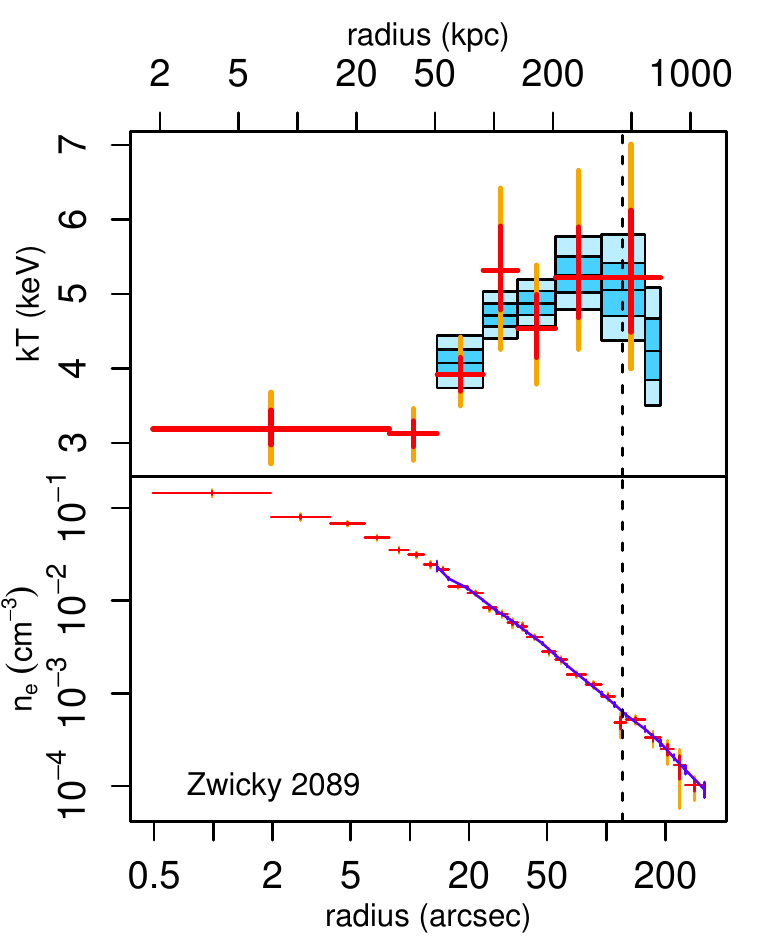}\bigskip\\
 \includegraphics[scale=0.75]{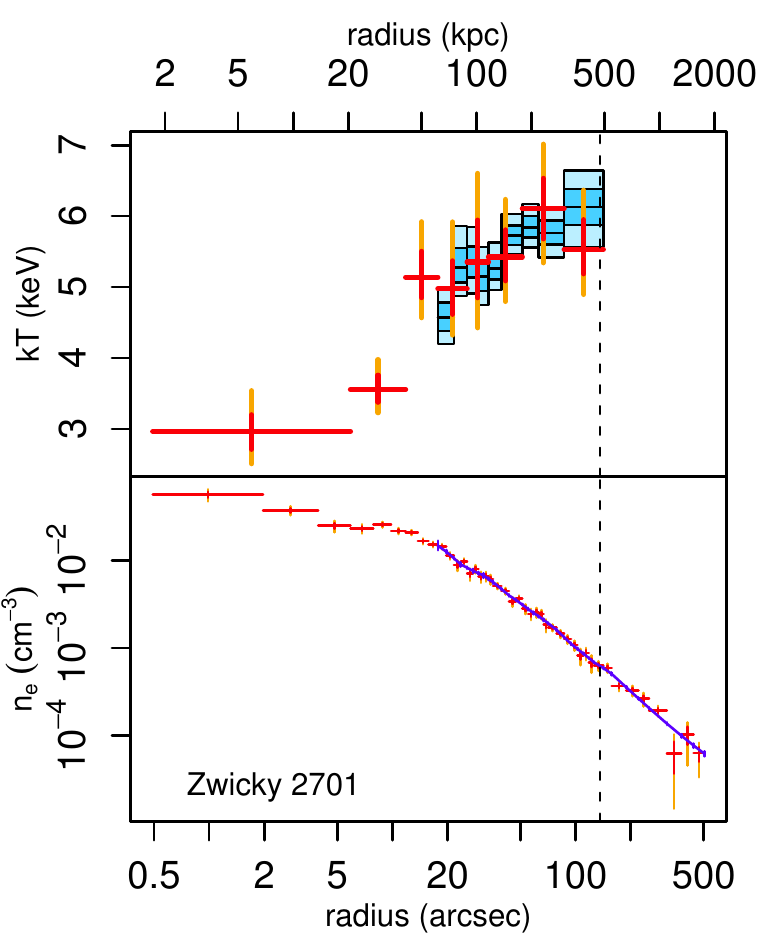}
 \contcaption{}
\end{figure*}

\label{lastpage}
\end{document}